\documentclass[12pt]{article}
\usepackage{graphicx}
\usepackage{graphics}
\usepackage[dvips]{color}
\usepackage{amssymb}
\usepackage{amsmath}
\usepackage{amsfonts}
\usepackage{mathbbol}
\usepackage{graphicx}
\usepackage{graphics}
\usepackage{amscd}
\usepackage{exscale}
\usepackage{booktabs}

\begin{document}
\setlength{\topmargin}{0.6in}
\setlength{\oddsidemargin}{0.2in}
\setlength{\evensidemargin}{0.5in}
\renewcommand{\thefootnote}{\fnsymbol{footnote}}
\centerline{\Large{\bf Anharmonic Oscillator Equations:}} 
\centerline{\Large{\bf Treatment Parallel to Mathieu Equation}} 

\vspace{1.0cm}
\begin{center}
{J.--Q. Liang\raisebox{0.8ex}{\small a}\footnote[1]
{ Email:jqliang@sxu.edu.cn}
and H. J.W. M\"uller-Kirsten\raisebox{0.8ex}{\small b}\footnote[2]
{Email:mueller1@physik.uni-kl.de}}\\

\vspace{0.5cm}

\raisebox{0.8 ex}{\small a)}{\it Institute  of Theoretical Physics, Shanxi
University, Taiyuan, Shanxi 030 006, P.R. China}

\raisebox{0.8 ex}{\small b)}{\it Department of Physics,
University of Kaiserslautern, D-67653 Kaiserslautern, Germany }

\end{center}
\vspace{2cm}

{\centerline {\bf Abstract}}
\noindent
The treatment of anharmonic oscillators (including
double wells)  by instanton
methods is wellknown. The alternative differential
equation method --- excepting various  
WKB eigenvalue approximations --- 
is not so wellknown. Here we reformulate
the latter completely  parallel to the
strong coupling case of the cosine
potential and  Mathieu equation
for which extensive literature and monographs
exist. The solutions and eigenvalues 
of the anharmonic oscillator equations are
obtained analogously as  asymptotic expansions,
and the exponentially small deviations
are shown to follow  from the
imposition of boundary conditions --- completely
parallel to the case of the Mathieu equation.    
In spite of the additional involvement
of WKB solutions, the exploitation
of parameter symmetries of the basic equation 
(as in the case of the Mathieu equation)
leads to a double well level splitting
which agrees  with the otherwise
Furry--factor improved WKB approximation
(and, of course, with calculations using
periodic instantons).

\section{Introduction}

In a recent paper on new ways to
solve the Schr\"odinger equation,
R. Friedberg and T. D. Lee \cite{1} 
described the anharmonic oscillator
case as a ``{\it long standing difficult problem
of a quartic potential with symmetric minima}''.
Quartic anharmonic oscillator potentials 
are for many  reasons
of fundamental interest in quantum mechanics and have
therefore  repeatedly been  the subject of
detailed investigations in diverse  directions.
In particular the investigations of Bender and 
Wu \cite{2}, \cite{3}
which related analyticity considerations to perturbation
theory and hence to the
large-order behaviour of the eigenvalue expansion
attracted  widespread  interest. 
An important  part of their work was
concerned with the calculation of the
imaginary part of the eigenenergy in the
non-selfadjoint  case
which permits tunneling.
Derivations of such 
a quantity are much less
familiar than calculations of
discrete bound state eigenenergies in quantum mechanical problems.
This lack of popularity of the calculation of
complex eigenvalues even in texts on quantum
mechanics may be attributed to the 
necessity of matching   various branches of eigenfunctions in
domains of overlap and to the necessary imposition of
suitable boundary conditions, both
of which make the calculation more difficult.

The development and recognition of
the significance of the path-integral method, 
particularly in connection with 
consideration of  pseudo-particle
configurations, displaced the  Schr\"odinger
equation into the  shadow of the path-integral
and suggested
that little else would be gained from
further studies of the differential equation.   
In addition, the equivalence of the
two methods (at least in the one loop
approximation) is not always  appreciated \cite{4}.

Our intention in the following is to present
 treatment of the three
different types of quartic 
anharmonic oscillators
(these types arising from different signs of
quadratic and quartic terms)
formulated  in parallel to the
strong coupling case of the cosine
potential \cite{5}, \cite{6},  the analogy
with the periodicity of the
latter entering only in the
formulation of the boundary conditions. 
Indeed the Schr\"odinger equation
with quartic oscillator potential
should be raised to the rank of a
standard differential equation
akin to the Mathieu and similar equations 
which are not  reducible to some hypergeometric
type. Thus compare the equations
\begin{eqnarray*}
&& y^{\prime\prime}(z)+[E-V(z)]y(z)=0\nonumber\\
{\rm for} \;\; V(z)&=&\left\{\begin{array}
{l@{\quad:\quad}l}
2h^2\cos\,2z=2h^2-4h^2z^2+\frac{4}{3}h^2z^4\cdots & {\rm Mathieu},\\
 az^2-bz^4 & {\rm quartic}.
\end{array}\right.
\end{eqnarray*}
Large-$h^2$ asymptotic expansions of the
Mathieu equation are wellknown,  and their
lifting of the degeneracy of the  asymptotically degenerate
oscillator levels (i.e. for $h^2\rightarrow\infty$) 
yields the boundaries
of the energy bands obtained with  the (convergent)
small-$h^2$ expansion \cite{7},\cite{8}. Books on Special Functions
of Mathematical Physics contain all sorts
of asymptotic expansions like the
large-$h^2$ expansion above, so that
there is no reason to see in the allegedly
``divergent perturbation series'' of the
anharmonic oscillator an ill-natured   problem. 
These series are completely analogous to 
asymptotic expansions of Special Functions.

The treatment of the double-well potential
by instanton methods and the calculation
of its lowest level splitting \cite{9}
constitute a  standard example of 
the path-integral method using instantons.  
The corresponding calculation
for higher levels 
requires a different method 
\cite{10}, \cite{11} \cite{12} and  is achieved  with the use
of periodic instantons \cite{13}
or other multi--instanton methods \cite{14}.  
For the inverted double-well potential
corresponding path-integral calculations 
using periodic instantons have been
carried out in Ref. \cite{15}.
It is natural therefore to compare the
Schr\"odinger and  path-integral methods  and it is reassuring
to obtain in  both cases  identical results.
Earlier WKB investigations of the eigenvalues
of the Schr\"odinger equation with double well
potential serve as valuable checks of the
results of other  methods, particularly since
these have in some cases  also been compared
with numerical methods. 
However pure WKB results do not always
agree with one-loop path-integral results \cite{9}.
Thus improved WKB 
approximations to double well eigenvalues  
have in particular
been investigated and results
in agreement with the instanton method
have been achieved. Such WKB method
investigations  can be found particularly in  
 Refs. \cite{16}, \cite{17},
\cite{18}, \cite{19} (regrettably with considerable
suppression of intermediate steps).

In the following we review the method of the 
differential equation \cite{20}, however reformulated
in parallel to the corresponding treatment of
the Mathieu equation \cite{5}
for which the perturbation method
in terms of a tunneling deviation
from an integer (in the following $q-q_0, q_0=1,3,5,\ldots,$)
was originally developed. In  our treatment here
 we leave  various detailed calculational aspects
to  appendices. 
A feature different from that of the Mathieu equation
is the additional involvement of WKB solutions
for the evaluation of the boundary conditions
at the central maximum in the
case of the double well potential. 
However, exploiting parametric  symmetries
of the original Schr\"odinger equation,
the level splitting obtained
is in complete agreement with
the Furry-factor corrected WKB result
and that of the one-loop approximation
in path-integral methods, the Furry-factor
being a correction factor to the normalisation 
constant of the WKB wave function \cite{21}.
More  detailed considerations are collected in our appendices 
which contain in addition to
calculational aspects,  some  WKB formulae for easy reference, since these
are not always available but are 
are  essential in this context and are
frequently
or  casually  referred to as
easily obtainable.

\section{The Three Types of Potenials}

In the   cases treated most frequently   in the literature
the anharmonic oscillator potential
is defined by the sum of
an  harmonic oscillator potential
and  a quartic contribution.   
These contributions may be given
different signs, and thus  lead  to very
different physical situations, which are
nonetheless linked as a consequence
of their common  origin which is for all
one and the same basic differential
equation.

\vspace{0.3cm}

\begin{figure}[ht]
\centering
\includegraphics[angle=0,totalheight=4.7cm]{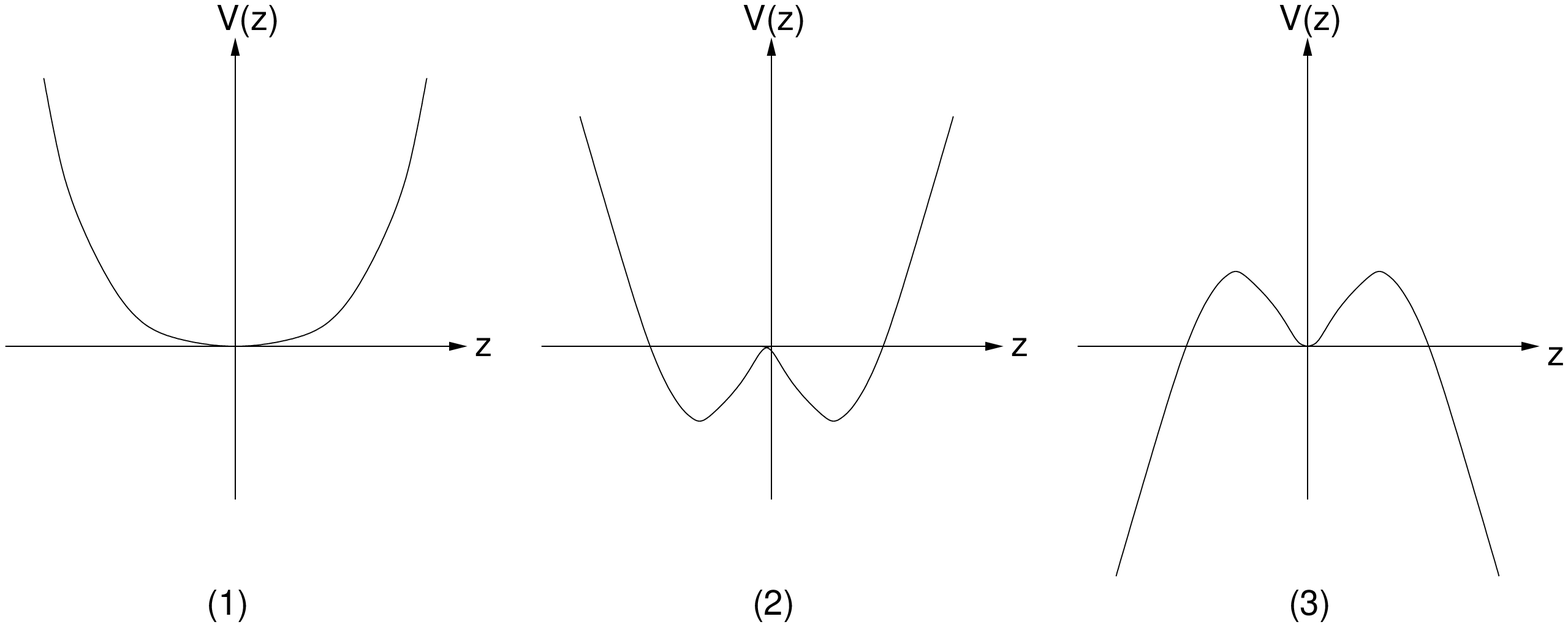}
\end{figure}
\centerline{\small Fig.~1 The three different types
of anharmonic potentials.}

\vspace{0.3cm}

\noindent  
To avoid confusion we specify first
the potential $V(z)$ in the Schr\"odinger
equation 
$$
\frac{d^2y(z)}{dz^2}+[E-V(z)]y(z)=0
$$
for  the different
cases which are possible and illustrate these
in Fig.~1. The three different cases are:

\noindent
(1)\,{\it Discrete eigenvalues with no tunneling}:
In this case
$$
V(z)=\frac{1}{4}|h^4|z^2+\frac{1}{2}|c^2|z^4,
$$

\noindent
(2)\, {\it Discrete eigenvalues with tunneling}: In this case,
described as the case of the double well potential,
$$
V(z)=-\frac{1}{4}|h^4|z^2+\frac{1}{2}|c^2|z^4,
$$

\noindent
(3)\, {\it Complex eigenvalues with tunneling}: In this case,
with the potential described as an inverted double well
potential,
$$
V(z)=\frac{1}{4}|h^4|z^2-\frac{1}{2}|c^2|z^4.
$$

Case (1) is obviously the simplest with the anharmonic
term implying simply a shift of the discrete harmonic
oscillator eigenvalues with similarly normalisable
wave functions. The shift of the eigenvalues is
easily calculated with
straightforward perturbation theory. The result is
an expansion in descending powers of $h^2$. It is
this expansion which led to a large number of
investigations culminating (so to speak) in the
work of Bender and Wu \cite{2},\cite{3} 
who established  the asymptotic
nature of the expansion\footnote{\scriptsize This is
shown by demonstrating that the $i$-th  late term  behaves
like a factorial in $i$  divided by an $i$-th  power.
See Ref.~\cite{22}.} and its Borel 
summability.\footnote{\scriptsize This is the case
when the late terms alternate in sign.} 
The resulting eigenvalues are given by Eq.~(34) below
with $q=q_0=2n+1, n=0,1,2,\ldots$ and  
$c^2$ replaced by $-c^2$,  i.e.
\begin{equation*}
E(q,h^2)=\frac{1}{2}q_0h^2+\frac{3c^2}{4h^4}(q^2_0+1)-
\frac{c^4}{h^{10}}(4q^3_0+29q_0)
+O\bigg(\frac{1}{h^{16}}\bigg).
\end{equation*}

Case (2) is also seen to allow only discrete eigenvalues
(the potential rising to infinity on either side), however
the central hump with troughs on either side permits
tunneling and hence (if the hump is sufficiently high)
a splitting of the asymptotically degenerate eigenvalues
in the wells on either side which vanishes in the
limit of an infinitely high central hump.
The resulting eigenvalues are given by Eq.~(145)
below, i.e.
\begin{equation*}
E(q_0,h^2) \simeq E_0(q_0,h^2) 
\mp 
\frac{2^{q_0+1}h^2(\frac{h^6}{2c^2})^{q_0/2}}{\sqrt{\pi}2^{q_0/4}
[\frac{1}{2}(q_0-1)]!}
e^{ -\frac{h^6}{6\sqrt{2}c^2}}, \;\; q_0=1,3,5,\ldots,
\end{equation*}
where 
$E_0(q_0,h^2)$ is given by Eq.~(\ref{95}) or   Eq.~(\ref{C.1}), i.e.
$$
E_0(q_0,h^2)=-\frac{h^8}{2^5c^2}+\frac{1}{\sqrt{2}}q_0h^2
-\frac{c^2(3q^2_0+1)}{2h^4}+O\bigg(\frac{1}{h^{10}}\bigg).
$$

Case (3) is seen to be very different from the first two cases,
since the potential decreases without limit on either side
of the centre. The boundary conditions are non-selfadjoint
and hence the eigenvalues are complex.  This type of potential
allows tunneling through the barriers and hence a passage
out to infinity so that a current can be defined. If the
barriers are sufficiently high we expect the states in
the trough to approximate  those of an harmonic oscillator,
however with decay  as a consequence of tunneling.      
The resulting complex eigenvalues are given by Eq.~(76)
together with  Eq.~({34}), i.e.
\begin{equation*}
E=E_0(q_0,h^2)\stackrel{+}{(-)}i\frac{2^{q_0}h^2
\bigg(\displaystyle \frac{h^6}{2c^2}\bigg)^
{{q_0}/2}}
{(2\pi)^{1/2}[\frac{1}{2}(q_0-1)]!}{\displaystyle e}^{
\displaystyle -\frac{h^6}{6c^2}}
\end{equation*}
with 
\begin{equation*}
E_0(q_0,h^2)=\frac{1}{2}q_0h^2-\frac{3c^2}{4h^4}(q^2_0+1)-
\frac{c^4}{h^{10}}(4q^3_0+29q_0)
+O\bigg(\frac{1}{h^{16}}\bigg).
\end{equation*}

The question is therefore:
  How does one calculate the eigenvalues in these
cases with the
help of the Schr\"odinger equation?
 This is  the questions we address here,
and we present a fairly complete treatment of the case
with large values of $h^2$ along lines 
parallel to those in the case  of the cosine
potential in Ref. \cite{5}.
We do not dwell on Case (1) since this is effectively
included in the first part of Case (3),  except for a change
of sign of $|c^2|$. Thus we are mainly concerned
with the double well potential and its 
inverted form.\footnote{\scriptsize In each of these
two cases we have to take two different boundary
conditions into account. In the cosine case
of Ref. \cite{5} only one type is required
at $z=\pi/2$. The
reason is that this is actually derived from
the boundary condition at $z=0$ using the
periodicity of this case; see Ref. \cite{7},
equations (6), p. 108, and the relations (9), 
p.100.}  
We begin with the second. In this case our aim is
to obtain the aforementioned complex eigenvalue.
In the case of the double well potential our aim
is to obtain the separation of harmonic oscillator
eigenvalues as a result of tunneling between the two wells.
Calculations of complex eigenvalues
(imaginary parts of eigenenergies) are rare in
texts on quantum mechanics.  We therefore consider
here
in detail a prominent example  and in such
a way, that the general applicability
of the method becomes evident.

\section{The Inverted Double--Well Potential}

\subsection{Defining the problem}
We consider the case of the inverted double-well
potential depicted as Case (3) in Fig.~1.
The potential in this case is given by
\begin{equation}
V(z)=-v(z), \;\;\; v(z)=-\frac{1}{4}h^4z^2+\frac{1}{2}c^2z^4,
\label{1}
\end{equation}
for $h^4$ and $c^2$ real and positive,
and the Schr\"odinger equation to be
considered is
\begin{equation}
\frac{d^2y}{dz^2}+[E+v(z)]y=0.
\label{2}
\end{equation} 

\vspace{0.3cm}
\begin{figure}[ht]
\centering
\includegraphics[angle=0,totalheight=7.5cm]{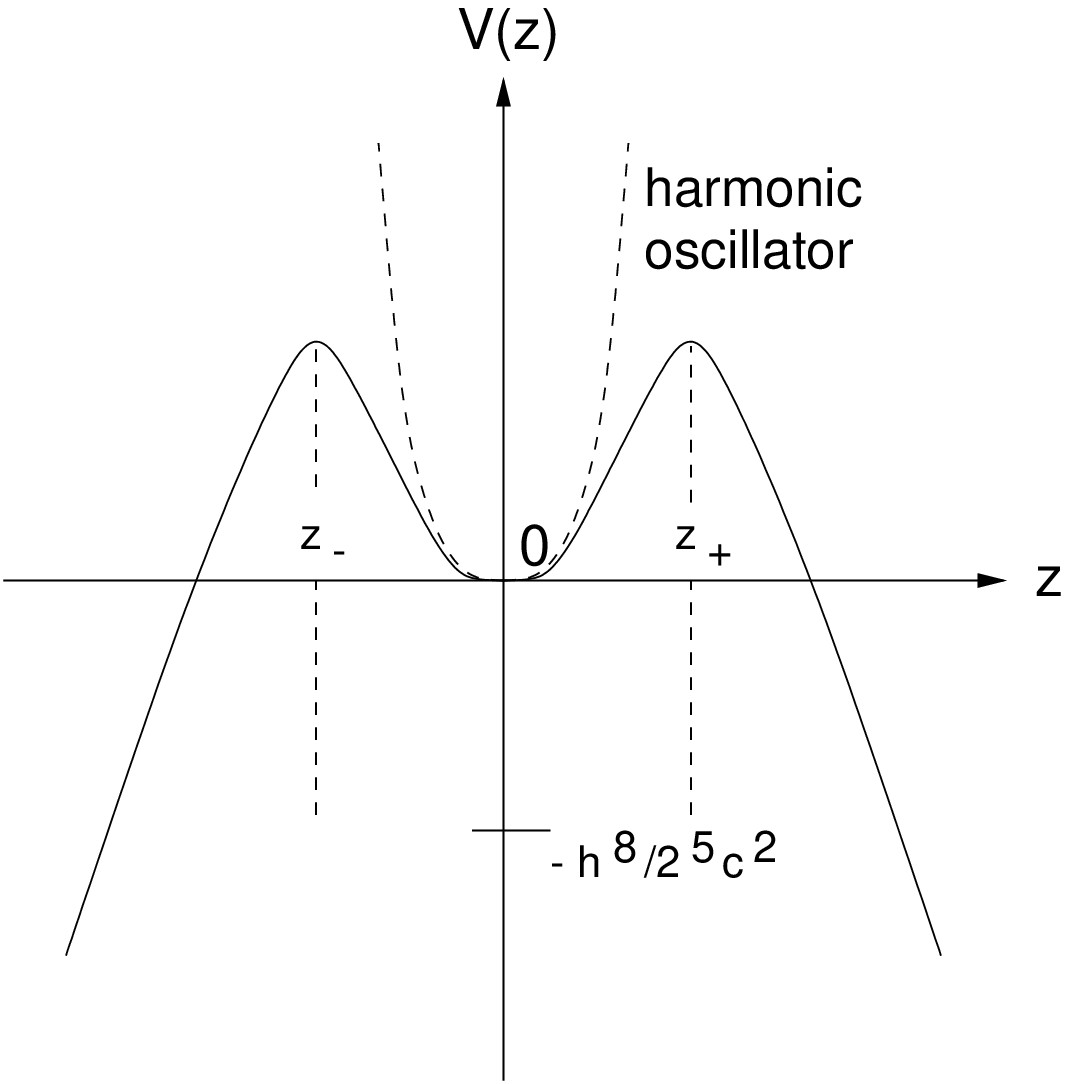}
\end{figure}
\centerline{\small Fig.~2 The inverted  double-well potential
with (hatched) oscillator potential.}

\vspace{0.3cm}

\noindent
We adopt the following conventions which it is essential to
state in order to assist comparison with other literature.
We take  $\hbar =1$ and the mass $m_0$ of the
particle $=1/2$.  This implies that
results for $m_0=1$ (a frequent convention in field
theory considerations) differ from those obtained here by 
factors of $2^{1/2}$, a point which has to be kept in mind
in comparisons. 
If suffixes $1/2,1$ refer to the two cases, we can pass from one
to the other by making the replacements:
\begin{equation}
E_{1/2}=2E_1, \;\; h^4_{1/2}=2h^4_1, \;\; c^2_{1/2}=2c^2_1.
\label{3}
\end{equation}

Introducing a parameter $q$ and a quantity $\triangle\equiv\triangle(q,h)$,
and a variable $w$ defined by setting 
\begin{equation}
E=\frac{1}{2}qh^2+\frac{\triangle}{2h^4}\;\;\;
{\rm and} \;\;\; w=hz,
\label{4}
\end{equation}
we can rewrite Eq.~(\ref{2}) as
\begin{equation}
{\cal D}_q(w)y(w)=-\frac{1}{h^6}(\triangle +c^2w^4)y(w)
\label{5}
\end{equation}
with
\begin{equation}
{\cal D}_q(w)=2\frac{d^2}{dw^2}+q-\frac{w^2}{2}.
\label{6}
\end{equation}
In the domain of $w$ finite, $|h^2|\rightarrow\infty$ and $ c^2$ finite,
the harmonic part of the
potential
dominates over the quartic contribution and Eq.~(\ref{5}) becomes
\begin{subequations}
\begin{equation}
{\cal D}_q(w)y(w)=O\bigg(\frac{1}{h^2}\bigg).
\label{7a}
\end{equation}
The problem then reduces to that of the pure harmonic 
oscillator 
with $y(w)$ a parabolic cylinder function, i.e.
\begin{equation}
y(w)\propto D_{\frac{1}{2}(q-1)}(w) 
\;\;\; {\rm and} \;\;\;
q=q_0=2n+1, \;\;\; n=0,1,2,\ldots .
\label{7b}
\end{equation}
\end{subequations}
The perturbation expansion in descending powers of $h$ suggested
by the above considerations is therefore an expansion around the central
minimum of $V(z)$ at $z=0$. The 
positions $z_{\pm}$ of the maxima of $V(z)$ on either side
of $z=0$ in the case $c^2>0$
are obtained from
\begin{equation}
v^{\prime}(z_{\pm})= 0 \;\;\; {\rm as} \;\;\;
z_{\pm}=\pm\frac{h^2}{2c}
\label{8}
\end{equation}
with
$$
v^{\prime\prime}(z_{\pm})=h^4 \;\;\; {\rm and}
\;\;\; V(z_{\pm})=\frac{h^8}{2^5c^2}.
$$
Thus for $c^2>0$ and relatively small, and $h^2$ large
the eigenvalues are essentially perturbatively
shifted eigenvalues of the harmonic oscillator
as is evident from Fig.~1.

The problem here is to obtain the solutions in
various domains of the variable, to match these
in domains of overlap, then  to specify the necessary
boundary conditions and finally to exploit the latter
for the derivation of the complex eigenvalue.
The result will be that derived originally 
by Bender and Wu \cite{2}, \cite{3}, although our method here 
(which parallels that used in the case of the
cosine potential) is different.

\subsection{Three  pairs  of solutions}

We are concerned with the equation
\begin{equation}
\frac{d^2y(z)}{dz^2}+\bigg[E-\frac{h^4z^2}{4}+\frac{c^2z^4}{2}
\bigg]y(z)=0
\label{9}
\end{equation}
where 
\begin{equation}
E=\frac{1}{2}qh^2+\frac{\triangle}{2h^4}.
\label{10}
\end{equation}
Here again $q$ is a parameter still to be determined
from boundary conditions, and $\triangle=\triangle(q,h)$ is
obtained from the perturbation expansion of the 
eigenvalue, as encountered and explained earlier. Inserting
(\ref{10}) into (\ref{9}) we obtain
\begin{equation}
\frac{d^2y}{dz^2}+\bigg[\frac{1}{2}qh^2+\frac{\triangle}{2h^4}
-\frac{h^4z^2}{4}+\frac{c^2z^4}{2}\bigg]y=0.
\label{11}
\end{equation}
The  solutions
 in terms of parabolic cylinder functions
are valid around $z=0$ and extend up to $z\simeq O(1/h^2)$, as
we shall see.
Before we return to these  solutions we derive a new pair which is
valid in the adjoining domains.
Thus these solutions are not valid around $z=0$.  In order
to arrive at these solutions we set in Eq.~(\ref{11})
\begin{equation}
y(z)=A(z)\exp\bigg[\pm i\int^z dz\bigg\{
-\frac{h^4z^2}{4}+\frac{c^2z^4}{2}\bigg\}^{1/2}\bigg].
\label{12}
\end{equation}
Then $A(z)$ is found to satisfy the following equation
\begin{eqnarray}
&& A^{\prime\prime}(z)\pm 2i\bigg\{-\frac{h^4z^2}{4}+
\frac{c^2z^4}{2}\bigg\}^{1/2}A^{\prime}(z)
\pm iA(z)
\frac{d}{dz}\bigg\{-\frac{h^4z^2}{4}+\frac{c^2z^4}{2}\bigg\}^{1/2}
\nonumber\\
&&\qquad\qquad \quad \quad +\bigg[\frac{1}{2}qh^2+
\frac{\triangle}{2h^4}\bigg]A(z)=0.
\label{13}
\end{eqnarray}
Later we will be interested in the construction of  wave functions
which are even or odd around $z=0$.  This construction is
simplified by  the
consideration of symmetry properties of our solutions which
arise at this point. We observe --- before touching the square
roots in Eq.~(\ref{13}) --- that one equation
(of the two alternatives) follows from the other by changing the
sign of $z$ throughout.  This observation allows us
to define the pair of solutions
\begin{subequations}
\begin{equation}
y_A(z)=A(z)\exp\bigg[+i\int^z dz 
\bigg\{-\frac{h^4z^2}{4}+\frac{c^2z^4}{2}\bigg\}^{1/2}\bigg],
\label{14a}
\end{equation}
\begin{equation}
{\overline y}_A(z)={\overline A}(z)\exp\bigg[-i\int^z dz 
\bigg\{-\frac{h^4z^2}{4}+\frac{c^2z^4}{2}\bigg\}^{1/2}\bigg],
\label{14b}
\end{equation}
\end{subequations}
with
\begin{equation}
{\overline A}(z)=A(-z)\;\; {\rm and} \;\;
{\overline y}_A(z)=y_A(-z),
\label{15}
\end{equation}
where $A(z)$ is the solution of the upper of Eqs.~(\ref{13})
and ${\overline A}(z)$ that of the lower of these equations.
We take the square root by setting
\begin{equation}
\bigg\{-\frac{z^2h^4}{4}\bigg\}^{1/2}=
\stackrel{+}{(-)}i\frac{zh^2}{2}.
\label{16}
\end{equation}
For large $h^2$ we can write the Eqs.~(\ref{13}) as
$$
\mp zA^{\prime}(z)\mp\frac{1}{2}A(z)+\frac{1}{2}qA(z)=O\bigg(
\frac{1}{h^2}\bigg).
$$
We define $A_q(z)$ as the solution of the equation
\begin{equation}
 zA^{\prime}_q(z) - \frac{1}{2}(q-1)A_q(z)=0, 
\label{17}
\end{equation}
i.e.
\begin{equation}
A_q(z)=z^{\frac{1}{2}(q-1)}
\equiv \frac{1}{(z^2)^{1/4}}\exp\bigg[\frac{1}{2}q\int^z\frac{dz}
{(z^2)^{1/2}}\bigg].
\label{18}
\end{equation}
We define correspondingly
\begin{equation}
{\overline A}_q(z)=
z^{-\frac{1}{2}(q+1)}=A_{-q}(z)
\equiv \frac{1}{(z^2)^{1/4}}\exp\bigg[-\frac{1}{2}q\int^z\frac{dz}
{(z^2)^{1/2}}\bigg].
\label{19}
\end{equation}
We see that one solution follows from the other by replacing
$z$ by $-z$.  
Clearly $A_q(z), {\overline A}_q(z)$  approximate the
solutions of Eqs.~(\ref{13}) and we can develop
a perturbation theory along the lines of our
method as employed in the case of periodic potentials.
One finds  that these solutions are associated
with the same asymptotic expansion for $\triangle$
--- to be derived in detail below  in Appendix~A  and
as a verification again 
in connection with the solution $y_B$ --- as the other solutions. 
 Since these higher order
contributions are of little interest for our present
considerations, we
do not pursue their calculation.  Thus we now have the
pair of solutions
\begin{subequations}
\begin{equation}
y_A(z)=\exp\bigg[i\int^z dz\bigg\{-\frac{1}{4}z^2h^4
+\frac{1}{2}c^2z^4\bigg\}^{1/2}\bigg]\bigg[A_q(z)
+O\bigg(\frac{1}{h^2}\bigg)\bigg],
\label{20a}
\end{equation}
\begin{equation}
{\overline y}_A(z)=\exp\bigg[-i\int^z dz\bigg\{-\frac{1}{4}z^2h^4
+\frac{1}{2}c^2z^4\bigg\}^{1/2}\bigg]\bigg[{\overline A}_q(z)
+O\bigg(\frac{1}{h^2}\bigg)\bigg].
\label{20b}
\end{equation}
\end{subequations}
These expansions are valid as decreasing
asymptotic expansions in the domain
$$
|z|>O\bigg(\frac{1}{h}\bigg),
$$
i.e. away from the central minimum.  
With proper care in selecting signs of square roots
we can use the  solutions  (\ref{20a}) and  (\ref{20b})
to construct solutions $y_{\pm}(z)$ which are
respectively even and odd under the parity
transformation $z\rightarrow -z$ (or equivalently
$q\rightarrow  -q, h^2\rightarrow -h^2$), i.e. we write
\begin{equation}
y_{\pm}(z)=\frac{1}{2}[y_A(q,h^2;z)\pm {\overline y}_A(q,h^2;z)].
\label{21}
\end{equation}

In the next  two pairs of solutions the exponential factor
of the above solutions of type $A$ is contained in the
parabolic cylinder functions (which are effectively exponentials
times Hermite functions).

We return to Eq.~(\ref{5}). 
The solutions $y_q(z)$ of the equation
\begin{equation}
{\cal D}_q(w)y_q(w)=0, \;\;\;\; w=hz, 
\label{22}
\end{equation}
are parabolic cylinder functions $D_{\frac{1}{2}(q-1)}(\pm w)$ and
 $D_{-\frac{1}{2}(q+1)}(\pm iw)$ (observe
that Eq.~(\ref{22}) is invariant under the combined  substitutions
$q\rightarrow -q, w\rightarrow\pm iw$)  
or functions
\begin{equation}
B_q(w)=\frac{D_{\frac{1}{2}(q-1)}(\pm w)}{[\frac{1}{4}(q-1)]!2^{\frac{1}{4}
(q-1)}}\;\;\;
{\rm and} \;\;\; 
C_q(w)=\frac{D_{-\frac{1}{2}(q+1)}(\pm iw) 2^{\frac{1}{4}
(q+1)}}{[-\frac{1}{4}(q+1)]!}.
\label{23}
\end{equation}
These solutions satisfy the following
recurrence relation (obtained from the basic
recurrence relation for parabolic
cylinder functions given in the literature\footnote{\scriptsize
Ref.~ \cite{23}, pp. 115 - 123. Comparison with our
notation is easier if this reference is used.})
\begin{equation}
w^2y_q=\frac{1}{2}(q+3)y_{q+4}+qy_q+\frac{1}{2}(q-3)y_{q-4}.
\label{24}
\end{equation} 
The extra factors in Eq.~(\ref{23})  have been inserted
to make this recurrence relation assume this  particularly
symmetric and appealing form.\footnote{\scriptsize As an
alternative to $B_q(w)$ in  Eq.~(\ref{23}) one can 
choose the solutions as ${\tilde B}_q(w)$ with
$$
{\tilde B}_q(w)=\frac{D_{\frac{1}{2}(q-1)}(w)}
{[\frac{1}{4}(q-3)]!2^{\frac{1}{4}(q-1)}}.
$$
These satisfy the recurrence relation
$$
w^2y_q(w)=\frac{1}{2}(q+1)y_{q+4}+qy_q
+\frac{1}{2}(q-1)y_{q-4}.
$$
}
For higher  even powers of $w$ we write
\begin{equation}
w^{2i}y_q=\sum^i_{j=-i}S_{2i}(q,4j)y_{q+4j},
\label{25}
\end{equation}
where in the case $i=2$
\begin{eqnarray}
S_4(q,\pm 8) &=& \frac{1}{4}(q\pm 3)(q\pm 7),\nonumber\\
S_4(q,\pm 4) &=& (q\pm 2)(q\pm 3),\nonumber\\
S_4(q,0) &=& \frac{3}{2}(q^2 +1).
\label{26}
\end{eqnarray}
The first approximation $y(w)=y^{(0)}(w)=y_q(w)$ therefore
leaves  uncompensated terms amounting to
\begin{equation}
R^{(0)}_q=-\frac{1}{h^6}(\triangle +c^2w^4)y_q(w)
\equiv-\frac{1}{h^6}\sum^2_{j=-2}[q,q+4j]y_{q+4j},
\label{27}
\end{equation}
where
\begin{equation}
[q,q]=\triangle +c^2S_4(q,0), \;\;
{\rm and \; for} \; j\neq 0: \;\; [q,q+4j]=c^2S_4(q,4j).
\label{28}
\end{equation}
Now, since ${\cal D}_qy_q=0$, we  also have 
${\cal D}_{q+4j}y_{q+4j}=0$, and so
\begin{equation}
{\cal D}_qy_{q+4j}(w)=-4jy_{q+4j}(w).
\label{29}
\end{equation}
Hence a term $\mu y_{q+4j}$ on the right hand side of
Eq.~(\ref{27}) can be removed by adding to $y^{(0)}$ 
the contribution $(-\mu/4j)y_{q+4j}$.  Thus the
next order contribution to $y_q$ is
\begin{equation}
y^{(1)}(w)=\frac{1}{h^6}\sum^2_{j=-2,j\neq 0}\frac{[q,q+4j]}{4j}y_{q+4j}.
\label{30}
\end{equation}
For the sum $y(w)=y^{(0)}(w)+y^{(1)}(w)$ to be a solution
to that order we must also have to that order 
\begin{equation}
[q,q]=0, \;\; {\rm i.e.} \;\; 
\triangle = -\frac{3}{2}(q^2+1)c^2
+O\bigg(\frac{1}{h^{6}}\bigg).
\label{31}
\end{equation}
Proceeding in this way we obtain the solution
\begin{equation}
y=y^{(0)}(w)+y^{(1)}(w) +y^{(2)}(w)+\cdots 
\label{32}
\end{equation}
with the corresponding equation from which $\triangle$ can be
obtained, i.e.
\begin{equation}
0=\frac{1}{h^6}[q,q]+\bigg(\frac{1}{h^6}\bigg)^2\sum_{j\neq 0}
\frac{[q,q+4j]}{4j}[q+4j,q]+\cdots .
\label{33}
\end{equation}
Evaluating this expansion  and inserting the result for $\triangle$ 
into Eq.~(\ref{10})
we obtain
\begin{equation}
E(q,h^2)=\frac{1}{2}qh^2-\frac{3c^2}{4h^4}(q^2+1)-
\frac{c^4}{h^{10}}(4q^3+29q)
+O\bigg(\frac{1}{h^{16}}\bigg).
\label{34}
\end{equation}
We observe that odd powers of $q$ arise in combination with
odd powers of $1/h^2$, and even powers of $q$ in
combination with even powers of $1/h^2$, so that the entire
expansion is invariant under the interchanges
$$
q\rightarrow -q, \;\; h^2 \rightarrow -h^2.
$$
This type of invariance is a property of a very large
class of eigenvalue problems.
Equation (\ref{34}) is the expansion of the 
eigenenergies $E$ of Case (1) with $q=q_0=2n+1, n=0, 1, 2, \ldots$ .
In Case (3) the parameter $q$ is only approximately
an odd integer, the difference $q-q_0$ arising from tunneling..

We can now write the solution $y(w)$ in the form
\begin{equation}
y(w)=y_q(w)+\sum^{\infty}_{i=1}\bigg(\frac{1}{h^6}\bigg)^i
\sum^{2i}_{j=-2i,j\neq 0}P_i(q,q+4j)y_{q+4j}(w),
\label{35}
\end{equation}
where for instance
\begin{eqnarray*}
P_1(q,q\pm 4)&=&\frac{[q,q\pm 4]}{\pm 4}
=c^2\frac{(q\pm 2)(q\pm 3)}{\pm 4}, \nonumber\\
P_2(q,q\pm 4)&=& \frac{[q,q\pm 4]}{\pm 4}\frac{[q\pm 4, q\pm 4]}
{\pm 4}
+\frac{[q,q\pm 8]}{\pm 8}\frac{[q\pm 8, q\pm 4]}
{\pm 4}\nonumber\\
&& +
\frac{[q,q\mp 4]}{\mp 4}\frac{[q\mp 4, q\pm 4]}
{\pm 4},
\end{eqnarray*}
and so on. 
Again we can write down a recurrence relation for the coefficients
$P_i(q,q+4j)$, i.e.
\begin{equation}
4tP_i(q,q+4t)=\sum^2_{j=-2}P_{i-1}(q,q+4j+4t)[q+4j+4t,q+4t]
\label{36}
\end{equation}
with the boundary conditions
\begin{eqnarray}
P_0(q,q)&=& 1,\;\; {\rm and}\; {\rm for}
\; j\neq 0  \; {\rm all \; other} \; P_0(q,q+4j)=0,
\nonumber\\
P_{i\neq 0}(q,q)&=& 0,\nonumber\\
 P_i(q,q+4j)&=& 0 \; {\rm for}
\;|j|>2i \; {\rm or} \; |j|\geq 2i+1.
\label{37}
\end{eqnarray} 
For further details concerning these coefficients,
their recurrence relations and  the solutions of
the latter we refer to Ref. \cite{20}.
Since our starting equation (\ref{11}) is
invariant under a change of sign of
$z$, we may infer that given one solution
$y(z)$, there is another solution $y(-z)$.
We thus have the following pair of solutions 
\begin{eqnarray}
y_B(z)&=&\bigg[B_q(w)
+\sum^{\infty}_{i=1}\bigg(\frac{1}{h^6}\bigg)^i
\sum^{2i}_{j=-2i,j\neq 0}P_i(q,q+4j)B_{q+4j}(w)
\bigg]_{w=hz, \arg z=0},\nonumber\\ 
{\overline y}_B(z)&=&[y_B(z)]_{\arg z=\pi} =[y_B(-z)]_{\arg z=0}.
\label{38}
\end{eqnarray}
These solutions are suitable in the sense
of decreasing asymptotic expansions in the domains
$$
|z|\lesssim  O\bigg(\frac{1}{h^2}\bigg), \arg z\sim 0, \pi.
$$
They are linearly independent there  as long
as $q$ is not an integer.

Our third pair of solutions is obtained
from the parabolic cylinder functions of
complex argument. We observed earlier that
these are obtained by making the replacements 
$$
q\rightarrow -q,\;\;  w\rightarrow\pm iw.
$$
These solutions  are therefore defined by the following
substitutions:
\begin{equation}
y_C(z)=[y_B(z)]_{q\rightarrow -q, h\rightarrow ih}, 
\;\;
{\overline y}_C(z)=[{\overline y}_B(z)]_{q\rightarrow -q, h\rightarrow ih}
\label{39}
\end{equation}
with the same coefficients $P_i(q,q+4j)$ as in $y_B$.
The solutions $y_C, {\overline y}_C$ are suitable
asymptotically decreasing expansions in one of the domains
$$
|z|\lesssim O\bigg(\frac{1}{h^2}\bigg), \;\;
\arg z \sim \mp \frac{\pi}{2}.
$$

We emphasise  again that all three pairs of solutions are associated
with the same expansion of the eigenvalue $E(q,h^2)$ in which
odd powers of $q$ are associated with odd powers
of $h^2$, so that the
eigenvalue expansion remains unaffected by the interchanges
$q\rightarrow -q, h^2\rightarrow -h^2$ as long as
corrections resulting from boundary conditions are ignored.

\subsection{Matching of solutions}

We saw that the solutions of types $B$ and $C$ are valid
around the central  minimum at $|z|=0$, the solutions of type
$A$ being valid away from the minimum. Thus in the transition
region some become proportional. In order to 
be able to extract the proportionality factor
between  two solutions,
one has to stretch  each  
by appropriate expansion to
the limit of its  domain of validity. In this bordering domain
the adjoining branches of the overall  solution  
then differ by a constant.\footnote{\scriptsize Variables
like those we use here for expansion about the minimum of
a potential (e.g. like $w$ of Eq.~(\ref{4}))  
are known in some mathematical literature as ``{\it stretching
variables}'' and are there discussed in connection with
matching principles, see e.g. Ref.~ \cite{24}.}

First we deal with the exponential factor in the solutions
of type $A$.
Since these are not valid around $z=0$
but $h^2$ is assumed to be large, we expand
the integrand  as follows:    
\begin{eqnarray}
&& \exp\bigg[i\int^z dz\bigg\{-\frac{1}{4}z^2h^4+\frac{1}{2}
c^2z^4\bigg\}^{1/2}\bigg]\nonumber\\
&=&\exp\bigg[-\frac{h^6}{8c^2}\int^z
d\bigg(\frac{2c^2z^2}{h^4}\bigg)
\bigg\{1-\frac{2c^2z^2}{h^4}\bigg\}^{1/2}\bigg]\nonumber\\
&=&\exp\bigg[\frac{h^6}{8c^2}\frac{2}{3}\bigg\{1-\frac{2c^2z^2}{h^4}
\bigg\}^{3/2}\bigg]\nonumber\\
&=&\exp\bigg[\frac{h^6}{12c^2}-\frac{h^2z^2}{4}+O\bigg(
\frac{z^4}{h^2}\bigg)\bigg].
\label{40}
\end{eqnarray}
Considering the pair of solutions $y_A(z), {\overline y}_A(z)$
we see that in the direction of $z=0$ (of course, not around
that point) 
\begin{eqnarray}
y_A(z) &=& e^{h^6/12c^2}e^{-\frac{1}{4}z^2h^2}\bigg[
z^{\frac{1}{2}(q-1)}+O\bigg(\frac{1}{h^2}\bigg)\bigg],\nonumber\\
{\overline y}_A(z) &=& e^{-h^6/12c^2}e^{\frac{1}{4}z^2h^2}\bigg[
z^{-\frac{1}{2}(q+1)}+O\bigg(\frac{1}{h^2}\bigg)\bigg].
\label{41}
\end{eqnarray}

The cases  of the solutions of types $B$ and $C$
require a careful look at the parabolic cylinder
functions since these differ in different regions of
the argument of the variable $z$. Thus from the
literature \cite{25} we obtain
\begin{equation}
D_{\frac{1}{2}(q-1)}(w)
=w^{\frac{1}{2}(q-1)}e^{-\frac{1}{4}w^2}
\sum^{\infty}_{i=0}\frac{[\frac{1}{2}(q-1)]!}
{i![\frac{1}{2}(q-4i-1)]!}\frac{1}{(-2w^2)^i}, \;\;\;
\;|\arg w|<\frac{3}{4}\pi,
\label{42}
\end{equation}
but
\begin{eqnarray}
D_{\frac{1}{2}(q-1)}(w)
&=&w^{\frac{1}{2}(q-1)}e^{-\frac{1}{4}w^2}
\sum^{\infty}_{i=0}\frac{[\frac{1}{2}(q-1)]!}
{i![\frac{1}{2}(q-4i-1)]!}\frac{1}{(-2w^2)^i} 
\nonumber\\
&& \;- \;
\frac{(2\pi)^{1/2}e^{-i\frac{\pi}{2}(q-1)}}{[-\frac{1}{2}(q+1)]!}
 \frac{e^{\frac{1}{4}w^2}}{w^{\frac{1}{2}(q+1)}} 
\sum^{\infty}_{i=0}\frac{[-\frac{1}{2}(q+1)]!}
{i![-\frac{1}{2}(q+4i+1)]!}\frac{1}{(2w^2)^i} 
\nonumber\\
&& {\rm with} \;\; \frac{5}{4}\pi>\arg w>\frac{1}{4}\pi.
\label{43}
\end{eqnarray}
The function $D_{\frac{1}{2}(q-1)}(w)$ has a
 similarly complicated expansion for
$$
-\frac{1}{4\pi} > \arg w > -\frac{5}{4}\pi.
$$
From (\ref{42}) we obtain for the solution $y_B(z), w=hz$:
\begin{equation}
y_B(z)\simeq B_q(w)=\frac{(h^2z^2)^{\frac{1}{4}(q-1)}}
{[\frac{1}{4}(q-1)]! 2^{\frac{1}{4}(q-1)}}e^{-\frac{1}{4}h^2z^2}
\bigg[1+O\bigg(\frac{1}{h^2}\bigg)\bigg].
\label{44}
\end{equation}
In  the solution ${\overline y}_B(z)$, with $z$ in $y_B(z)$
replaced by $-z$, we would have to substitute correspondingly
the  expression (\ref{43})
(since $z\rightarrow -z$ implies
$\arg z=\pm\pi$). We do not require this at present.
Comparing the solution  $y_A(z)$ of Eq.~(\ref{41})
with the solution $y_B(z)$ of Eq.~(\ref{44}) we see that
in their
common domain of validity
\begin{equation}
y_A(z)=\frac{1}{\alpha}y_B(z)
\label{45}
\end{equation}
with
\begin{equation}
\alpha = \frac{(h^2)^{\frac{1}{4}(q-1)}e^{-\frac{h^6}{12c^2}}}
{[\frac{1}{4}(q-1)]!2^{\frac{1}{4}(q-1)}}\bigg[1
+O\bigg(\frac{1}{h^2}\bigg)\bigg].
\label{46}
\end{equation}
However, the ratio of ${\overline y}_A(z), {\overline y}_B(z)$
is not a constant.

We proceed similarly with the solutions $y_C(z), {\overline y}_C(z)$.
Inserting the expansion (\ref{42}) into 
${\overline y}_C(z)$ we obtain
\begin{equation}
{\overline y}_C(z)=
\frac{(-h^2z^2)^{-\frac{1}{4}(q+1)}e^{\frac{1}{4}h^2z^2}}
{[-\frac{1}{4}(q+1)]!2^{-\frac{1}{4}(q+1)}}
\bigg[1+O\bigg(\frac{1}{h^2}\bigg)\bigg].
\label{47}
\end{equation}
Comparing this behaviour of the solution 
${\overline y}_C(z)$ with that of solution ${\overline y}_A(z)$
of Eq.~(\ref{41}), we see that in their common 
domain of validity
\begin{equation}
{\overline y}_A(z) = \frac{1}{\overline \alpha}{\overline y}_C(z),
\label{48}
\end{equation}
where
\begin{equation}
{\overline \alpha}= 
\frac{(-h^2)^{-\frac{1}{4}(q+1)}e^{\frac{h^6}{12c^2}}}
{[-\frac{1}{4}(q+1)]!2^{-\frac{1}{4}(q+1)}}
\bigg[1+O\bigg(\frac{1}{h^2}\bigg)\bigg].
\label{49}
\end{equation}
Again there is no such simple relation between
$y_A(z)$ and $y_C(z)$.

\subsection{Boundary conditions at the origin}
\vspace{-0.3cm}
$\qquad\quad $ {\bf (A)$\,$  Formulation of the  boundary conditions}
\vspace{0.3cm}

The really difficult part of the problem
is to recognise the boundary conditions we
have to impose.
Looking at the potential we are considering here ---
as depicted in Fig.~2 --- we see
that near the origin the potential behaves
like that of the harmonic oscillator,  in fact, 
our large-$h^2$ solutions require this for large $h^2$.  Thus
the boundary conditions to be imposed there
are the same as in the case of the harmonic oscillator
for alternately even and odd wave functions. 
Recalling the solutions $y_{\pm}(z)$  which we defined
with  Eq.~(\ref{21}) as even and odd about $z=0$, we see
that at the origin we have to demand the conditions
\begin{equation}
y_+^{\prime}(0) = 0 \;\;\; {\rm and} \;\;\;
y_-(0)=0
\label{50}
\end{equation}
and $y_+(0)\neq 0, y_-^{\prime}(0) \neq 0$.
The first of the conditions (\ref{50}) will be seen to
imply $q_0\equiv 2n+1 = 1,5,9,\ldots$ and the second
$q_0=3,7,11,\ldots$. For instance $q_0=1$ (or $n=0$) 
implies a ground state wave function with
the shape of a Gauss curve above $z=0$, i.e. large
probability for the particle to be found thereabouts. 
At $z=0$ the solutions of type $A$ are invalid; hence
we have to use the proportionalities just derived 
in order to match these to the solutions valid
around the origin.
Then imposing the above boundary conditions we obtain
\begin{eqnarray}
0&=& y_+^{\prime}(0)=\lim_{z\rightarrow 0}\frac{1}{2}[y_A^{\prime}(z)
+{\overline y}_A^{\prime}(z)]\nonumber\\
&=& \frac{1}{2}\bigg[\frac{1}{\alpha}y_B^{\prime}(0)
+\frac{1}{\overline \alpha}{\overline y}_C
^{\prime}(0)\bigg]
\label{51}
\end{eqnarray}
and
\begin{eqnarray}
0&=& y_-(0)=\lim_{z\rightarrow 0}\frac{1}{2}[y_A(z)
-{\overline y}_A(z)]\nonumber\\
&=&\frac{1}{2}\bigg[ \frac{1}{\alpha}y_B(0)-
\frac{1}{\overline \alpha}{\overline y}_C
(0)\bigg].
\label{52}
\end{eqnarray}
Thus we obtain the equations
\begin{equation}
\frac{y_B^{\prime}(0)}{{\overline y}_C^{\prime}(0)}
=-\frac{\alpha}{\overline \alpha}\;\;\;
{\rm and} \;\;\; 
\frac{y_B(0)}{{\overline y}_C(0)}
=\frac{\alpha}{\overline \alpha}.
\label{53}
\end{equation}
Clearly  we now have to evaluate the solutions involved
and their derivatives at the origin. We leave details  
to Appendix~B.

$\qquad\quad $ {\bf (B)$\,$ Evaluation of the boundary conditions}

\vspace{0.3cm}

We now evaluate Eqs.~(\ref{53}) in dominant order
and insert from Eqs.~(\ref{46}), (\ref{49}) the
appropriate expressions for $\alpha$ and ${\overline \alpha}$.
Starting with the derivative expression we obtain
(apart from contributions of order $1/h^2$)
$$
-\frac{1}{i}\frac{\sin\{\frac{\pi}{4}(q+3)\}}{\sin\{\frac{\pi}{4}(q-3)\}}
=-\frac{(h^2)^{\frac{1}{4}(q-1)}[-\frac{1}{4}(q+1)]!
2^{-\frac{1}{4}(q+1)}}{[\frac{1}{4}(q-1)]!
2^{\frac{1}{4}(q-1)}(-h^2)^{-\frac{1}{4}(q+1)}}e^{-\frac{h^6}{6c^2}}.
$$
We rewrite the left hand side as
$$
-\frac{1}{i}\frac{\sin\{\frac{\pi}{4}(q+3)\}}
{\sin\{\frac{\pi}{4}(q+3-6)\}}=i\,{\rm tan}\bigg\{\frac{\pi}{4}(q+3)\bigg\}.
$$
We rewrite the right hand side of the derivative
equation again with the help of the inversion
and duplication formulae and obtain
$$
-\pi\frac{(\frac{h^4}{4})^{q/4}(-1)^{\frac{1}{4}(q+1)}
e^{-h^6/6c^2}}{[\frac{1}{4}(q-1)]![\frac{1}{4}(q-3)]!
\sin\{\frac{\pi}{4}(q+1)\}}
=\sqrt{\frac{\pi}{2}}\frac{(h^4)^{q/4}(-1)^{\frac{1}{4}(q+1)}}
{[\frac{1}{2}(q-1)]!\cos\{\frac{\pi}{4}(q+3)\}}
e^{-\frac{h^6}{6c^2}}.
$$
Then the derivative relation of (\ref{53}) becomes
\begin{equation}
\sin\bigg\{\frac{\pi}{4}(q+3)\bigg\}
=-i\sqrt{\frac{\pi}{2}}
\frac{(h^4)^{q/4}(-1)^{\frac{1}{4}(q+1)}}
{[\frac{1}{2}(q-1)]!}e^{-\frac{h^6}{6c^2}}.
\label{54}
\end{equation}
Proceeding similarly with the second of relations (\ref{53}),
we obtain
\begin{equation}
\cos\bigg\{\frac{\pi}{4}(q+3)\bigg\}
=-\sqrt{\frac{\pi}{2}}
\frac{(h^4)^{q/4}(-1)^{\frac{1}{4}(q+1)}}
{[\frac{1}{2}(q-1)]!}e^{-\frac{h^6}{6c^2}}.
\label{55}
\end{equation}
In each of Eqs.~(\ref{54}) and (\ref{55}) the
right hand side is an exponentially small
quantity. In fact the left hand side of (\ref{54})  vanishes
for $q=q_0=1,5,9,\ldots $ and the left hand side of 
(\ref{55}) for $q=q_0=3,7,11,\ldots $.
With a Taylor
expansion  about $q_0$
 the left hand side of (\ref{54}) becomes
 $$
(q-q_0)\frac{\pi}{4}\cos\bigg\{\frac{\pi}{4}(q_0+3)\bigg\}+\cdots
\simeq (q-q_0)\frac{\pi}{4}(-1)^{-\frac{1}{4}(q_0+3)}.
$$
It follows that we obtain for the even function with
$q=q_0=1,5,9,\ldots $ 
\begin{equation}
(q-q_0)\simeq \pm \frac{2\sqrt{2}}{\sqrt\pi}\frac{(h^2)^{q_0/2}}
{[\frac{1}{2}(q_0-1)]!}e^{-\frac{h^6}{6c^2}}.
\label{56}
\end{equation}
Expanding  similarly the left hand side of  Eq.~(\ref{55})
about $q_0=3,7,11,\ldots$, we again obtain (\ref{56})
but now for the odd function with these values of $q_0$.
We have thus obtained the conditions resulting from
the
boundary conditions at $z=0$.  Our next task is to
extend the solution all the way to the
region  beyond the shoulders of
the inverted double well potential and to impose
the necessary boundary conditions there.
Thus we have to determine these conditions first.

\subsection{Boundary conditions at infinity}
\vspace{-0.3cm}
$\qquad\quad $ {\bf (A)$\,$  Formulation of the boundary conditions}
\vspace{0.3cm}

We explore first the conditions we have to impose
at $|z|\rightarrow\infty$.  Recall the original
Schr\"odinger equation (\ref{2}) with 
potential (\ref{1}). For $c^2<0$ and the solution
$y(z)$ square integrable in $-\infty< \Re z< \infty$,
the energy $E$ is real; this is the case of the purely
discrete spectrum (the differential operator
being selfadjoint for the appropriate boundary
conditions, i.e. the vanishing of the
wave functions at infinity). This is
 Case (1) of Fig.~1. The analytic continuation
of one case to 
the other is accomplished by replacing $\pm c^2$ by $\mp c^2$
or, equivalently, by the rotations
$$
E\rightarrow e^{i\pi}E=-E, \;\; z\rightarrow e^{i\pi/2}z, \;\; 
z^2\rightarrow -z^2.
$$
One can therefore retain $c^2$ as it is and perform these
rotations.  It is then necessary to insure that when one rotates
to the case of the purely discrete spectrum without tunneling, the resulting
wave functions vanish at infinity and thus are
square integrable. Thus in our case here the behaviour
of the solutions at infinity has to be chosen such that 
this condition is satisfied. Now, for $\Re z\rightarrow\pm\infty$
we have
\begin{equation}
y(z)\sim\exp\bigg\{\pm i\int^z dz\bigg[\frac{c^2}{2}z^4\bigg]^{1/2}
\bigg\}=\exp\bigg\{\pm i\bigg(\frac{c^2}{2}\bigg)^{1/2}\frac{z^3}{3}
\bigg\}.
\label{57}
\end{equation}
In order to decide which solution or combination of
solutions is
compatible with the square integrability in the rotated
($c^2$ reversed) case, we set 
$$
z=|z|e^{i\theta}.
$$
Then
\begin{eqnarray*}
\exp\bigg\{- i\bigg(\frac{c^2}{2}\bigg)^{1/2}\frac{z^3}{3}
\bigg\}
&=& 
\exp\bigg\{-i\bigg(\frac{c^2}{2}\bigg)^{1/2}\frac{|z|^3}{3}
e^{3i\theta})\bigg\}\nonumber\\
&=&
\exp\bigg\{ -i\bigg(\frac{c^2}{2}\bigg)^{1/2}\frac{|z|^3}{3}
(\cos\,3\theta + i\sin\,3\theta\bigg\}
\nonumber\\
&\propto & 
\exp\bigg\{ \bigg(\frac{c^2}{2}\bigg)^{1/2}\frac{|z|^3}{3}
\sin\,3\theta\bigg\}.
\end{eqnarray*}
This expression vanishes for $|z|\rightarrow +\infty$ if the angle $\theta$
lies in the range $-\pi < 3\theta <0$, i.e. if $-\pi/3<\theta<0$.
Thus
\begin{eqnarray}
\exp\bigg\{-i
\bigg(\frac{c^2}{2}\bigg)^{1/2}\frac{z^3}{3}
\bigg\}& \rightarrow &\!\!\!\! 0\;\;\; {\rm for}\; \Re z\rightarrow +\infty
\;{\rm in} \; \arg z\in \; \bigg(-\frac{\pi}{3},0\bigg),\nonumber\\
\exp\bigg\{+i\bigg(\frac{c^2}{2}\bigg)^{1/2}\frac{z^3}{3}
\bigg\}& \rightarrow & \!\!\!\!0 \;\;\;{\rm for}\; \Re z\rightarrow -\infty
\;{\rm in} \; \arg z\in \; \bigg(0,\frac{\pi}{3}\bigg).
\label{58}
\end{eqnarray}
Rotating $z$ by $\pi/2$, i.e. replacing $\sin\,3\theta$ by
$$
\sin\,3\bigg(\theta+\frac{\pi}{2}\bigg)=-\cos\,3\theta
$$
we see that the solution with the exponential factor is
exponentially decreasing for $|z|\rightarrow\infty$  provided that
$$
\cos\,3\theta > 0,
$$
i.e. in the domain $-\pi/2< 3\theta< \pi/2$, or
$$
-\frac{\pi}{6}<\theta<\frac{\pi}{6}.
$$
In the case of the inverted double well potential
under consideration here (i.e. the case of complex $E$), 
we therefore demand that for $\Re z\rightarrow +\infty$
and $-\pi/3<\arg z <0$ the wave functions have
decreasing phase, i.e.
\begin{equation}
y(z)\sim\exp\bigg\{
-i\bigg(\frac{c^2}{2}\bigg)^{1/2}\frac{z^3}{3}
\bigg\}, \;\; c^2>0.
\label{59}
\end{equation}
This is the boundary condition also used by Bender and Wu \cite{2}.
For $c^2<0$ we have correspondingly
$$
y(z)\sim\exp\bigg\{
\pm\bigg( \frac{|c^2|}{2}\bigg)^{1/2}\frac{z^3}{3}
\bigg\}, \;\; c^2<0, \;\; {\rm for} \; z\rightarrow\mp\infty.
$$ 
This is not the asymptotic behaviour of a wave function
of the simple harmonic oscillator. We have to
remember that we have various branches of the solutions $y(z)$ in different
domains of $z$.

Our procedure now is to continue
(in the sense of matched asymptotic
expansions)
the even and
odd solutions (\ref{21}) to $+$ infinity
and to demand that they  satisfy the condition
(\ref{59}) for $c^2>0$.
Equating to zero the coefficient of
 the term with
sign opposite to that in the exponential
of Eq.~(\ref{59})
will
lead to our second  condition  
which together with the first
obtained from boundary conditions at the origin
 determines the imaginary
part of the eigenvalue  $E$.

\vspace{0.3cm}
\begin{figure}[ht]
\centering
\includegraphics[angle=0,totalheight=6.5cm]{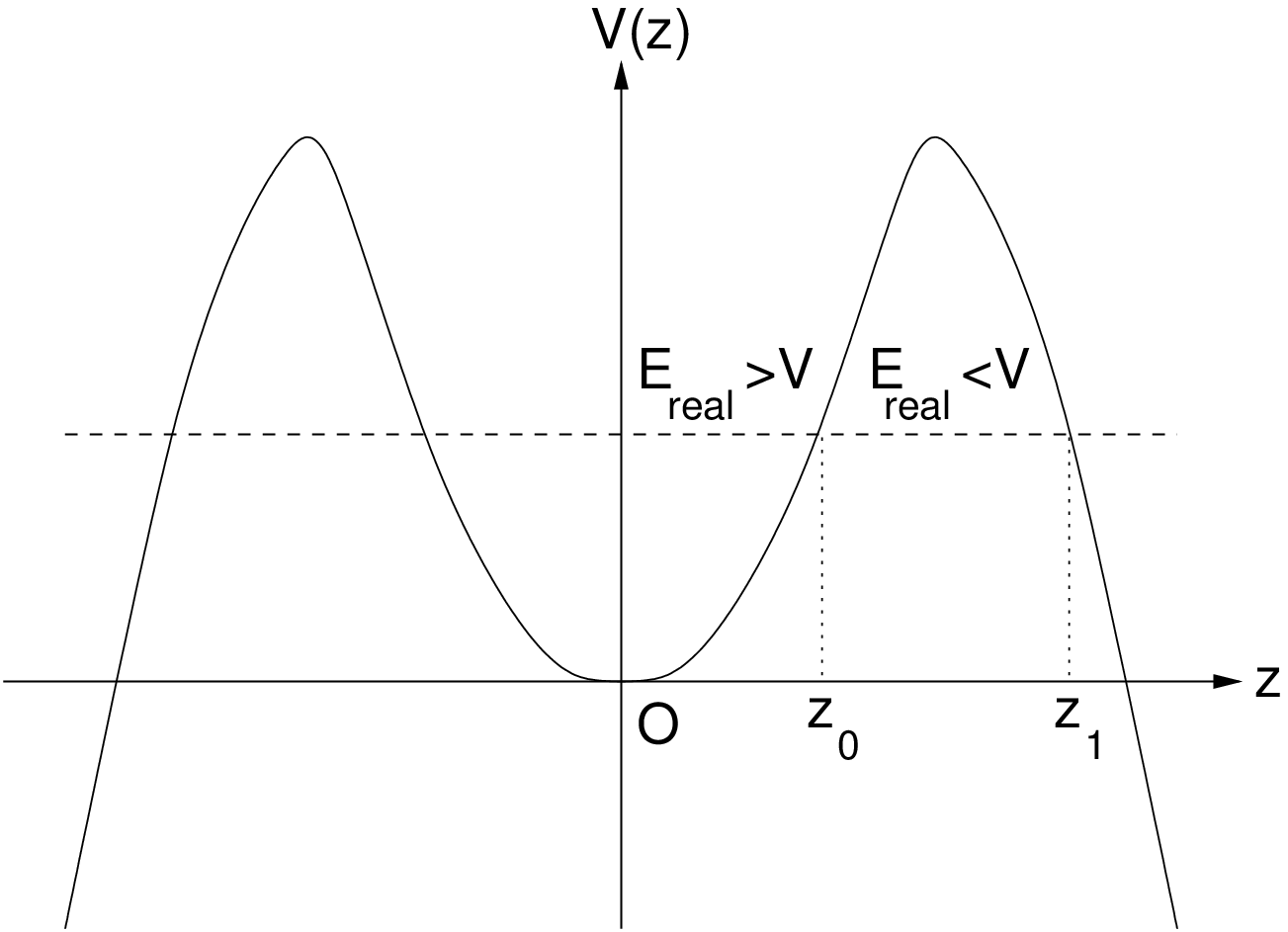}
\end{figure}
\centerline{\small Fig.~3 The inverted double-well potential
with turning points $z_0,z_1$.}

\vspace{0.3cm}

$\qquad\quad ${\bf (B)$\,$  Evaluation of the boundary conditions}
\vspace{0.3cm}

The following  considerations 
(usually for real $z$)
require  some  algebraic steps   which could obscure the
basic procedure. We therefore explain our procedure first.
Our even and odd solutions $y_{\pm}(z)$ 
(cf. Eq.~(\ref{21}))  were defined in
terms of solutions  of type $A$
which have a wide domain of validity.
Looking at Fig.~3 we see that at
a given energy $E$
and to the right of $z=0$
 (which is the only
region we consider for reasons of symmetry)
 there are two turning points $z_0, z_1$.
Thus we have to match the solutions of type $A$
first to solutions to the left of $z_1$ and then
extend  these to solutions to the right,  and 
there impose the  boundary condition (\ref{59})
on $y_{\pm}(z)$ (by demanding that the
coefficient of the solutions  with other behaviour be zero). 
We do this extension with the help of
WKB solutions, i.e. we match the WKB solutions
to the left of $z_1$ to the solutions of type $A$,
and then use the WKB procedure (called ``linear
matching'' across the turning point)
to obtain the dominant WKB solutions beyond $z_1$.

The distant turning point at $z_1$ as indicated
in Fig.~3 is given by (using Eq.~(\ref{4})
for $E$ and ignoring nondominant terms)
$$
-\frac{1}{4}z^2h^4+\frac{1}{2}c^2z^4\simeq -\frac{1}{2}qh^2,
$$
i.e.
\begin{equation}
z_1\simeq \frac{h^2}{\sqrt{2c^2}}\bigg(1-\frac{2qc^2}{h^6}\bigg)
\simeq  \frac{h^2}{\sqrt{2c^2}}.
\label{60}
\end{equation} 
The WKB solutions have been discussed in 
the literature.\footnote{\scriptsize Ref.~ \cite{22},
p. 291, equations  (21), (22) or Ref.~\cite{26}, Vol. I, 
Sec.6.2.4.} We obtain in the domain $V>\Re E$ to the left
of $z_1$ as the dominant terms of the WKB solutions
\begin{eqnarray}
y^{(l,z_1)}_{\rm WKB}(z)
&=&\bigg[-\frac{1}{2}qh^2+\frac{1}{4}z^2h^4-\frac{1}{2}c^2z^4\bigg]^{-1/4}
\nonumber\\
&&\!\!\!\!\!\!\!\!\!\!\!\!\!
\times \exp\bigg\{\int^{z_1}_z\,dz
\bigg[-\frac{1}{2}qh^2+\frac{1}{4}z^2h^4-\frac{1}{2}c^2z^4\bigg]^{1/2}\bigg\},
\nonumber\\
{\overline y}^{(l,z_1)}_{\rm WKB}(z)
&=&\bigg[-\frac{1}{2}qh^2+\frac{1}{4}z^2h^4-\frac{1}{2}c^2z^4\bigg]^{-1/4}
\nonumber\\
&&\!\!\!\!\!\!\!\!\!\!\!\!\! \times \exp\bigg\{-\int^{z_1}_z\,dz
\bigg[-\frac{1}{2}qh^2+\frac{1}{4}z^2h^4-\frac{1}{2}c^2z^4\bigg]^{1/2}\bigg\},
\label{61}
\end{eqnarray}
where $z<z_1$, i.e. $z^2h^4/4>c^2z^4/2$. In using these
expressions it has to be remembered that the moduli
of the integrals have to be taken.\footnote{\scriptsize 
Ref.~\cite{22}, p. 291.} 
To the right of the turning point at $z_1$ these solutions match on to
\begin{eqnarray}
y^{(r,z_1)}_{\rm WKB}(z)
&=&\bigg[\frac{1}{2}qh^2-\frac{1}{4}z^2h^4+\frac{1}{2}c^2z^4\bigg]^{-1/4}
\nonumber\\
&&\!\!\!\!\!\!\!\!\!\!\!\!\!
\times \cos\bigg\{\int^{z_1}_z\,dz
\bigg[\frac{1}{2}qh^2-\frac{1}{4}z^2h^4+\frac{1}{2}c^2z^4\bigg]^{1/2}
+\frac{\pi}{4}\bigg\},
\nonumber\\
{\overline y}^{(r,z_1)}_{\rm WKB}(z)
&=&2\bigg[\frac{1}{2}qh^2-\frac{1}{4}z^2h^4+\frac{1}{2}c^2z^4\bigg]^{-1/4}
\nonumber\\
&&\!\!\!\!\!\!\!\!\!\!\!\!\! \times \sin\bigg\{-\int^{z_1}_z\,dz
\bigg[\frac{1}{2}qh^2-\frac{1}{4}z^2h^4+\frac{1}{2}c^2z^4\bigg]^{1/2}
+\frac{\pi}{4}\bigg\}.\nonumber\\
\label{62}
\end{eqnarray}

We now come to the algebra of evaluating the integrals in the
above solutions. We begin with the exponential factors 
occurring in Eqs.~(\ref{61}), i.e.
\begin{eqnarray}
&& E_{\pm}\nonumber\\ 
&=&\exp\bigg\{\pm\int^{z_1}_z\,dz
\bigg[-\frac{1}{2}qh^2+\frac{1}{4}z^2h^4-\frac{1}{2}c^2z^4\bigg]^{1/2}\bigg\}
\nonumber\\
&=&\exp\bigg[\pm\frac{h^6}{8c^2}\int^{z_1}_z
\,d\bigg(\frac{2c^2z^2}{h^4}\bigg)
\bigg\{1-\frac{2c^2z^2}{h^4}\bigg\}^{1/2}
\bigg\{1-\frac{2q}{z^2h^2}\frac{1}{[1-\frac{2c^2z^2}{h^4}]}\bigg\}^{1/2}
\bigg]\nonumber\\
&\simeq &
\exp\bigg[\mp\frac{h^6}{8c^2}\bigg\{\frac{2}{3}\bigg[
1-\frac{2c^2z^2}{h^4}\bigg]^{3/2}\bigg\}^{z_1}_z\mp\frac{qh^4}{8c^2}
\int^{z_1}_z\frac{\frac{1}{z^2}d(\frac{2c^2z^2}{h^4})}
{[1-\frac{2c^2z^2}{h^4}]^{1/2}}
\bigg]\nonumber\\
&=&\exp\bigg[\pm\frac{h^6}{8c^2}\frac{2}{3}
\bigg\{1-\frac{2c^2z^2}{h^4}\bigg\}^{3/2}
\mp\frac{q}{2}\int^{z_1}_z\frac{dz}{z[\frac{h^4}{2c^2}-z^2]^{1/2}}
\bigg(\frac{h^4}{2c^2}\bigg)^{1/2}\bigg].\nonumber\\
\label{63}
\end{eqnarray}
Here the first part is the exponential factor contained in 
${y}_A(z), {\overline y}_A(z)$ respectively
(cf. Eq.~(\ref{40})).
In the remaining factor we have
(looking up Tables of Integrals)
\begin{eqnarray*}
&& \int^{z_1}_z
\frac{dz}{z[\frac{h^4}{2c^2}-z^2]^{1/2}}\nonumber\\
&=& \bigg\{-\bigg(\frac{2c^2}{h^4}\bigg)^{1/2}
\ln\bigg|\frac{1}{z}\bigg\{\bigg(\frac{h^4}{2c^2}\bigg)^{1/2}
+\bigg(\frac{h^4}{2c^2}-z^2\bigg)^{1/2}\bigg\}\bigg|
\bigg\}^{z_1}_z\nonumber\\
&=& +\bigg(\frac{2c^2}{h^4}\bigg)^{1/2}
\ln\bigg|\frac{1}{z}\bigg\{\bigg(\frac{h^4}{2c^2}\bigg)^{1/2}
+\bigg(\frac{h^4}{2c^2}-z^2\bigg)^{1/2}\bigg\}\bigg|.
\end{eqnarray*}
Since we are interested in determining
the proportionality of two solutions
in their common domain of validity we require
only the dominant $z$-dependence  contained in this expression.  
We obtain this factor by expanding the expression
in powers of $z/z_1$ (since in the integral $z<z_1$).
Thus the above factor yields
$$
\bigg(\frac{2c^2}{h^4}\bigg)^{1/2}
\ln\bigg|\frac{2}{z}{\bigg(\frac{h^4}{2c^2}\bigg)^{1/2}}\bigg|,
$$
so that (cf. Eq.~(\ref{40}))
\begin{eqnarray}
E_{\pm} &=& \exp\bigg[\pm\frac{h^6}{8c^2}
\frac{2}{3}\bigg\{1-\frac{2c^2z^2}{h^4}\bigg\}^{3/2}
\mp\frac{q}{2}\ln\bigg|2\bigg(\frac{h^4}{2c^2}\bigg)^{1/2}\bigg|\bigg]
z^{\pm q/2}\nonumber\\
&=& z^{\pm q/2}\bigg[2\bigg(\frac{h^4}{2c^2}\bigg)^{1/2}\bigg]^
{\mp q/2}\exp\bigg[\pm i\int^z\,dz\bigg\{
-\frac{1}{4}z^2h^4+\frac{1}{2}c^2z^4\bigg\}^{1/2}\bigg].
\nonumber\\
\label{64}
\end{eqnarray}
Thus at the left end of the domain of validity of the WKB
solutions we have
\begin{equation}
y^{(l,z_1)}_{\rm WKB}(z) \simeq \frac{E_+}
{\bigg[\frac{1}{4}z^2h^4\bigg]^{1/4}} \;\;\;
{\rm and} \;\;\; 
{\overline y}^{(l,z_1)}_{\rm WKB}(z) \simeq \frac{E_-}
{\bigg[\frac{1}{4}z^2h^4\bigg]^{1/4}}.
\label{65}
\end{equation}
Comparing these solutions now with the solutions 
(\ref{20a}) and (\ref{20b}), we see that in their
common domain of validity
\begin{equation}
y_A(z)=\beta y^{(l,z_1)}_{\rm WKB}(z), \;\;\;
{\overline  y}_A(z)={\overline \beta} {\overline y}^{(l,z_1)}_{\rm WKB}(z),
\label{66}
\end{equation}
where
\begin{subequations}
\begin{equation}
\beta=\bigg[\frac{h^2}{2}\bigg]^{1/2}\bigg[
2\frac{(h^2)}{(2c^2)^{1/2}}\bigg]^{q/2}\;\;\;
{\rm and} \;\;\;
{\overline \beta}=\bigg[-\frac{h^2}{2}\bigg]^{1/2}\bigg[
2\frac{(-h^2)}{(2c^2)^{1/2}}\bigg]^{-q/2}
\label{67a}
\end{equation}
or
\begin{equation}
\frac{\beta}{\,{\overline \beta}\,}=\bigg[\frac{2h^2}{(2c^2)^{1/2}}\bigg]^q
\frac{(-1)^{q/2}}{\sqrt{-1}}
\label{67b}
\end{equation}
\end{subequations}
apart from factors $[1+O(1/h^2)]$. 
In these expressions we have chosen the 
signs of  square roots of $h^4$  so that the 
conversion symmetry
under replacements $q\rightarrow -q, h^2\rightarrow -h^2$
is maintained. 

Returning to the even and odd solutions defined by
Eqs.~(\ref{21}) we now have
\begin{eqnarray}
y_{\pm}(z)&=&\frac{1}{2}[y_A(z)\pm {\overline y}_A(z)]\nonumber\\
&=&\frac{1}{2}[ \beta y^{(l,z_1)}_{\rm WKB}(z) \pm 
{\overline \beta} {\overline y}^{(l,z_1)}_{\rm WKB}(z)]
\nonumber\\
&=&\frac{1}{2}[ \beta y^{(r,z_1)}_{\rm WKB}(z) \pm 
{\overline \beta} {\overline y}^{(r,z_1)}_{\rm WKB}(z)].
\label{68}
\end{eqnarray}
Now in the domain $z\rightarrow \infty$ we have
\begin{eqnarray}
\int^z_{z_1}\bigg[\frac{1}{2}qh^2-\frac{1}{4}z^2h^4
+\frac{1}{2}c^2z^4\bigg]^{1/2}
&\simeq & \int^z_{z_1}\, dz\frac{cz^2}{\sqrt 2}
\bigg[1-\frac{h^4}{4c^2z^2}\bigg]\nonumber\\
&\simeq & \bigg[ \frac{cz^3}{3\sqrt{2}}\bigg]^z_{z_1}
=\frac{cz^3}{3\sqrt{2}}-\frac{h^6}{12c^2}.
\label{69}
\end{eqnarray}
Inserting this into the solutions
(\ref{62}) and these into 
(\ref{68}) we can rewrite the even and odd solutions
for $\Re z\rightarrow\infty$ as (by separating cosine and
sine into their exponential components)
\begin{eqnarray}
y_{\pm}(z) &\simeq &
\frac{1}{2}\bigg[\frac{1}{2}c^2z^4\bigg]^{-1/4}\bigg[S_+(\pm)
\exp\bigg\{i\bigg(\frac{cz^3}{3\sqrt{2}}-\frac{h^6}{12c^2}\bigg)
\bigg\}\nonumber\\
&& \qquad +S_-(\pm)\exp
\bigg\{-i\bigg(\frac{cz^3}{3\sqrt{2}}-\frac{h^6}{12c^2}\bigg)
\bigg\}\bigg],
\label{70}
\end{eqnarray}
where
\begin{eqnarray}
S_+(\pm) &=& \bigg(\frac{1}{2}\beta\pm\frac{1}{i}
{\overline \beta}\bigg)\exp\bigg(i\frac{\pi}{4}\bigg),
\nonumber\\
S_-(\pm) &=& \bigg(\frac{1}{2}\beta\mp \frac{1}{i}
{\overline \beta}\bigg)\exp\bigg(-i\frac{\pi}{4}\bigg).
\label{71}
\end{eqnarray}
Imposing the boundary condition that the even and odd
solutions have the asymptotic behaviour given by Eq.~(\ref{59}),
we see that we have to demand that
\begin{equation}
S_+(\pm)=0, \;\;\; {\rm i.e.} \;\;\;
\frac{1}{2}\beta\pm\frac{1}{i}{\overline \beta}=0.
\label{72}
\end{equation}
Inserting expressions (\ref{67a}) for $\beta$ and
${\overline \beta}$,  this equation can be rewritten as
\begin{subequations}
\begin{equation}
(-h^2)^{q/2}=(-h^2)^{q/2}
\frac{i\beta}{2{\overline \beta}}=
 i\frac{2^{q-1}(-1)^q}{\sqrt{-1}}\bigg(\frac{h^6}{2c^2}\bigg)^
{q/2}.
\label{73a}
\end{equation}
Thus we impose this second
boundary condition by making
in Eq.~(\ref{56}) the replacement 
\begin{equation}
(-h^2)^{q_0/2}\Rightarrow \stackrel{+}{(-)}2^{q_0-1}
\bigg(\frac{h^6}{2c^2}\bigg)^{q_0/2}.
\label{73b}
\end{equation}
\end{subequations}

Inserting this  into the latter  equation we obtain
(the factor ``$i$'' arising from the minus sign
on the left of Eq.~(\ref{73b}))
\begin{equation}
(q-q_0)=  \pm i\sqrt{\frac{2}{\pi}}\frac{2^{q_0}
\bigg(\displaystyle \frac{h^6}{2c^2}\bigg)^
{{q_0}/2}}
{[\frac{1}{2}(q_0-1)]!}e^{-\frac{h^6}{6c^2}}.
\label{74}
\end{equation}
with $q_0=1,3,5\ldots$.

\subsection{The complex eigenvalues}

We now return  to the expansion of the eigenvalues, i.e.
Eq.~(\ref{34}),
$$
E(q,h^2)=\frac{1}{2}qh^2-\frac{3c^2}{4h^4}(q^2+1) +O\bigg(
\frac{1}{h^{10}}\bigg).
$$
Expanding about $q=q_0$ we obtain
\begin{eqnarray}
E(q,h^2) &=& E_0(q_0,h^2) + (q-q_0)\bigg(\frac{dE}{dq}\bigg)_{q_0}+\cdots
\nonumber\\
&=& E_0(q_0,h^2)+(q-q_0)\frac{h^2}{2}+\cdots.
\label{75}
\end{eqnarray}
Clearly the expression for $(q-q_0)$ has to be inserted here giving
in the dominant approximation
\begin{equation}
E=E_0(q_0,h^2)\stackrel{+}{(-)}i\frac{2^{q_0}h^2
\bigg(\displaystyle \frac{h^6}{2c^2}\bigg)^
{{q_0}/2}}
{(2\pi)^{1/2}[\frac{1}{2}(q_0-1)]!}{\displaystyle e}^{
\displaystyle -\frac{h^6}{6c^2}}.
\label{76}
\end{equation}
The imaginary part of this expression agrees with the result
of Bender and Wu (see formula (3.36) of Ref. \cite{3})
for $\hbar=1$ and in their notation
$$
q_0=2K+1, \;\;\; \frac{h^6}{2c^2}=\epsilon.
$$
For comparison with the case of the double well below we 
note here that the ratio $\beta/{\overline \beta}$ which
we required for the derivation of the result is the
ratio of matching coefficients. Hence the result  does not
involve a specific normalisation of the WKB solutions.
This is different in the case of the double well
potential that we consider below.

\section{The Double-Well Potential}

\subsection{Defining the problem}

In dealing with the case
of the symmetric  double-well potential, we shall employ basically
the same procedure as above.  But there are
significant differences.

We consider the following equation
\begin{equation}
\frac{d^2 y(z)}{dz^2}+[E-V(z)]y(z)=0
\label{77}
\end{equation}
with double-well potential
\begin{equation}
V(z)=v(z)=-\frac{1}{4}z^2h^4+\frac{1}{2}c^2z^4
\;\;\; {\rm for} \;\;\; c^2>0, \; h^4>0.
\label{78}
\end{equation}

\vspace{0.3cm}
\begin{figure}[ht]
\centering
\includegraphics[angle=0,totalheight=6.5cm]{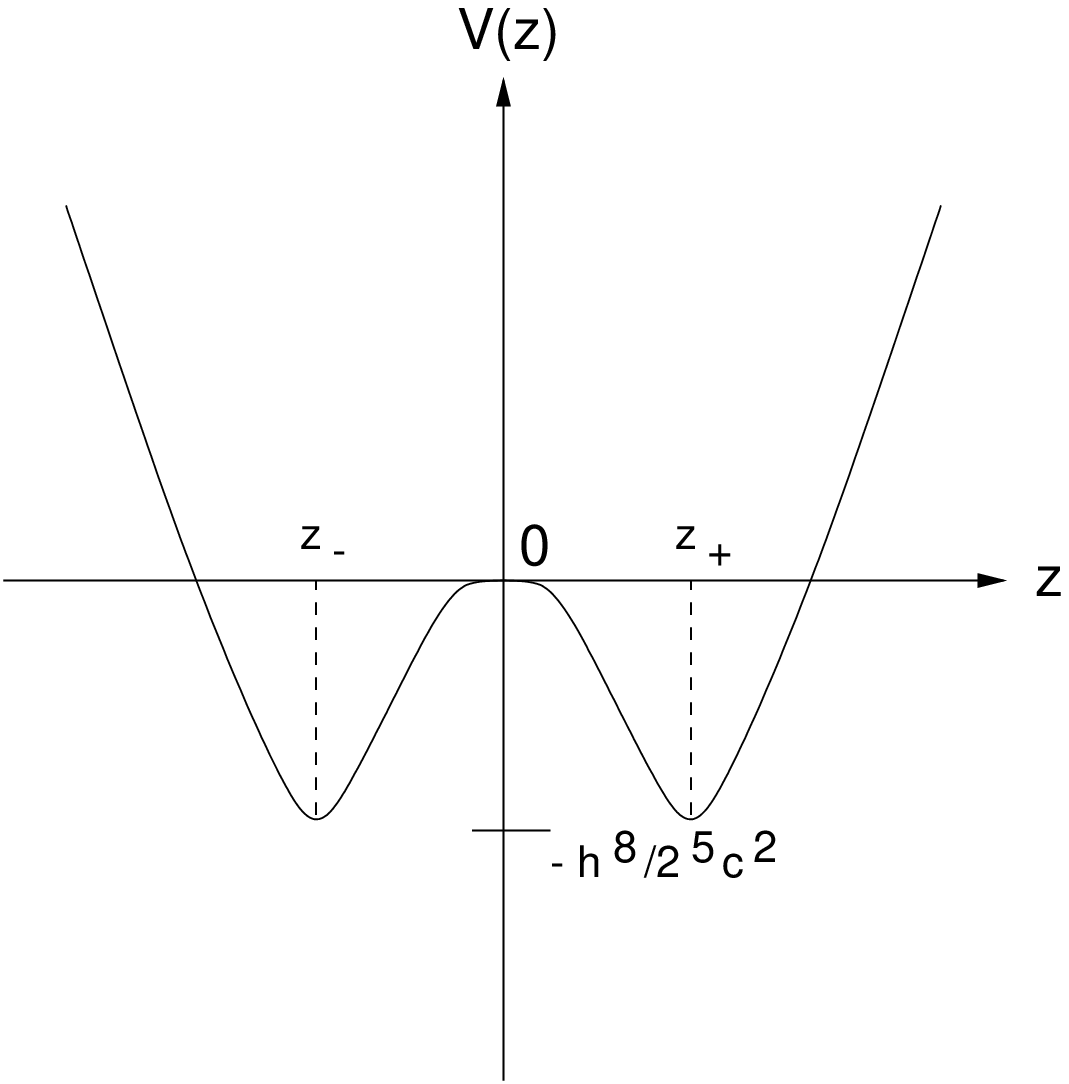}
\end{figure}
\centerline{\small Fig.~4 The double-well potential.}

\vspace{0.3cm}

\noindent
The minima of $V(z)$ on either side of the central
maximum at $z=0$ are located at
\begin{equation}
z_{\pm}=\pm \frac{h^2}{2c}
\label{79}
\end{equation}
with
\begin{eqnarray}
V(z_{\pm}) &=& -\frac{h^8}{2^5c^2},\nonumber\\
V^{(2)}(z_{\pm})&=& h^4,\;\; \;\;\;  V^{(3)}(z_{\pm})=\pm 6ch^2,
\nonumber\\
V^{(4)}(z_{\pm})&=& 12c^2, \;\;\; V^{(i)}(z_{\pm})= 0, \; i\geq 5.
\label{80}
\end{eqnarray}

In order to obtain a rough approximation of the eigenvalues
we expand the potential about the minima at
$z_{\pm}$ and obtain
\begin{equation}
\frac{d^2y}{dz^2}+\bigg[E-V(z_{\pm})-\frac{1}{2}(z-z_{\pm})^2h^4
+O[(z-z_{\pm})^3]\bigg]y=0.
\label{81}
\end{equation}
We set
\begin{equation}
\frac{1}{4}h^4_{\pm}\equiv\frac{1}{2}h^4, \;\;\; h^2_{\pm}=\sqrt{2}h^2
\label{82}
\end{equation}
(thus we sometimes use $h^4$ and sometimes $h^4_{\pm}$)
and 
\begin{equation}
E-V(z_{\pm})=\frac{1}{2}q_{\pm}h^2_{\pm}+\frac{\triangle}{h_{\pm}^4}.
\label{83}
\end{equation}
With the further substitution
\begin{equation}
\omega_{\pm}=h_{\pm}(z-z_{\pm})
\label{84}
\end{equation}
Eq.~ (\ref{81}) becomes
\begin{subequations}
\begin{equation}
{\cal D}_{q_{\pm}}(\omega_{\pm})y(\omega_{\pm})
=O\bigg(\frac{1}{h^3_{\pm}}\bigg)y,
\label{85a}
\end{equation}
where
\begin{equation}
{\cal D}_{q_{\pm}}(\omega_{\pm})
=2\frac{d^2}{d\omega^2}+q_{\pm}-\frac{1}{2}\omega^2_{\pm}.
\label{85b}
\end{equation}
\end{subequations}
By comparison of Eqs.~(\ref{85a}), (\ref{85b})  with the
equation  of parabolic cylinder
functions $u(z)\equiv D_{\nu}(z)$, i.e.
$$
\frac{d^2u(z)}{dz^2}+\bigg[\nu+\frac{1}{2}-\frac{1}{4}z^2\bigg]u(z)=0,
$$
 we conclude that in the dominant
approximation $q_{\pm}$ is an  odd integer,
$q_0=2n+1, n=0,1,2,\ldots$.  
Inserting the expression (\ref{83})  for $E$ into 
Eq.~(\ref{77}) we obtain
\begin{equation}
\frac{d^2y}{dz^2}+\bigg[\frac{1}{2}q_{\pm}h^2_{\pm}
+\frac{\triangle}{h_{\pm}^4}-\frac{1}{4}h^4_{\pm}U(z)\bigg]y=0,
\label{86}
\end{equation}
where
\begin{subequations}
\begin{equation}
U(z)=\frac{4}{h^4_{\pm}}[V(z)-V(z_{\pm})],
\label{87a}
\end{equation}
and near a minimum at $z_{\pm}$
\begin{equation}
U(z)=(z-z_{\pm})^2+O[(z-z_{\pm})^3].
\label{87b}
\end{equation}
\end{subequations}
Our basic equation, Eq.~(\ref{86}), is again
seen to be invariant under a change of sign of $z$.  Thus
again, given one solution, we obtain another by replacing $z$ by $-z$.

\setcounter{footnote}{0}           
\renewcommand{\thefootnote}{\fnsymbol{footnote}}

\subsection{Three pairs  of solutions}

We define our first pair of solutions $y(z)$ as
solutions with the proportionality
\begin{equation}
y(z)\propto \exp\bigg[\pm\frac{1}{2}h^2_{\pm}\int^zU^{1/2}(z)dz\bigg].
\label{88}
\end{equation}
Evaluating the exponential we can define these as the pair
\begin{eqnarray}
y_A(z)&=&A(z)\exp\bigg[-\frac{1}{\sqrt 2}
\bigg\{\frac{c}{3}z^3-\frac{h^4}{4c}z\bigg\}\bigg],
\nonumber\\
{\overline y}_A(z)&=&{\overline A}(z)\exp\bigg[+\frac{1}{\sqrt 2}
\bigg\{\frac{c}{3}z^3-\frac{h^4}{4c}z\bigg\}\bigg].
\label{89}
\end{eqnarray}
The equation for $A(z)$ is given by the following equation
with upper signs and the equation for ${\overline A}(z)$
by the following equation with lower signs:
\begin{equation}
A^{\prime\prime}(z)\mp\sqrt{2}
\bigg\{cz^2-\frac{h^4}{4c}\bigg\}A^{\prime}(z)
\mp\sqrt{2}czA(z)
+\bigg(\frac{1}{2}q_{\pm}h^2_{\pm}+\frac{\triangle}{h_{\pm}^4}\bigg)
A(z)=0.
\label{90}
\end{equation}
Since
$$
z_+=\frac{h^2}{2c},\;\;\; h^2_{\pm}=\sqrt{2}h^2, 
$$
and selecting $z_+$ with $q_+=q$, 
these equations can be rewritten as
\begin{subequations}
\begin{equation}
(z^2_+-z^2)A^{\prime}(z)
+(qz_+-z)A(z)
=-\frac{\sqrt{2}}{2c}\bigg[A^{\prime\prime}(z)+
\frac{\triangle}{h_{+}^4}A(z)\bigg],
\label{91a}
\end{equation}
\begin{equation}
(z^2_+-z^2){\overline  A}^{\prime}(z)
-(qz_++ z){\overline  A}(z)
=\frac{\sqrt{2}}{2c}\bigg[{\overline  A}^{\prime\prime}(z)+
\frac{\triangle}{h_{+}^4}{\overline A}(z)\bigg].
\label{91b}
\end{equation}
\end{subequations}
To a first approximation for large $h^2=2cz_+$ we can neglect the
right hand side.  The dominant approximation to $A$ is
then the function $A_q$ given by the solution of the first
order differential equation
\begin{equation}
{\cal D}_qA_q(z)=0, \;\;\;\; {\cal D}_q=(z^2_+ -z^2)\frac{d}{dz}
+(qz_+ -  z).
\label{92}
\end{equation}
We observe that a change of sign of $z$
in this equation is equivalent to
a change of sign of $q$, but the solution
is a different one, i.e.
${\overline A}_q(z)=A_q(-z)=A_{-q}(z)$.
Integration of Eq.~(\ref{92}) yields the following expression
\begin{equation}
A_q(z)=\frac{1}{|z^2-z^2_+|^{1/2}}
\bigg|\frac{z-z_+}{z+z_+}\bigg|^{q/2} = 
\frac{|z-z_+|^{\frac{1}{2}(q-1)}}{|z+z_+|^{\frac{1}{2}(q+1)}}.
\label{93}
\end{equation}
Looking at Eqs.~(\ref{91a}), (\ref{91b})  we  
observe  that the solution ${\overline y}_A(q, h^2;z)$
may be obtained from the solution $y_A(q,h^2;z)$ by  either
changing the sign of $z$ throughout or --- alternatively ---
the signs of both $q$ and $h^2$ (and/or $c$), i.e.
\begin{equation}
{\overline y}_A(q,h^2;z)= y_A(q,h^2;-z)=y_A(-q,-h^2;z).
\label{94}
\end{equation}
Both solutions $y_A(z), {\overline y}_A(z)$ are
associated with one and the
same expansion for $\triangle$ and hence $E$.  We 
leave the calculation of  $\triangle$ to Appendix~C.
The result is given by Eq.~(\ref{C.1}), i.e.
\begin{eqnarray}
{\triangle}&=& -c^2(3q^2+1) -\frac{\sqrt{2}c^4}{4h^6}q(17q^2+19)+\cdots
,\nonumber\\
E(q,h^2)&=&-\frac{h^8}{2^5c^2} +\frac{1}{\sqrt{2}}q h^2
 -\frac{c^2(3q^2+1)}{2h^4}
\nonumber\\
&& -\frac{\sqrt{2}c^4}{8h^{10}}q(17q^2+19)
+O\bigg(\frac{1}{h^{16}}\bigg).
\label{95}
\end{eqnarray}

The solutions $y_A(z), {\overline y}_A(z)$ derived above
are valid  in the domains away
from the minima,
$$
|z-z_{\pm}|>O\bigg(\frac{1}{h^2_{+}}\bigg).
$$
We can define solutions which are even or odd
about $z=0$ as
\begin{eqnarray}
y_{\pm}(z)&=&\frac{1}{2}[y_A(q,h^2;z)\pm  {\overline y}_A(q,h^2;z)]\nonumber\\
&=&\frac{1}{2}[ y_A(q,h^2;z)\pm  {y}_A(q,h^2;-z)]\nonumber\\
&=&\frac{1}{2}[ y_A(q,h^2;z)\pm  {y}_A(-q,-h^2;z)].
\label{96}
\end{eqnarray}
Considering only the leading approximations of
the unnormalised solutions   considered
explicitly above, we have 
(since $A_q(0)=1/z_+={\overline A}_q(0),
A^{\prime}_q(0)=-q/z^2_+$)
\begin{eqnarray}
y_+(q,h^2;0)=\frac{2c}{h^2}, \;\;\; y_-(q,h^2;0)=0, \nonumber\\
y^{\prime}_+(q,h^2;0)= 0, \;\;  
y^{\prime}_-(q,h^2;0)=\frac{h^2}{2\sqrt{2}} -\frac{4qc^2}{h^4}.
\label{97}
\end{eqnarray}

Our second pair of solutions, $y_B(z), {\overline y}_B(z)$
is obtained around a minimum of the potential. We see already
from   Eqs.~(\ref{85a})
and (\ref{85b}),  that
the solution there is of parabolic cylinder type.
This means, in this case we use the Schr\"odinger equation
with the potential $V(z)$ expanded about $z_{\pm}$ as in
Eq.~(\ref{81}). Inserting (\ref{83}) 
and setting $\omega_{\pm}=h_{\pm}(z-z_{\pm})$,
the equation is ---  with differential operator ${\cal D}_q$ as defined
by Eq.~(\ref{85b})) ---
\begin{equation*}
{\cal D}_q(\omega_{\pm})y(\omega_{\pm})=\frac{1}{h^6_{\pm}}
\bigg[\pm 2^{5/4}ch^3\omega^3_{\pm}
+c^2\omega^4_{\pm}-2\triangle\bigg]y(\omega_{\pm}).
\end{equation*}
Thus we can write a first solution
\begin{subequations}
\begin{equation}
y_B(z)=B_q[w_{\pm}(z)]+O\bigg(\frac{1}{h^2_{\pm}}\bigg), \;\;
B_q[w_{\pm}(z)]=\frac{D_{\frac{1}{2}(q-1)}(w_{\pm}(z))}
{[\frac{1}{4}(q-1)]!2^{\frac{1}{4}(q-1)}},
\label{98a}
\end{equation}
and another
\begin{eqnarray}
{\overline y}_B(z)&=&y_B(-z)={\overline B}_q[w_{\pm}(z)]
+O\bigg(\frac{1}{h^2_{\pm}}\bigg)\nonumber\\
&=&B_q[w_{\pm}(-z)]
+O\bigg(\frac{1}{h^2_{\pm}}\bigg).
\label{98b}
\end{eqnarray}
\end{subequations}
Again the higher order terms along with the eigenvalue
expansion in terms of $q$ are obtained perturbatively.

It is clear that correspondingly we have solutions
$y_C(z), {\overline y}_C(z)$ with complex variables
and  $C_q(w)$ given by Eq.~(\ref{23})  with 
appropriate change of parameters
to those of the present case.
Thus
\begin{subequations}
\begin{eqnarray}
y_C(z)&=&C_q[w_{\pm}(-z)]+O\bigg(\frac{1}{h^2_{\pm}}\bigg),\nonumber\\
 C_q[w_{\pm}(-z)]&=&\frac{D_{-\frac{1}{2}(q+1)}(iw_{\pm}(-z))
2^{\frac{1}{4}(q+1)}}
{[-\frac{1}{4}(q+1)]!},
\label{99a}
\end{eqnarray}
\begin{equation}
{\overline y}_C(z)=y_C(-z)
={\overline C}_q[w_{\pm}(-z)]
=C_q[w_{\pm}(+z)]
+O\bigg(\frac{1}{h^2_{\pm}}\bigg).
\label{99b}
\end{equation}
\end{subequations}
These are solutions again
around a minimum and with  
$y_B(z), {\overline y}_C(z)$
providing a pair of decreasing asymptotic
solutions there (or correspondingly
${\overline y}_B(z), y_C(z)$).
We draw attention to two additional points.  Since
$w_{\pm}(-z)=-h_{\pm}(z+z_{\pm})$, we have
$w_{\pm}(-z_{\pm})=-2h_{\pm}z_{\pm}$, but
$w_{\pm}(z_{\pm})=0$.  Moreover, in view of the
factor ``$i$'' in the argument of $C_q$ the solutions
${\overline y}_A, {\overline y}_C$ have the same exponential behaviour
near a minimum.

\subsection{Matching of solutions}

Next we consider the proportionality of solutions
$y_A(z)$ and $y_B(z)$.  
Evaluating the exponential factor contained in $y_A(z)$ of
Eq.~(\ref{89}) for $z\rightarrow z_{\pm}$,
 we have (cf. (\ref{87b}))
\begin{eqnarray*}
\exp\bigg[-\frac{1}{2}h^2_{\pm}\int^z U^{1/2}(z)dz\bigg]
&\simeq &\exp\bigg[-\frac{1}{2}h^2_{\pm}\int(z-z_{\pm})dz\bigg]
\nonumber\\
&=& \exp\bigg[-\frac{1}{2}h^2_{\pm}\bigg(\frac{1}{2}z^2-zz_{\pm}
\bigg)\bigg]\nonumber\\
&=&\exp\bigg[-\frac{1}{4}h^2_{\pm}(z-z_{\pm})^2\bigg]
\exp\bigg[\frac{1}{4}h^2_{\pm}z^2_{\pm}\bigg].
\nonumber\\
\end{eqnarray*}
Allowing $z$ to approach $z_+$ in the solution $y_A(z)$, we
have
\begin{equation}
y_A(z)\simeq \frac{|z-z_+|^{\frac{1}{2}(q-1)}}{|2z_+|^{\frac{1}{2}(q+1)}}
e^{\frac{1}{4}h^2_+z^2_+}e^{-\frac{1}{4}h^2_+(z-z_+)^2}
=
\frac{(\frac{w_+}{h_+})^{\frac{1}{2}(q-1)}}{(2z_+)^{\frac{1}{2}(q+1)}}
e^{\frac{1}{4}h^2_+z^2_+}e^{-\frac{1}{4}w^2_+}.
\label{100}
\end{equation}
Recalling that around $\arg w_{\pm}\sim 0$,  the dominant term
in the power expansion of the parabolic cylinder function
for large values of its argument  is given by
$$
D_{\nu}(w_{\pm})\simeq w^{\nu}_{\pm}e^{-w^2_{\pm}/4},
$$
and comparing with Eq.~(\ref{98a}) we see that (considering only
dominant contributions) in their common domain of validity
\begin{equation}
y_A(z)=\frac{1}{\alpha} y_B(z), \;\;
\alpha
=\frac{(h_+)^{\frac{1}{2}(q-1)}(2z_+)^{\frac{1}{2}(q+1)}}
{2^{\frac{1}{4}(q-1)}[\frac{1}{4}(q-1)]!}e^{-\frac{1}{4}h^2_+z^2_+}
\bigg[1+O\bigg(\frac{1}{h^2_+}\bigg)\bigg].
\label{101}
\end{equation}
Similarly we obtain in approaching $z_+$:
\begin{equation}
{\overline y}_A(z)\simeq
\frac{(2z_+)^{\frac{1}{2}(q-1)}}
{(z-z_+)^{\frac{1}{2}(q+1)}}e^{-\frac{1}{4}h^2_+z^2_+}
e^{\frac{1}{4}h^2_+(z-z_+)^2}
\label{102}
\end{equation}
and
\begin{eqnarray}
{\overline y}_C(z) &\simeq &
\frac{D_{-\frac{1}{2}(q+1)}(iw_+(z))2^{\frac{1}{4}(q+1)}}
{[-\frac{1}{4}(q+1)]!}\nonumber\\
&\simeq &
\frac{2^{\frac{1}{4}(q+1)}e^{\frac{1}{4}h^2_+(z-z_+)^2}}
{[-\frac{1}{4}(q+1)]![ih_+(z-z_+)]^{\frac{1}{2}(q+1)}}.
\label{103}
\end{eqnarray}
Therefore in their common domain of validity
\begin{equation}
{\overline y}_A(z)=\frac{1}{\overline \alpha} 
{\overline y}_C(z), \;\;
{\overline \alpha}
=\frac{2^{\frac{1}{4}(q+1)}
e^{\frac{1}{4}h^2_+z^2_+}}
{(2z_+)^{\frac{1}{2}(q-1)}
(-h^2_+)^{\frac{1}{4}(q+1)}[-\frac{1}{4}(q+1)]!}
\bigg[1+O\bigg(\frac{1}{h^2_+}\bigg)\bigg].
\label{104}
\end{equation}

We have thus found three pairs of solutions: The two solutions
of type $A$ are valid in regions away from the minima,
and are both
in their parameter-dependence
 asymptotically decreasing 
there
and permit
us therefore to define the 
extensions of the solutions $y_{\pm}$ which
are respectively  even and odd about $z=0$
to the minima.  The pair
of solutions of type $B$ is defined around
$\arg z =0, \pi $ and the solutions of type $C$
around $\arg z =\pm \pi/2$. The next aspect to be
considered is that of boundary conditions. We
have to impose boundary conditions at the minima
and at the origin. 
The solutions in terms of parabolic cylinder functions
have a wide range of validity, even above  the
turning points, but it is clear that none of the above 
solutions can be used at the top of the central
barrier.  Thus it is unavoidable to appeal
to (other and that means) WKB methods to apply
the necessary boundary conditions at
that point. The involvement  of these
WKB solutions leads to problems since, 
 basically, they assume  large
quantum numbers. Various investigations,
such as those of Refs. \cite{17}, \cite{18}, \cite{19}, 
\cite{21}, therefore struggle to overcome this
to a good approximation. 
We achieve the same goal (i.e. approximation)
as such corrections
here by demanding our basic perturbation
solutions to be interconvertible 
on the basis of the parameter symmetries
of the original equation.

\subsection{Boundary conditions at the minima}
\vspace{-0.3cm}
$\qquad\quad $ {\bf (A)$\,$  Formulation of the  boundary conditions}
\vspace{0.3cm}

The present case of the double-well potential
differs from that of the simple harmonic oscillator
potential in having two minima instead of one.

\vspace{0.3cm}
\begin{figure}[ht]
\centering
\includegraphics[angle=0,totalheight=8.0cm]{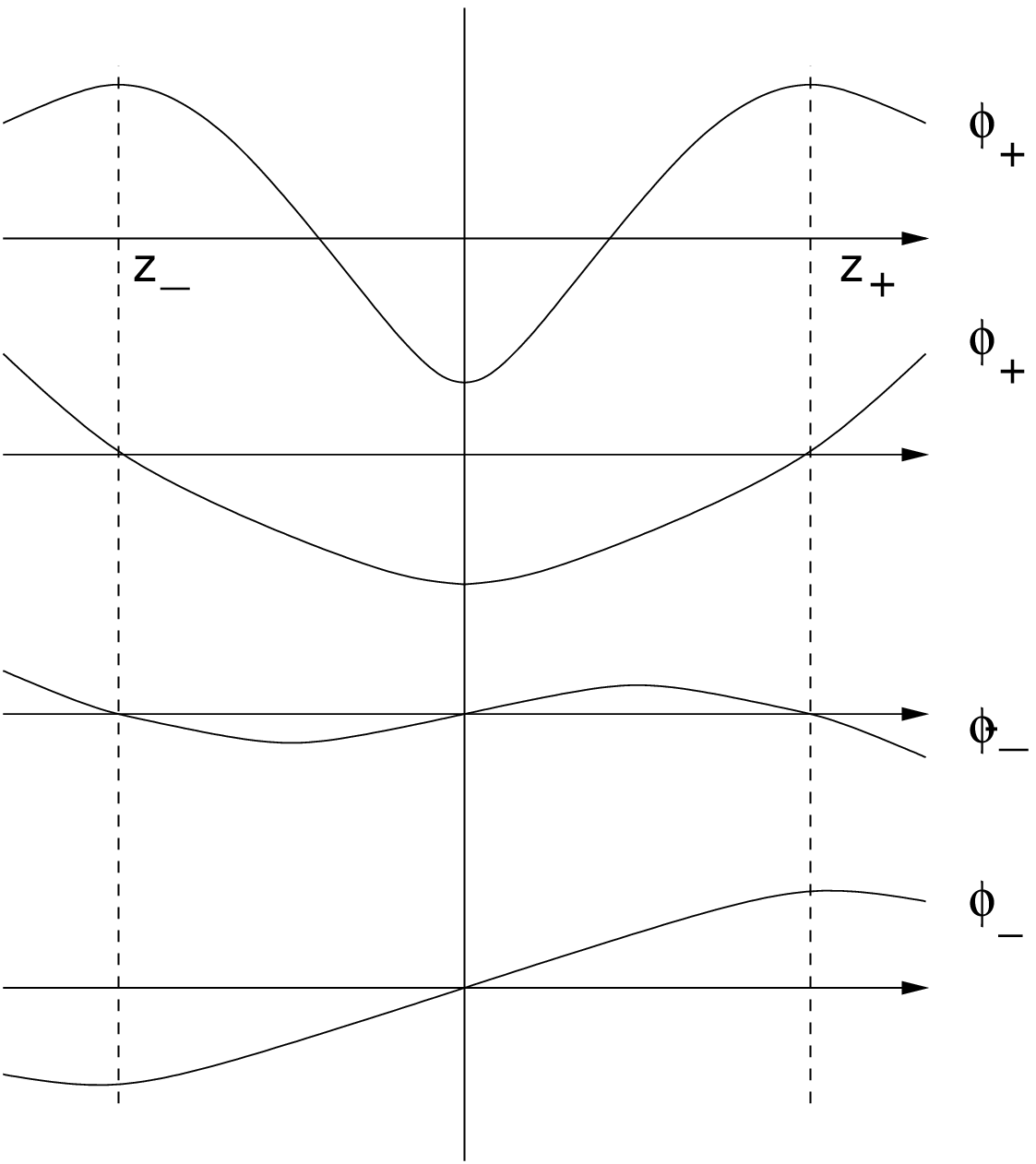}
\end{figure}
\centerline{\small Fig.~5 Behaviour of fundamental wave functions.}

\vspace{0.3cm}

\noindent
Since it is more probable to find  a particle
in the region of a minimum than elsewhere, we
naturally expect the wave function
there to be similar to that of the harmonic
oscillator, and this means at both minima.
Thus the most basic solution would be even with
maxima at $z_{\pm}$, as indicated in Fig.~5.
However, an even wave function can also
pass through zero at these points, as indicated
in Fig.~5. The odd wave function  then
exhibits a correspondingly opposite behaviour,
as indicated there.  
At $\Re z \rightarrow \pm \infty$ we require the wave functions
to vanish so that they are square integrable. 
We have therefore the
following two sets of boundary conditions at the local minima
of the double well potential:
\begin{eqnarray}
y^{\prime}_+(z_{\pm}) &=& 0, \;\;\; y_+(z_{\pm}) \neq 0,\nonumber\\
y_-(z_{\pm})\, &=& 0, \;\;\; y^{\prime}_-(z_{\pm})\neq 0
\label{105}
\end{eqnarray}
and
\begin{eqnarray}
y_+(z_{\pm}) &=& 0, \;\;\; y^{\prime}_+(z_{\pm}) \neq 0,\nonumber\\
y^{\prime}_-(z_{\pm})\, &=& 0, \;\;\; y_-(z_{\pm})\neq 0.
\label{106}
\end{eqnarray}
We have
\begin{eqnarray}
y_{\pm}(z_{\pm})&=& \frac{1}{2}[y_A(z)\pm{\overline y}_A(z)]_{z
\rightarrow z_{\pm}}
\nonumber\\
&=& \frac{1}{2}\bigg[\frac{1}{\alpha}y_B(z_{\pm})\pm\frac{1}{\overline \alpha}
{\overline y}_C(z_{\pm})\bigg]
\label{107}
\end{eqnarray}
and
\begin{equation}
y^{\prime}_{\pm}(z_{\pm})
=\frac{1}{2}\bigg[\frac{1}{\alpha}y^{\prime}_B(z_{\pm})
\pm\frac{1}{\overline \alpha}{\overline y}^{\prime}_C(z_{\pm})\bigg].
\label{108}
\end{equation}
Hence the conditions (\ref{105}), (\ref{106}) imply
\begin{equation}
\frac{y_B(z_{\pm})}{{\overline  y}_C(z_{\pm})}
=\mp\frac{\alpha}{\overline \alpha}, \;\; {\rm and} \;\;
\frac{y^{\prime}_B(z_{\pm})}{{\overline  y}^{\prime}_C(z_{\pm})}
=\mp\frac{\alpha}{\overline \alpha}.
\label{109}
\end{equation}

\vspace{0.3cm}
$\qquad\quad $ {\bf (B)$\,$  Evaluation
 of the  boundary conditions}
\vspace{0.3cm}

Inserting into the first of Eqs.~(\ref{109}) the
dominant approximations we obtain (cf. also Eqs.~(\ref{B.1a})
and (\ref{B.1b}))
\begin{eqnarray}
1&=&\mp\frac{\alpha}{\overline \alpha}=
\mp (-1)^{\frac{1}{4}(q+1)}\frac{(h^2_+)^{q/2}(2z_+)^q[-\frac{1}{4}(q+1)]!}
{2^{q/2}[\frac{1}{4}(q-1)]!}e^{-\frac{1}{2}h^2_+z^2_+}
\nonumber\\
&=&
\mp (-1)^{\frac{1}{4}(q+1)}\frac{\pi(h^2_+)^{q/2}(2z_+)^q}
{2^{q/2}[\frac{1}{4}(q-1)]![\frac{1}{4}(q-3)]!
\sin\{\frac{\pi}{4}(q+1)\}}
e^{-\frac{1}{2}h^2_+z^2_+}
\nonumber\\
&=&
\mp (-1)^{\frac{1}{4}(q+1)}\sqrt{\frac{\pi}{2}}
\frac{(h^2_+)^{q/2}(2z_+)^q}
{[\frac{1}{2}(q-1)]!\sin\{\frac{\pi}{4}(q+1)\}}
e^{-\frac{1}{2}h^2_+z^2_+},
\label{110}
\end{eqnarray}
where we used first the reflection formula and
then the duplication formula.
Thus 
\begin{equation}
\sin\bigg\{\frac{\pi}{4}(q+1)\bigg\}=
\mp (-1)^{\frac{1}{4}(q+1)}\sqrt{\frac{\pi}{2}}
\frac{(h^2_+)^{q/2}(2z_+)^q}
{[\frac{1}{2}(q-1)]!}
e^{-\frac{1}{2}h^2_+z^2_+},
\label{111}
\end{equation}

Using formulae derived in Appendix~B  we can rewrite the second
of Eqs.~(\ref{109}) as
\begin{equation}
\frac{y^{\prime}_B(z_{\pm})}{{\overline  y}^{\prime}_C(z_{\pm})}
=-i\, {\rm cot}\bigg\{\frac{\pi}{4}(q-3)\bigg\}
\equiv -i\,  {\rm cot}\bigg\{\frac{\pi}{4}(q+1)\bigg\}
=\mp \frac{\alpha}{\overline \alpha}.
\label{112}
\end{equation}
Using Eq.~(\ref{110}) this equation can be written
\begin{equation}
-i\cos\bigg\{\frac{\pi}{4}(q+1)\bigg\}
=
\mp (-1)^{\frac{1}{4}(q+1)}\sqrt{\frac{\pi}{2}}
\frac{(h^2_+)^{q/2}(2z_+)^q}
{[\frac{1}{2}(q-1)]!}
e^{-\frac{1}{2}h^2_+z^2_+}.
\label{113}
\end{equation}
Now
\begin{eqnarray}
\sin\bigg\{\frac{\pi}{4}(q+1)\bigg\}
&\simeq & \sin\bigg\{\frac{\pi}{4}(q_0+1)\bigg\}
+\frac{\pi}{4}(q-q_0)\cos\bigg\{\frac{\pi}{4}(q_0+1)\bigg\}
+\cdots
\nonumber\\
&\simeq& (-1)^{\frac{1}{4}(q_0+1)}(q-q_0)\frac{\pi}{4}\;\;\;
{\rm for} \;\;\; q_0=3,7,11,\ldots\nonumber\\
\label{114}
\end{eqnarray}
and
\begin{eqnarray}
\cos\bigg\{\frac{\pi}{4}(q+1)\bigg\}
&\simeq & \cos\bigg\{\frac{\pi}{4}(q_0+1)\bigg\}
-\frac{\pi}{4}(q-q_0)\sin\bigg\{\frac{\pi}{4}(q_0+1)\bigg\}
+\cdots 
\nonumber\\
&\simeq& -(-1)^{\frac{1}{4}(q_0-1)}(q-q_0)\frac{\pi}{4}\;\;\;
{\rm for} \;\;\; q_0=1,5,9,\ldots
\nonumber\\
&=& -(q-q_0)\frac{\pi}{4}(-1)^{\frac{1}{4}(q_0+1)}(-1)^{1/2}.
\label{115}
\end{eqnarray}
Thus altogether we obtain
from the boundary conditions at $z_+$
(and correspondingly from those at $z_-$)
\begin{equation}
(q-q_0)\simeq \mp 4\sqrt{\frac{1}{2\pi}}\frac{(h^2_+)^{q_0/2}
(2z_+)^{q_0}}{[\frac{1}{2}(q_0-1)]!}e^{-\frac{1}{2}h^2_+z^2_+}, \;\;
\; q_0=1,3,5,\ldots .
\label{116}
\end{equation}
In Appendix~E (after calculation of 
turning points) we rederive this relation 
using the WKB solutions from above the turning points
matched (linearly) to their counterparts
below the turning points and then evaluated
at  the minimum.

In summary: We needed the solutions of type $A$ for the
definition of even and odd solutions. Since the
type $A$ solutions  are not
valid at the minima, we matched them to the solutions of
types $B$ and $C$ which are valid there
and hence permit the imposition of
boundary conditions at the minima.

\subsection{Boundary conditions at the origin}
\vspace{-0.3cm}
$\qquad\quad $ {\bf (A)$\,$  Formulation of the  boundary conditions}
\vspace{0.3cm}

Since our even and odd solutions are defined to be 
even or odd with respect to the origin, we must also
demand this behaviour here  along with a nonvanishing
Wronskian.

\vspace{0.3cm}
\begin{figure}[ht]
\centering
\includegraphics[angle=0,totalheight=8.0cm]{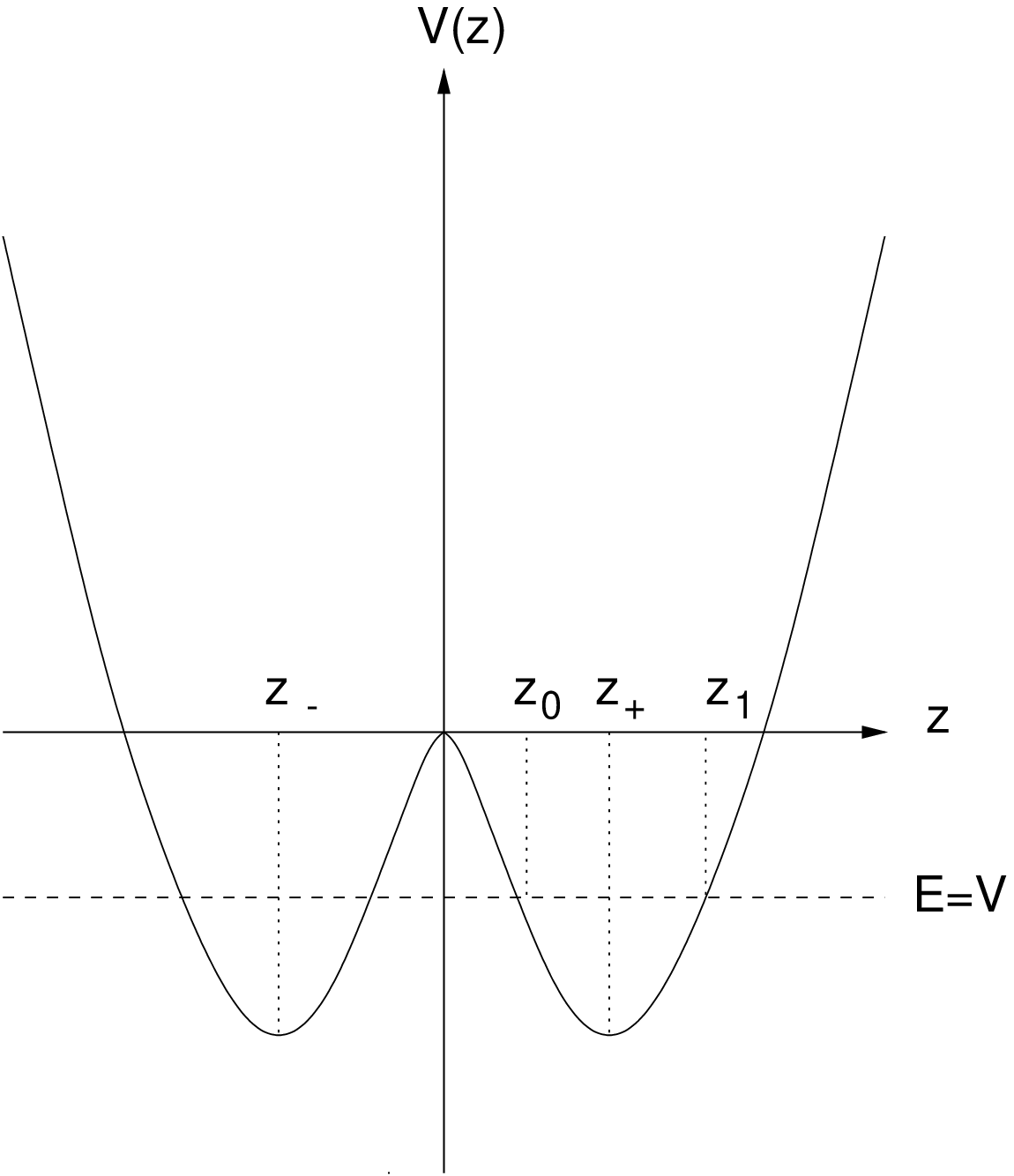}
\end{figure}
\centerline{\small Fig.~6 Turning points $z_0, z_1$.}

\vspace{0.3cm}

\noindent
Hence we have to impose at $z=0$ the conditions
\begin{eqnarray}
y_-(0)&=& 0, \;\;\; y^{\prime}_-(0) \neq 0,\nonumber\\
y_+(0) &\neq & 0, \;\;\; y^{\prime}_+(0) = 0.
\label{117}
\end{eqnarray}
Thus we require the extension of our solutions to
the region around the local maximum at $z=0$.  
We do this with the help of WKB solutions.
We deduce from Eq.~(\ref{97}) that the two
turning points at $z_0$ and $z_1$
to the right of the origin 
are given by
\begin{equation}
\frac{1}{2}qh^2_++\frac{\triangle}{h_+^4}-\frac{1}{4}h^4_+U(z)=0,
\label{118}
\end{equation}
i.e.
$$
\frac{1}{2}qh^2_++\frac{1}{4}z^2h^4-\frac{1}{2}c^2z^4
-\frac{h^8}{2^5c^2}=O\bigg(\frac{\triangle}{h_+^4}\bigg).
$$
Using $z_+=h^2/2c$, one finds that
\begin{eqnarray}
z_0&=&\bigg\{\frac{h^4}{4c^2}-\sqrt{\frac{qh^2_+}{c^2}}
+O\bigg(\frac{\triangle}{h_+^4}\bigg)\bigg\}^{1/2}
\nonumber\\
&=&\frac{h^2}{2c}-\frac{(2q^2)^{1/4}}{h}
+\frac{2^{1/2}cq}{h^4}+O\bigg(\frac{1}{h^5}\bigg)
\label{119}
\end{eqnarray}
and
\begin{equation}
z_1=\bigg\{\frac{h^4}{4c^2}+\sqrt{\frac{qh^2_+}{c^2}}
+O\bigg(\frac{\triangle}{h_+^4}\bigg)\bigg\}^{1/2}
\simeq \frac{h^2}{2c}+\frac{(2q^2)^{1/4}}{h}.
\label{120}
\end{equation}
The height of the potential at these points is
$$
V(z)|_{z_0,z_1}\simeq -\frac{h^8}{2^5c^2}+\frac{qh}{2^{1/2}}.
$$
Thus for large values of $h^2$ the turning points
are very close to minima of the
potential for nonasymptotically large values of $q$.

\setcounter{footnote}{0}           
\renewcommand{\thefootnote}{\fnsymbol{footnote}}

\vspace{0.3cm}

$\qquad\quad $ {\bf (B)$\,$  Evaluation
 of the  boundary conditions}
\vspace{0.3cm}

We now proceed to evaluate the boundary conditions
(\ref{117}). In the domain $0<z<z_0$, i.e. to the 
left of $z_0$ where $V>E$, the dominant
terms of the WKB solutions are\footnote{\scriptsize The
superscript $(l,z_0)$  means  ``to the left of
$z_0$''.}
\begin{eqnarray}
y^{(l,z_0)}_{\rm WKB}(z)
&=&\bigg[-\frac{1}{2}qh^2_++\frac{1}{4}h^4_+U(z)\bigg]^{-1/4}\nonumber\\
&& \times \exp\bigg(-\int^{z_0}_z\, dz
\bigg[-\frac{1}{2}qh^2_++\frac{1}{4}h^4_+U(z)\bigg]^{1/2}\bigg)
\label{121}
\end{eqnarray}
and
\begin{eqnarray}
{\overline  y}^{(l,z_0)}_{\rm WKB}(z)
&=&\bigg[-\frac{1}{2}qh^2_++\frac{1}{4}h^4_+U(z)\bigg]^{-1/4}\nonumber\\
&&\times \exp\bigg(\int^{z_0}_zdz
\bigg[-\frac{1}{2}qh^2_++\frac{1}{4}h^4_+U(z)\bigg]^{1/2}\bigg).
\label{122}
\end{eqnarray}
In order to be able to extend the even and odd solutions
to $z=0$, we have to match $y_A(z), {\overline y}_A(z)$
to $y^{(l,z_0)}_{\rm WKB}(z)$ and $ {\overline  y}^{(l,z_0)}_{\rm WKB}(z) $. 
We therefore have to consider the exponential
factors occurring in (\ref{121}) and (\ref{122})
and consider both types of solutions in a domain
approaching but not reaching the minimum of the
potential at $z_+$.
Thus we consider in the domain $|z-z_+|>(2q/h^2_+)^{1/2}$:
\begin{eqnarray}
I_1 &=& \int^{z_0}_z\, dz
\bigg[-\frac{1}{2}qh^2_++\frac{1}{4}h^4_+U(z)\bigg]^{1/2}\nonumber\\
&\simeq & \pm\frac{1}{2}h^2_+\int^{z_0}_z\, dz\bigg[(z-z_+)^2
-\frac{2q}{h^2_+}\bigg]^{1/2}\nonumber\\
&=& \pm\frac{1}{2}h^2_+\int^{z_0-z_+}_{z-z_+}\, dz^{\prime}\bigg(
{z^{\prime}}^2-\frac{2q}{h^2_+}\bigg)^{1/2}
\nonumber\\
&=&\pm\frac{1}{2}h^2_+\bigg[\frac{1}{2}z^{\prime}
\bigg({z^{\prime}}^2-\frac{2q}{h^2_+}\bigg)^{1/2}
\nonumber\\
&& -\frac{q}{h^2_+}\ln\bigg|z^{\prime}+
\bigg({z^{\prime}}^2-\frac{2q}{h^2_+}\bigg)^{1/2}
\bigg|\bigg]^{z_0-z_+}_{z-z_+}.
\label{123}
\end{eqnarray}
Evaluating this we have (with $z_0-z_+=-(2q^2)^{1/4}/h=-(2q/h^2_+)^{1/2}$)
\begin{eqnarray}
\pm I_1 &\simeq & 
\frac{1}{2}h^2_+\bigg[\frac{1}{2}\bigg\{{z^{\prime}}^2
\bigg(1-\frac{q}{h^2_+{z^{\prime}}^2}\bigg)\bigg\}_{z^{\prime}
=-(2q/h^2_+)^{1/2}}
\nonumber\\
&& -\frac{q}{h^2_+}\ln\bigg|
-\bigg(\frac{2q}{h^2_+}\bigg)^{1/2}+O\bigg(\frac{1}{h^5}\bigg)\bigg|
-\frac{1}{2}(z-z_+)\bigg\{(z-z_+)^2-\frac{2q}{h^2_+}\bigg\}^{1/2}
\nonumber\\
&& +\frac{q}{h^2_+}\ln\bigg|(z-z_+)+\bigg\{(z-z_+)^2
-\frac{2q}{h^2_+}\bigg\}^{1/2}\bigg|\bigg]\nonumber\\
&\simeq & \frac{1}{4}q
+\frac{1}{4}q\ln\bigg|\frac{h^2_+}{2q}\bigg|
-\frac{1}{4}h^2_+(z-z_+)^2
+\frac{1}{2}q\ln|2(z-z_+)|.
\label{124}
\end{eqnarray}
In identifying the WKB exponentials we recall that
${\overline y}_{\rm WKB}$ is exponentially
increasing and $y_{\rm WKB}$ exponentially decreasing.
Hence we have for $y_{\rm WKB}$ the exponential factor
\begin{subequations}
\begin{eqnarray}
&& \exp\bigg(-\int^{z_0}_z\, dz\bigg[
-\frac{1}{2}qh^2_++\frac{1}{4}h^4_+U(z)\bigg]^{1/2}\bigg)\nonumber\\
&\simeq & |z-z_+|^{q/2}e^{\frac{1}{4}q}\bigg(\frac{1}{2}h^2_+\bigg)^{q/4}
\bigg(\frac{q}{4}\bigg)^{-q/4}e^{-\frac{1}{4}(z-z_+)^2h^2_+}.
\label{125a}
\end{eqnarray}
We observe here that the WKB solution involves
unavoidably the quantum-number-dependent factors
$\exp(q/4)$ and $(q/4)^{q/4}$ which do not appear
as such in the perturbation
solutions. The only way to relate these solutions is with
the help of the Stirling formula which converts the product
or ratio of such factors into
factorials such as those inserted from the beginning
into the unperturbed solutions (\ref{98a}) and (\ref{99a}).
However, Stirling's formula is the dominant term of
the asymptotic expansion of a factorial or
gamma function and thus assumes the
argument ($\propto q\sim 2n+1$) to be large (it is known,
of course, that the Stirling approximation is
amazingly good even for small values of the argument).
Thus, using 
 the Stirling formula
$z!\simeq e^{-(z+1)}(z+1)^{z+\frac{1}{2}}\sqrt{2\pi}$,
we can rewrite the exponential as
\begin{eqnarray}
&& \exp\bigg(-\int^{z_0}_z\, dz\bigg[
-\frac{1}{2}qh^2_++\frac{1}{4}h^4_+U(z)\bigg]^{1/2}\bigg)\nonumber\\
&\simeq & (2\pi)^{1/2}\frac{|z-z_+|^{q/2}}{[
\frac{1}{4}(q-1)]!}\bigg(\frac{1}{2}h^2_+\bigg)^{q/4}
e^{-\frac{1}{4}(z-z_+)^2h^2_+}.
\label{125b}
\end{eqnarray}
\end{subequations}
Here $q/4$ was assumed to be large but we write
$[\frac{1}{4}(q-1)]!$ since this is the factor appearing in
the solution (\ref{98a}). We see therefore, since there is no way to 
obtain
an exact leading order
approximation with Stirling's formula for small values of $q$,
the results necessarily require adjustment or
normalisation there in the $q$-dependence.  This is the
aspect investigated by Furry \cite{21}. 
Since correspondingly
\begin{equation*}
\bigg[
-\frac{1}{2}qh^2_++\frac{1}{4}h^4_+U(z)\bigg]^{-1/4}
\simeq \frac{2^{1/2}}{(z-z_+)^{1/2}(h^4_+)^{1/4}},
\end{equation*}
we obtain
\begin{eqnarray}
y^{(l,z_0)}_{\rm WKB}(z)
&\simeq &
\frac{2\sqrt{\pi}}{(h^4_+)^{1/4}}
\frac{|z-z_+|^{\frac{1}{2}(q-1)}}{
[\frac{1}{4}(q-1)]!}\bigg(\frac{1}{2}h^2_+\bigg)^{q/4}
e^{-\frac{1}{4}(z-z_+)^2h^2_+}\nonumber\\
&& {\rm for} \;\;\; |z-z_+|>\bigg(\frac{2q}{h^2_+}\bigg)^{1/2}.
\label{126}
\end{eqnarray}
This expression is valid to the left of the turning point
at $z_0$ above the minimum at $z_+$.

In a corresponding manner     
--- i.e. using Stirling's formula (and not the inversion
relation) --- 
we have
\begin{equation}
{\overline y}^{(l,z_0)}_{\rm WKB}(z)  
= \frac{1}{\sqrt{\pi}(h^4_+)^{1/4}}\frac{[\frac{1}{4}(q-3)]!}
{|z-z_+|^{\frac{1}{2}(q+1)}}\bigg(\frac{1}{2}h^2_+\bigg)^{-q/4}
e^{\frac{1}{4}(z-z_+)^2h^2_+},
\label{127}
\end{equation}
where $[\frac{1}{4}q]!$ was written as $[\frac{1}{4}(q-3)]!$
for $q$ large.
Comparing for $z\rightarrow z_+$
 the WKB solutions (\ref{126}), (\ref{127})
with the type-A solutions (\ref{100}), 
(\ref{104}) we obtain in leading order
(i.e. multiplied by $(1+O(1/h^2_+))$)
 the proportionality
constants $\gamma, {\overline \gamma}$ of the 
matching relations
\begin{equation}
y^{(l,z_0)}_{\rm WKB}(z)=\gamma y_A(z),\; \;\;\;
{\overline y}^{(l,z_0)}_{\rm WKB}(z)
={\overline \gamma}\,{\overline y}_A(z),
\label{128}
\end{equation}
i.e.
\begin{subequations}
\begin{equation}
\gamma =\frac{2\sqrt{\pi}}{(h^4_+)^{1/4}[\frac{1}{4}(q-1)]!}
\bigg(\frac{1}{2}h^2_+\bigg)^{q/4}
(2z_+)^{\frac{1}{2}(q+1)}e^{-\frac{1}{4}h^2_+z^2_+} 
\label{129a}
\end{equation}
and
\begin{equation}
{\overline \gamma} =\frac{[\frac{1}{4}(q-3)]!}
{\sqrt{\pi}(h^4_+)^{1/4}}
\bigg(\frac{1}{2}h^2_+\bigg)^{-q/4}
(2z_+)^{-\frac{1}{2}(q-1)}e^{\frac{1}{4}h^2_+z^2_+}. 
\label{129b}
\end{equation}
\end{subequations}
Using again the duplication formula\footnote{\scriptsize In
the present case as
$$
[\frac{1}{4}(q-1)]![\frac{1}{4}(q-3)]!=\sqrt{\pi}
2^{-\frac{1}{2}(q-1)}[\frac{1}{2}(q-1)]!.
$$
}
the ratio of these
constants becomes
\begin{equation}
\frac{\gamma}{\overline \gamma}
=\sqrt{2\pi}\frac{(h^2_+)^{q/2}(2z_+)^q}{[\frac{1}{2}(q-1)]!}
e^{-\frac{1}{2}h^2_+z^2_+}.
\label{130}
\end{equation}
Since the factorials $[\frac{1}{4}(q-1)]!,  [\frac{1}{4}(q-3)]!$
are really correct replacements of $ [\frac{1}{4}(q)]!$
only for $q$ large, this result
is somewhat imprecise. However, it is our
philosophy here that the factorials with factors
occurring in the perturbation solutions
are the more natural and hence correct expressions,
as the results also seem to support.
The  relation (\ref{130})  is used  in Appendix~E  for the
calculation of
the tunneling deviation  $q-q_0$ by using the usual 
(i.e. linearly matched)  WKB solutions,
and is shown to reproduce correctly
the result (\ref{116}) which was
obtained with our perturbation  solutions from the
boundary conditions at a minimum.

Returning to the even and odd solutions (\ref{96})
we have
\begin{equation}
y_{\pm}(z)=\frac{1}{2}[y_A(z)\pm{\overline y}_A(z)]
=\frac{1}{2\gamma}y^{(l,z_0)}_{\rm WKB}(z)\pm
\frac{1}{2{\overline \gamma}}
{\overline y}^{(l,z_0)}_{\rm WKB}(z).
\label{131}
\end{equation}
Applying the boundary conditions (\ref{117}) we obtain
\begin{equation}
\frac{y^{(l,z_0)}_{\rm WKB}(0)}{{\overline y}^{(l,z_0)}_{\rm WKB}(0)}
=\frac{\gamma}{\overline \gamma}, \;\;\; 
\frac{y^{(l,z_0)\, {\prime}}_{\rm WKB}(0)}
{{\overline y}^{(l,z_0)\, {\prime}}_{\rm WKB}(0)}
=-\frac{\gamma}{\overline \gamma}.
\label{132}
\end{equation}
Thus we have to consider the behaviour of the integrals
occurring in the solutions (\ref{121}), (\ref{122})
near $z=0$. We have
\begin{eqnarray*}
I_2(z) &\equiv & \int^{z_0}_z dz \bigg[-\frac{1}{2}qh^2_++\frac{1}{4}
h^4_+U(z)\bigg]^{1/2}\nonumber\\
&=&  
\int^{z_0}_z dz \bigg[-\frac{1}{2}qh^2_+-\frac{1}{4}
z^2h^4+\frac{1}{2}c^2z^4+\frac{h^8}{2^5c^2}\bigg]^{1/2}\nonumber\\
&=&\int^{z_0}_z dz\bigg[\frac{1}{2}\bigg(\frac{h^4}{2^2c}-cz^2\bigg)^2
-\frac{1}{2}qh^2_+\bigg]^{1/2}\nonumber\\
&=&\frac{c}{\sqrt{2}}\int^{z_0}_zdz\bigg[\bigg(
\frac{h^4}{4c^2}-\sqrt{q}\frac{h_+}{c}-z^2\bigg)
\bigg(\frac{h^4}{4c^2}+\sqrt{q}\frac{h_+}{c}-z^2\bigg)\bigg]^{1/2}.
\end{eqnarray*}
Setting 
\begin{equation}
b^2=\frac{h^4}{4c^2}-\sqrt{q}\frac{h_+}{c}\simeq z^2_0, \;\;\;
a^2=\frac{h^4}{4c^2}+\sqrt{q}\frac{h_+}{c}, \;\;\; b^2<a^2,
\label{133}
\end{equation}
we can rewrite the integral as
\begin{equation*}
I_2(z)=\frac{c}{\sqrt{2}}\int^b_z\, dz\sqrt{(a^2-z^2)(b^2-z^2)}.
\end{equation*}
The integral appearing here is an elliptic integral
which can be looked up in 
Ref. \cite{29}.\footnote{\scriptsize
See Ref. \cite{29}, formulae 220.05,  p. 60 and
361.19, p. 213.}
The  elliptic modulus $k$ (with ${k^{\prime}}^2=
1-k^2$) and an expression $u$  
appearing in the integral are
defined by
\begin{subequations}
\begin{equation}
k^2=\frac{b^2}{a^2}\equiv \frac{1-u}{1+u}, \;\;\;
u=\frac{8c}{h^3_+}\sqrt{q}\equiv G\sqrt{2q}, 
\;\;\;G^2=\frac{8\sqrt{2}c^2}{h^6},
\label{134a}
\end{equation}
so that
\begin{equation}
a=\frac{h^2}{2c}[1+u]^{1/2}.
\label{134b}
\end{equation}
\end{subequations}
The integral $I_2(z)$ evaluated at $z=0$ is then
given in Ref. \cite{29}  by
\begin{subequations}
\begin{eqnarray}
I_2(0)&=&\frac{ca^3}{3\sqrt{2}}[(1+k^2)E(k)-
{k^{\prime}}^2K(k)]\nonumber\\ 
&=&\frac{h^6}{12\sqrt{2}c^2}\sqrt{1+u}[E(k)-uK(k)],
\nonumber\\
&=& \frac{2}{3G^2}\sqrt{1+u}[E(k)-uK(k)],
\label{135a}
\end{eqnarray}
where $K(k)$ and $E(k)$ are the complete elliptic
integrals of the first and second kinds  respectively. 
Here we are interested in the behaviour of the integral
in the domain of large $h^6/c^2$ which implies small
$G^2$. The nontrivial expansions are derived  in Appendix~F,
where the result is shown to be (with  $q\simeq q_0=2n+1, 
n=0,1,2,\ldots$)
\begin{eqnarray}
I_2(0)&=&
\mp\frac{q}{4}\pm\frac{1}{2^{1/2}}\frac{h^6}{12c^2}
\mp\frac{q}{2}\ln\bigg|\frac{2 h^3}{2^{1/4}cq^{1/2}}\bigg|
\nonumber\\
&=&\pm \frac{2}{3G^2}\pm\bigg(n+\frac{1}{2}\bigg)\ln\bigg(
\frac{G}{4}\bigg)\pm\frac{1}{2}\bigg(n+\frac{1}{2}\bigg)
\ln\bigg(n+\frac{1}{2}\bigg)
\nonumber\\
&& \qquad \mp\frac{1}{2}\bigg(n+\frac{1}{2}\bigg).
\label{135b}
\end{eqnarray}
\end{subequations}
in agreement with Ref. \cite{10}.\footnote{\scriptsize The expansion of
$I_2$ used in Ref. \cite{20} misses  an n-dependent  power
of $2$ in the result.}
Since
the integral is to be positive  (as required in the WKB solutions)
we have to take the upper signs. It then follows that
(again in each case in leading order)
\begin{equation}
y^{(l,z_0)}_{\rm WKB}(z)|_{z=0}
\simeq \frac{
(\frac{2h^6}{c^2})^{q/4}}
{2^{3q/8}(\frac{h^8}{2^5c^2})^{1/4}(\frac{q}{4})^{q/4}
e^{-q/4}}e^{-\frac{h^6}{12\sqrt{2}c^2}},
\label{136}
\end{equation}
where $h^8/2^5c^2=-V(z_{\pm})$.
Using Stirling's formula we can write this
\begin{equation}
y^{(l,z_0)}_{\rm WKB}(z)|_{z=0}
\simeq \frac{\sqrt{2\pi}
(\frac{2h^6}{c^2}\frac{1}{2^{3/2}})^{q/4}}
{(\frac{h^8}{2^5c^2})^{1/4}[\frac{1}{4}(q-1)]!}
e^{-\frac{h^6}{12\sqrt{2}c^2}}.
\label{137}
\end{equation}
Correspondingly we find
\begin{equation}
{\overline y}^{(l,z_0)}_{\rm WKB}(z)|_{z=0}
\simeq \sqrt{\frac{1}{2\pi}}\frac{[\frac{1}{4}(q-3)]!}{
(\frac{2h^6}{c^2}\frac{1}{2^{3/2}})^{q/4}
(\frac{h^8}{2^5c^2})^{1/4}}
e^{+\frac{h^6}{12\sqrt{2}c^2}}
\label{138}
\end{equation}
and
\begin{equation}
\frac{d}{dz}{y}^{(l,z_0)}_{\rm WKB}(z)\bigg|_{z=0}
\simeq 
\bigg(\frac{h^8}{2^5c^2}\bigg)^{1/4}
\frac{(2\pi)^{1/2}
(\frac{2h^6}{c^2}\frac{1}{2^{3/2}})^{q/4}}
{[\frac{1}{4}(q-1)]!}
e^{-\frac{h^6}{12\sqrt{2}c^2}}
\label{139}
\end{equation}
and
\begin{equation}
\frac{d}{dz}{\overline y}^{(l,z_0)}_{\rm WKB}(z)\bigg|_{z=0}
\simeq 
-\bigg(\frac{h^8}{2^5c^2}\bigg)^{1/4}
\frac{[\frac{1}{4}(q-3)]!}
{(2\pi)^{1/2}
(\frac{2h^6}{c^2}\frac{1}{2^{3/2}})^{q/4}}
e^{\frac{h^6}{12\sqrt{2}c^2}}.
\label{140}
\end{equation}
With these expressions we obtain now from
Eqs.~(\ref{132}) --- on using once  again the
duplication formula in the same form as above ---
\begin{equation}
\sqrt{2\pi}\frac{2^{q/2}}{[\frac{1}{2}(q-1)]!}
\bigg(\frac{2h^6}{c^2}\frac{1}{2^{3/2}}\bigg)^{q/2}
e^{-\frac{h^6}{6\sqrt{2}c^2}}
=\frac{\gamma}{\,{\overline \gamma}}.
\label{141}
\end{equation}
Comparing this result with that of
Eq.~(\ref{130}), we can therefore
impose the boundary conditions at $z=0$
by making in Eq.~(\ref{116})
the replacement
\begin{equation}
(h^2_+)^{q/2}(2z_+)^q e^{-\frac{1}{2}h^2_+z^2_+}
\Rightarrow 2^{q}\bigg(\frac{h^6}{2^{3/2}c^2}\bigg)^{q/2}
e^{-\frac{h^6}{6\sqrt{2}c^2}}.
\label{142}
\end{equation}
A corresponding relation holds for $h_+$ and
$z_+$ replaced by $h_-$ and $z_-$.
Inserting the expressions for $h_+$ and $z_+$ in terms
of $h$,  we see that the relation is
really an identity (the pre-exponential factors on the
left and on the right being equal) 
with the exponential on the left
being an approximation of the exponential on the right
(which contains the full action of the instanton).
This is our second condition  in the
present case of the double well potential.

\subsection{Eigenvalues and level splitting}

We now insert the relation (\ref{142}) into Eq.~(\ref{116})
with $q_0=2n+1, n=0,1,2,\ldots$,
and obtain
\begin{equation}
(q-q_0)\simeq \mp 4\sqrt{\frac{1}{2\pi}}
\frac{2^{q_0}(\frac{h^6}{2c^2})^{q_0/2}}{2^{q_0/4}
[\frac{1}{2}(q_0-1)]!}
e^{-\frac{h^6}{6\sqrt{2}c^2}}, \;\; q_0=1,3,5,\ldots .
\label{143}
\end{equation}
We obtain the energy $E(q,h^2)$  and hence the splitting of
asymptotically degenerate energy levels 
with the help of Eq.~(\ref{106}). Expanding
this around an odd integer $q_0$ we have
\begin{eqnarray}
E(q,h^2)&\simeq & E_0(q_0,h^2)
+(q-q_0)\bigg(\frac{\partial E}{\partial q}\bigg)_{q_0}
\nonumber\\
&\simeq & E_0(q_0,h^2)+(q-q_0)\frac{h^2}{\sqrt 2}.
\label{144}
\end{eqnarray}
Inserting here the result (\ref{143}) we 
obtain for $E(q_0,h^2)$ the split expressions
\begin{equation}
E_{\pm}(q_0,h^2) \simeq E_0(q_0,h^2) 
\mp 
\frac{2^{q_0+1}h^2(\frac{h^6}{2c^2})^{q_0/2}}{\sqrt{\pi}2^{q_0/4}
[\frac{1}{2}(q_0-1)]!}
e^{ -\frac{h^6}{6\sqrt{2}c^2}}, \;\; q_0=1,3,5,\ldots,
\label{145}
\end{equation}
where $E_0(q_0,h^2)$ is given by 
Eq.~(\ref{C.1}), i.e.\footnote{\scriptsize In Ref. \cite{18}
 the ``usual WKB approximation'' of this  expression
is --- with replacements  
$h^4/4\leftrightarrow k, c^2/2\leftrightarrow\lambda$ --- given as
$$
E_0=-\frac{k^2}{4\lambda}+(2k)^{1/2}q-\frac{3}{4}\frac{\lambda}{k}q^2,
$$
i.e. the following expression for large $q\sim 2n+1$, which 
evidently supplies some correction terms. }
$$
E_0(q_0,h^2)=-\frac{h^8}{2^5c^2}+\frac{1}{\sqrt{2}}q_0h^2
-\frac{c^2(3q^2+1)}{2h^4}+O\bigg(\frac{1}{h^{10}}\bigg).
$$
Thus
\begin{eqnarray}
\triangle E(q_0,h^2)
&=& E_-(q_0,h^2) - E_+(q_0,h^2)\nonumber\\
& \simeq &
\frac{2^{q_0+2}h^2}{\sqrt{\pi}2^{{q_0}/4}[\frac{1}{2}(q_0-1)]!}
\bigg(\frac{h^6}{2c^2}\bigg)^{{q_0}/2}
e^{ -\frac{h^6}{2^{1/2}6c^2}}, \;\; \bigg({\rm mass}\; 
m_0=\frac{1}{2}\bigg)\nonumber\\
&=&
2^{(2n+9)/4}\frac{h^2}{\sqrt{\pi}n!}\bigg(\frac{h^6}{c^2}\bigg)^{n+1/2}
e^{-\frac{h^6}{2^{1/2}6c^2}}.
\label{146}
\end{eqnarray}
Combining Eqs.~(\ref{144}) and  (\ref{E.14}) with (\ref{132}),  
the level splitting, i.e. the difference between the
eigenenergies of even and odd states with (here) 
$q_0=2n+1, n=0,1,2,\ldots$,
for finite $h^2$,  can be  given by
\begin{eqnarray}
\triangle E(q_0,h^2)&\simeq &
\frac{4}{\pi}\bigg(\frac{\partial E}{\partial q}\bigg)_{\!\! q_0}
\frac{y^{(l,z_0)}_{\rm WKB}(0)}{{\overline y}^{(l,z_0)}_{\rm WKB}(0)}
\nonumber\\
&=& 
\frac{2}{\pi}\bigg(\frac{\partial E}{\partial (n+\frac{1}{2})}
\bigg)
\frac{y^{(l,z_0)}_{\rm WKB}(0)}{{\overline y}^{(l,z_0)}_{\rm WKB}(0)}.
\label{147}
\end{eqnarray}
In Ref.~\cite{18}
the  WKB level splitting is
effectively 
(i.e. in the WKB restricted sense)
defined by
\begin{equation}
\triangle^{\rm WKB}_n=\frac{1}{\pi}\frac{\partial E}
{\partial (n+\frac{1}{2})}
\frac{y^{(l,z_0)}_{\rm WKB}(0)}{{\overline y}^{(l,z_0)}_{\rm WKB}(0)}
=\frac{1}{2}\triangle E(q_0,h^2).
\label{148}
\end{equation}
Here $\partial E/\partial n$ corresponds to the usual oscillator
frequency.
The result 
(\ref{146}) is in  Ref.~\cite{18}
 described as the ``{\it modified well
and barrier}''  result $\triangle^{\rm MWB}_n$,
 the pure  WKB result of Ref.~\cite{16}
(i.e. that without the use of Stirling's formula and so 
left in terms of $e$ and $n^n$) 
being this divided
by the Furry factor
$$
f_n:=\bigg[\frac{1}{2\pi}\bigg(\frac{e}{n+\frac{1}{2}}\bigg)^{n+\frac{1}{2}}
n!\bigg]^{-1},
$$
which is unity for $n\rightarrow\infty$ with Stirling's
formula.\footnote{\scriptsize  Ref. \cite{21}, Eq.~(66).}
Of course,  these derivations do not
exploit the symmetry of the original equation
under the interchanges $q\leftrightarrow -q, h^2\leftrightarrow -h^2$,
as we do here. We see therefore, that if this
symmetry is taken into account from the
very beginning (with the appropriate use
of the Stirling formula) the Furry factor corrected
result follows automatically.
The Furry factor represents
effectively a correction factor  to the normalisation 
constants of WKB wave functions 
(which are normally $1/\sqrt{2\pi}$
and independent of $n$  as explained in Ref.~\cite{21})
to yield an improvement of WKB results  for small values of $n$,
as is also  explained 
in Ref.~\cite{18}.\footnote{\scriptsize Comparing 
the present work with 
that of Ref. \cite{18},  we identity the parameter $k$ there
with $h^4/4$ here, and $\lambda$ there with $c^2/2$ here.
Then the splitting $\triangle^{\rm MWB}_n$ of  Eq.~(24) there  is
\begin{eqnarray*}
\triangle^{\rm MWB}_n
&=&\sqrt{\frac{k}{\pi}}\frac{2^{(10n+13)/4}}{n!}
\bigg(\frac{k^{3/2}}{\lambda}\bigg)^{n+1/2}
\exp\bigg[-\frac{\sqrt{2}}{3}\frac{k^{3/2}}{\lambda}\bigg]
\nonumber\\
&=&\frac{h^2}{\sqrt{\pi}n!}2^{(2n+5)/4}\bigg(\frac{h^6}{c^2}\bigg)^{n+1/2}
\exp\bigg[-\frac{h^6}{6\sqrt{2}c^2}\bigg]
\end{eqnarray*}
in agreement with $\triangle E(q_0,h^2)/2$ of Eq.~(\ref{146}). 
}

\setcounter{footnote}{0}                       
\renewcommand{\thefootnote}{\fnsymbol{footnote}}

Had we taken the mass $m_0$ of the particle in  the 
symmetric double-well potential equal to $1$ (instead of $1/2$),
we would have obtained the result with
$E, h^4$ and $c^2$ replaced by $2E, 2h^4$ and $2c^2$ respectively
(see Eq.~(\ref{3})).  Then
\begin{equation}
E(q_0,h^2)=E_0(q_0,h^2)
\mp\frac{2^{q_0+\frac{1}{2}}h^2(\frac{h^6}{\sqrt{2}c^2})^{q_0/2}}
{\sqrt{\pi}2^{q_0/4}[\frac{1}{2}(q_0-1)]!}
e^{-\frac{h^6}{6c^2}}.
\label{149}
\end{equation}
and  $\triangle E$ becomes
\begin{equation}
\triangle_1 E(q_0,h^2) \simeq 
2^{q_0}\sqrt{\frac{2}{\pi}}\frac{2h^2}
{2^{{q_0}/4}[\frac{1}{2}(q_0-1)]!}
\bigg(\frac{h^6}{2^{1/2}c^2}\bigg)^{q_0/2}
e^{-h^6/6c^2}, \;\;(m_0=1)
\label{150}
\end{equation}
with
$$
E_0(q_0,h^2)|_{m_0=1}=-\frac{h^8}{2^4c^2}+q_0h^2-\cdots .
$$
If in addition the potential is written in the form
\begin{equation}
V(z)=\frac{\lambda}{4}\bigg(z^2-\frac{\mu^2}{\lambda}\bigg)^2,
 \;\; \lambda >0, 
\label{151}
\end{equation}
a form frequently used in field theoretic applications, 
so that by comparison with Eq.~(\ref{78})
$ h^4\equiv 2\mu^2, c^2=\lambda/2$, the level splitting
is
\begin{equation}
\triangle_1E(q_0,h^2)\simeq
\frac{2^{q_0+2}\mu}{{\pi}^{1/2}2^{q_0/4}[\frac{1}{2}(q_0-1)]!}
\bigg(\frac{4\mu^3}{\lambda}\bigg)^{q_0/2}
e^{ - 8^{1/2}\mu^3/3\lambda}.
\label{152}
\end{equation}
This result agrees with the ground state ($n=0$)  result of
Ref.  \cite{9} 
using the path-integral method for the
evaluation of pseudoparticle (instanton)
contributions.\footnote{\scriptsize See 
Ref.  \cite{9}, formula (4.11).
The definition as $\triangle^{\rm WKB}_n$ is used, so that
the result of Eq.~(\ref{152}) differs by the factor 2
in Eq.~(\ref{148}).}
Equation (\ref{150}) agrees also with the
result of Ref.~\cite{10} for arbitrary levels
obtained with the use of periodic 
instantons\footnote{\scriptsize To help the 
comparison note that in Ref.~\cite{10} the potential is
written as
$$
V(z)=\frac{\eta^2}{2}\bigg[z^2-\frac{m^2}{\eta^2}\bigg]^2   
$$
for $m_0=1$.  The  comparison with Eq.~(\ref{149})
therefore implies the correspondence
$ \eta^2\leftrightarrow  \lambda/2, m^2\leftrightarrow \mu^2/2$,
and $g^2$ is given as $g^2=\eta^2/m^3$.}
and with the results  of Ref.~\cite{14} using multi-instanton 
methods.\footnote{\scriptsize Ref.~\cite{14}, 
first paper, Eqs.~(2.34) and (E.15).}

\section{Concluding  Remarks}

In the above we have presented a fairly complete
treatment of the large-$h^2$ case of the quartic
anharmonic oscillator,  carried out along the lines
of the corresponding calculations for the cosine
potential and thus of the well-established Mathieu
equation. In principle one could also consider the
case of small values of $h^2$ and obtain
convergent instead of asymptotic expansions; 
however,  these are presumably  not of much interest
in physics. We considered only the
symmetric two-minima potential. The asymmetric
case can presumably also be dealt with in a 
similar way since various references  point out
that the asymmetric double well case  can be
transformed into a symmetric 
one (see e.g. Refs.~\cite{1}. \cite{14}, \cite{16}).

Every now and then literature appears which
purports to overcome the allegedly ill-natured
``divergent perturbation series'' of the
anharmonic oscillator problem,  and reasons
and even numerical studies
are presented to support this  
claim. \cite{30}
It is  clear   --- as
also demonstrated by  the work of Bender and Wu
\cite{1}, \cite{2} --- that the expansions
considered above are asymptotic. Tables of
properties of Special Functions are filled with 
such expansions derived from
differential equations for all the well-known and
less well-known Special Functions. There is no
reason to view the anharmonic oscillator solutions 
differently.  In fact, in principle the
Schr\"odinger equation
of the quartic oscillator  is an equation akin
to equations like the Mathieu or modified
Mathieu equations which lie outside
the range  of equations of hypergeometric type.  The
immense amount of literature
meanwhile  accumulated for instance in
the case of the Mathieu equation 
can indicate   what else can be
achieved along parallel lines in the
case of special types of Schr\"odinger
equations, like that for anharmonic 
oscillators. Conversely, new methods discovered
for dealing with the quartic oscillator could
similarly be applied to the periodic
potential and tested there.  

The ground state splitting of the symmetric
double well potential  has been considered
in a countless number of investigations. 
A reasonable, though incomplete list
of references in this direction 
has been given by Garg \cite{31}
beginning with  the well-known
though nonexplicit (and hence
not really useful)  ground state
formula in the book of Landau and 
Lifshitz \cite{32}. 
Very illuminating discussions
of double-wells  and periodic potentials, mostly
in connection with instantons, can be found in an  article by
Coleman \cite{33}.
The wide publicity given to the work of
Bender and Wu \cite{2},
\cite{3} made pure mathematicians
aware of the subject; as some relevant 
references  with their approach   we cite 
Refs.~\cite{34}, \cite{35}, \cite{36},\cite{37},
\cite{38}, \cite{39}.
The double-well potential, in both symmetric and asymmetric
form, has also been the subject of numerous numerical
studies.  As references in this direction, though
not exclusively numerical, we cite papers of the
Uppsala group \cite{40}.
Perturbation theoretical aspects have also been  employed in
Ref.~\cite{41}.
Wave functions of symmetric and asymmetric 
double-well potentials have been considered in
Ref.~\cite{42}, in which  it is demonstrated that
actual physical tunneling takes place only into
those states which have significant overlap with
the false  vacuum eigenfunction.

\newpage

\vspace{3cm}
\centerline{\Large \bf  Appendices}

\setcounter{footnote}{0}                       
\renewcommand{\thefootnote}{\fnsymbol{footnote}}

\begin{appendix}

\setcounter{equation}{0}
\renewcommand{\theequation}{A.\arabic{equation}}

\section{Calculation of eigenvalues
with the  solutions of type ${\boldsymbol A}$
(inverted double well)}

Here we use the solutions of type $A$, i.e.
(\ref{20a}), (\ref{20b}),  to obtain in leading
order the eigenvalues $E$, i.e.
\begin{equation}
E(q,h^2)=\frac{1}{2}qh^2 +\frac{\triangle}{2h^4}, \;\;\;
\triangle =-\frac{3}{2}c^2(q^2+1)+O\bigg(\frac{1}{h^6}\bigg).
\label{A.1}
\end{equation}

 We rewrite the upper of Eqs.~(\ref{13}) in the following
form with power expansion of the square root quantities and
division by $h^2$:
\begin{eqnarray}
&&  -zA^{\prime}(z) +\frac{1}{2}(q-1)A(z)
\nonumber\\
&=&-\frac{1}{h^2}A^{\prime\prime}(z)
-\frac{\triangle}{2h^6}A(z)
 +\sum^{\infty}_{i=1}\bigg(\frac{2c^2z^2}{h^4}\bigg)^i
[\alpha_izA^{\prime}(z)
\nonumber\\
&&  \qquad +\frac{1}{2}\beta_iA(z)],
\label{A.2}
\end{eqnarray}
where the expansion coefficients are given by
$$
\alpha_0=1,\;\alpha_1=-\frac{1}{2}, \; \alpha_2=-\frac{1}{8}, \;
\alpha_3=-\frac{1}{16},\;\alpha_4=-\frac{5}{128}, \ldots,
$$
$$
 \beta_0=1, \; \beta_1=-\frac{3}{2}, \; \beta_2=-\frac{5}{8},
\;
\beta_3=-\frac{7}{16}, \ldots\, . 
$$
Using Eqs.~(\ref{17}) and (\ref{18}) we obtain
$$
A^{\prime}_q(z)=\frac{1}{2}\frac{(q-1)}{z}A_q(z)=\frac{1}{2}(q-1)A_{q-2}(z)
$$
and
\begin{eqnarray}
A^{\prime\prime}_q(z)&=&\bigg[\frac{1}{2}\frac{(q-1)}{z}\bigg]^2A_q(z)
-\frac{1}{2}\frac{(q-1)}{z^2}A_q(z)
\nonumber\\
&=&\frac{1}{4}(q-1)(q-3)A_{q-4}(z).
\label{A.3}
\end{eqnarray}
The lowest order solution $A^{(0)}=A_q(z)$ therefore leaves
uncompensated on the right hand side of the equation for $A$
the terms amounting to
\begin{eqnarray*}
R^{(0)}_q&=&-\frac{1}{4h^2}(q-1)(q-3)A_{q-4}-\frac{\triangle}{2h^6}A_q
-\frac{c^2}{2h^4}(q+2)A_{q+4}
\nonumber\\
&& -\bigg(\frac{c^2}{2h^4}\bigg)^2(q+4)A_{q+8}
 +\frac{1}{2}\sum^{\infty}_{i=3}\bigg(\frac{2c^2z^2}{h^4}\bigg)^i
[(q-1)\alpha_i+\beta_i]A_q.
\end{eqnarray*}
Clearly one now uses the relation
$$
z^{2i}A_q(z)=A_{q+4i}(z).
$$
In this way $R^{(0)}_q$ is expressed as a linear combination
of functions $A_{q+4i}(z)$. As always in the procedure, one
observes that 
\begin{equation}
{\cal D}_q:=- z\frac{d}{dz}+\frac{1}{2}(q-1), \;\;\;
{\cal D}_{q+4i}={\cal D}_q +2i, \;\; {\cal D}_q\frac{\mu A_{q+4i}}{-2i}
=\mu A_{q+4i}.
\label{A.4}
\end{equation}
Thus a term $\mu A_{q+4i}$ in $R^{(0)}_q$ can be taken care of
by adding to $A^{(0)}$ the contribution 
$\frac{\mu A_{q+4i}}{-2i}$ except, of course, when
$i=0$.  In this way we obtain
the next order contribution $A^{(1)}$ to $A^{(0)}$ and the 
coefficient of terms with $i=0$, i.e. those in $A_q(z)$, 
give an equation  from which $\triangle$  is 
determined. Hence
in the present case
\begin{eqnarray}
A^{(1)}&=&-\frac{1}{4h^2}(q-1)(q-3)\frac{A_{q-4}}{2}
-\frac{c^2}{2h^4}(q+2)\frac{A_{q+4}}{-2}-
\nonumber\\
&&\bigg(\frac{c^2}{2h^4}\bigg)^2(q+4)\frac{A_{q+8}}{-4}+\cdots.
\label{A.5}
\end{eqnarray}
In its turn $A^{(1)}$ leaves uncompensated terms amounting to
\begin{eqnarray*}
R^{(1)}_q&=&-\frac{1}{4h^2}(q-1)(q-3)\frac{1}{2}
\bigg\{\bigg(-\frac{1}{4h^2}\bigg)(q-5)(q-7)A_{q-8}
\nonumber\\
&& -\frac{c^2}{2h^4}(q-2)A_q
 -\frac{\triangle}{2h^6}A_{q-4}
-\bigg(\frac{2c^2}{h^4}\bigg)^2\frac{q}{16}A_{q+4}+\cdots
\bigg\}\nonumber\\
&& -\frac{c^2}{2h^4}\frac{(q+2)}{-2}\bigg\{
-\frac{1}{4h^2}(q+3)(q+1)A_q-\frac{c^2}{2h^4}(q+6)A_{q+8}
\nonumber\\
&&
-\frac{\triangle}{2h^6}A_{q+4}+\cdots\bigg\}+\cdots.
\end{eqnarray*}
The sum of terms with  $A_q$ in  
$R^{(0)}_q, R^{(1)}_q, \ldots $ must then be set equal to zero. Hence
to the order we are calculating here
\begin{eqnarray}
0&=&\frac{c^2}{(4h^2)(4h^4)}[(q-1)(q-2)(q-3)-(q+1)(q+2)(q+3)]
\nonumber\\
&&+O\bigg(\frac{1}{h^6}\bigg)-\frac{\triangle}{2h^6},
\label{A.6}
\end{eqnarray}
i.e.
$$
0=-\frac{c^2}{2^4h^6}12(q^2+1)-
\frac{\triangle}{2h^6}+O\bigg(\frac{1}{h^6}\bigg).
$$
It follows that
$$
\triangle = -\frac{3}{2}c^2(q^2+1)+O\bigg(\frac{1}{h^6}\bigg).
$$
The same result is obtained   in connection with
the solution of type $B$, as explained in the text.

\setcounter{equation}{0}
\renewcommand{\theequation}{B.\arabic{equation}}

\section{Evaluation of
$ \;\;{\boldsymbol y_{\boldsymbol B}{(\boldsymbol 0)}},
\; {\boldsymbol {\overline y}}_{\boldsymbol C}{(\boldsymbol 0)},\; 
{\boldsymbol y}_{\boldsymbol B}^{\prime}{(\boldsymbol 0)},\;
 {\boldsymbol {\overline y}}_{\boldsymbol C}^{\prime}{(\boldsymbol
0)}$}

We show that --- with $w=hz$ ---
 the leading terms of the quantities listed
are given by
\begin{subequations}
\begin{eqnarray}
B_q(0)&=& \frac{1}{\sqrt\pi}\frac{[\frac{1}{4}(q-3)]!} 
{[\frac{1}{4}(q-1)]!}\sin\{\frac{\pi}{4}(q+1)\},
\nonumber\\
{\overline C}_q(0)&=& \frac{\sqrt\pi}{[-\frac{1}{4}(q+1)]! 
[\frac{1}{4}(q-1)]!},
\nonumber\\
\bigg[\frac{d}{dw}B_q(w)\bigg]_0&=&-\sqrt{\frac{2}{\pi}}
\sin\{\frac{\pi}{4}(q+3)\}, 
\nonumber\\
\bigg[\frac{d}{dw}{\overline C}_q(w)\bigg]_0 &=&
i\sqrt{\frac{2}{\pi}}
\sin\{\frac{\pi}{4}(q-3)\}.
\label{B.1a}
\end{eqnarray}
In fact, with the reflection formula
 $(-z)!(z-1)!=\pi/\sin\pi z$, one finds that
\begin{equation}
[-\frac{1}{4}(q+1)]!=\frac{\pi}{[\frac{1}{4}(q-3)]!
\sin\{\frac{\pi}{4}(q+1)\}}
\;\;\; {\rm and \; hence}\;\;\;
\frac{B_q(0)}{{\overline C}_q(0)}=1.
\label{B.1b}
\end{equation}
\end{subequations}
From the literature, e.g. Ref.~ \cite{25}, we obtain
\begin{equation}
D_{\frac{1}{2}(q-1)}(0)=\frac{\sqrt{\pi}\,2^{\frac{1}{4}(q-1)}}
{[-\frac{1}{4}(q+1)]!},\;\;
D^{\prime}_{\frac{1}{2}(q-1)}(0)
=-
\frac{\sqrt{\pi}\,2^{\frac{1}{4}(q+1)}}
{[-\frac{1}{4}(q+3)]!}.
\label{B.2}
\end{equation}
Thus with the help of the reflection formula
cited above:
\begin{eqnarray}
B_q(0) & \stackrel{(\ref{23}), (\ref{98a})}{=}&
\frac{D_{\frac{1}{2}(q-1)}(0)}{[\frac{1}{4}(q-1)]!2^{\frac{1}{4}(q-1)}}
=\frac{\sqrt\pi}
{[\frac{1}{4}(q-1)]![-\frac{1}{4}(q+1)]!}
\nonumber\\
&=&
\frac{1}{\sqrt\pi}\frac{[\frac{1}{4}(q-3)]!} 
{[\frac{1}{4}(q-1)]!}\sin\{\frac{\pi}{4}(q+1)\}
\label{B.3}
\end{eqnarray}
and
\begin{eqnarray}
\bigg[\frac{d}{dw}B_q(w)\bigg]_0
&=&-\frac{\sqrt{2\pi}}
{[\frac{1}{4}(q-1)]![-\frac{1}{4}(q+3)]!}\nonumber\\ 
&=&-\sqrt{\frac{2}{\pi}}
\sin\{\frac{\pi}{4}(q+3)\},
\label{B.4}
\end{eqnarray}
Expressions for ${\overline C}_q(0), [{\overline C}^{\prime}_q(w)]
_{w=0}$ follow with the help of the ``{\it circuit relation}'' 
of parabolic cylinder functions given in the literature as
\begin{eqnarray*}
D_{\frac{1}{2}(q-1)}(w)
&=&e^{-i\frac{\pi}{2}(q-1)}D_{\frac{1}{2}(q-1)}(-w)
\nonumber\\
&+&\frac{\sqrt{2\pi}e^{-i\frac{\pi}{4}(q+1)}} 
{[-\frac{1}{2}(q+1)]!}
D_{-\frac{1}{2}(q+1)}(-iw).
\end{eqnarray*}
From this relation we obtain
$$
D_{-\frac{1}{2}(q+1)}(0)
=\sqrt{\frac{\pi}{2}}\frac{D_{\frac{1}{2}(q-1)}(0)}
{[\frac{1}{2}(q-1)]! 
\cos\{\frac{\pi}{4}(q-1)\}}.
$$
Inserting from (\ref{B.2}),  using the
above reflection  formula and the duplication
formula $\sqrt{\pi}(2z)!=2^{2z}z!(z-1/2)!$, we obtain
$$
D_{-\frac{1}{2}(q+1)}(0)=\frac{\sqrt\pi}
{2^{\frac{1}{4}(q+1)}[-\frac{1}{4}(q-1)]!}.
$$
From this we derive
\begin{eqnarray}
{\overline C}_q(0)&=&\frac{D_{-\frac{1}{2}(q+1)}(0)}
{2^{-\frac{1}{4}(q+1)}[-\frac{1}{4}(q+1)]!}
=\frac{\sqrt\pi}
{[-\frac{1}{4}(q+1)]![\frac{1}{4}(q-1)]!}
\nonumber\\
&=&\frac{[\frac{1}{4}(q-3)]!}{\sqrt{\pi}[\frac{1}{4}
(q-1)]!}\sin\{\frac{\pi}{4}(q+1)\}.
\label{B.5}
\end{eqnarray}
Similarly we obtain
\begin{equation}
[D^{\prime}_{-\frac{1}{2}(q+1)}(iw)]_{w=0}
=\frac{-i\sqrt{\pi}}{2^{\frac{1}{4}(q-1)}
[\frac{1}{4}(q-3)]!}
\label{B.6}
\end{equation}
and
\begin{eqnarray}
[{\overline C}^{\prime}_q(w)]_0& =&
\frac{D^{\prime}_{-\frac{1}{2}(q+1)}(0)}
{[-\frac{1}{4}(q+1)]!2^{-\frac{1}{4}(q+1)}}
=\frac{-i\sqrt{2\pi}}{[-\frac{1}{4}(q+1)]![\frac{1}{4}(q-3)]!}
\nonumber\\
&=&-i\sqrt{\frac{2}{\pi}}\sin\{\frac{\pi}{4}(q+1)\}
=i\sqrt{\frac{2}{\pi}}\sin\{\frac{\pi}{4}(q-3)\}.
\label{B.7}
\end{eqnarray}

\setcounter{equation}{0}
\renewcommand{\theequation}{C.\arabic{equation}}

\section{Calculation of 
eigenvalues with solutions of type 
${\boldsymbol A}$ (double well)}

We show  in conjunction with the derivation of  solution $y_A$ 
that the leading terms of $ \triangle$ and hence $E$
are given by
\begin{eqnarray}
{\triangle}&=& -c^2(3q^2+1) -\frac{\sqrt{2}c^4}{4h^6}q(17q^2+19)+\cdots
,\nonumber\\
E(q,h^2)&=&-\frac{h^8}{2^5c^2} +\frac{1}{\sqrt{2}}q h^2
 -\frac{c^2(3q^2+1)}{2h^4}
\nonumber\\
&& -\frac{\sqrt{2}c^4}{8h^{10}}q(17q^2+19)
+O\bigg(\frac{1}{h^{16}}\bigg).
\label{C.1}
\end{eqnarray}

The structure of the solution (\ref{93}) for $A_q(z)$  
is very similar to that of  the corresponding solution in 
considerations of other potentials, such as periodic
potentials. Thus it is natural to explore analogous
steps.  The first such step would be to
reexpress the right hand side of Eq.~(\ref{91a}) 
with $A$ replaced by $A_q$ as a linear combination  of
terms $A_{q+2i}, i=0, \pm 1, \pm 2, \ldots$. First, however,
we reexpress $A^{\prime\prime}_q$ in terms of functions
of $z$ multiplied by $A_q$. We know the first derivative
of $A_q$ from Eq.~(\ref{92}), i.e.
\begin{equation}
A^{\prime}_q(z)=\frac{qz_+-z}{z^2-z^2_+}A_q(z).
\label{C.2}
\end{equation}
Differentiation yields
\begin{eqnarray}
A^{\prime\prime}_q(z) &=& 
\frac{A_q(z)}{(z^2-z^2_+)^2}\bigg[(z-qz_+)^2
-(z^2-z^2_+)+2z(z-qz_+)\bigg]\nonumber\\
&=&    
\frac{A_q(z)}{(z^2-z^2_+)^2}[2z^2
-4zz_+q+z^2_+(q^2+1)].
\label{C.3}
\end{eqnarray}
We wish to rewrite this expression as a sum 
$$
\sum_i{\rm coefficient}_{i}A_{q+2i}(z).
$$
We also note at this stage the derivative of the entire
solution $y_A(z)$ taking into account only the dominant
contribution:
\begin{equation}
y^{\prime}_A(z)\simeq
\bigg[\frac{c}{\sqrt{2}}(z^2-z^2_+)
+\frac{(qz_+-z)}{(z^2-z^2_+)}\bigg]A_q(z)
\exp\bigg[\frac{1}{\sqrt 2}
\bigg\{\frac{c}{3}z^3-\frac{h^4}{4c}z\bigg\}\bigg].
\label{C.4}
\end{equation}
We observe some properties of the function $A_q(z)$ given by
Eq.~(\ref{93}):
\begin{equation}
A_{q+2i}A_{q+2j}=A_{q+2i+2j}A_q,\;\;\;
\frac{A_{q+2}}{A_q}=\frac{z-z_+}{z+z_+}=\frac{A_q}{A_{q-2}},
\label{C.5}
\end{equation}
\begin{equation}
\frac{A_{q+2}+A_{q-2}}{A_q}=2\frac{z^2+z^2_+}{z^2-z^2_+},
\;\;\;
\frac{A_{q+2}-A_{q-2}}{A_q}=-4\frac{zz_+}{z^2-z^2_+},
\label{C.6}
\end{equation}
\begin{equation}
\frac{(A_{q+2}-A_{q-2})^2}{A_q}=({4zz_+})^2
\frac{A_q}{(z^2-z^2_+)^2}=A_{q+4}-2A_q+A_{q-4},
\label{C.7}
\end{equation}
From these we obtain, for instance by componendo et dividendo,
\begin{eqnarray}
-\frac{z}{z_+}&=&\frac{A_{q+2}+A_q}{A_{q+2}-A_q}\nonumber\\
&=&-\bigg(1+\frac{A_{q+2}}{A_q}\bigg)\bigg[1+
\bigg(\frac{A_{q+2}}{A_q}\bigg)
+\bigg(\frac{A_{q+2}}{A_q}\bigg)^2
+\cdots \bigg]\nonumber\\
&=& -\bigg[1+2\frac{A_{q+2}}{A_q}+2\frac{A_{q+4}}{A_q}
+2\frac{A_{q+6}}{A_q}+\cdots \bigg]
\label{C.8}
\end{eqnarray}
Similarly we obtain
\begin{subequations}
\begin{eqnarray}
-\frac{z_+}{z} &=&\frac{A_{q+2}-A_q}{A_{q+2}+A_q}\nonumber\\
&=&-\bigg[1-2\frac{A_{q+2}}{A_q}+2\frac{A_{q+4}}{A_q}-
2\frac{A_{q+6}}{A_q}+\cdots \bigg],
\label{C.9a}
\end{eqnarray}
and from this 
\begin{equation}
\frac{z^2_+}{z^2}=1-4\frac{A_{q+2}}{A_q}+8\frac{A_{q+4}}{A_q}
-12\frac{A_{q+6}}{A_q}+16\frac{A_{q+8}}{A_q}-\cdots .
\label{C.9b}
\end{equation}
\end{subequations}
Hence with Eq.~(\ref{C.7})
\begin{eqnarray}
&& \frac{z}{(z^2-z^2_+)^2}A_q\nonumber\\
&=&\frac{z_+}{z_+z}
\bigg(\frac{1}{4z_+}\bigg)^2[A_{q+4}-2A_q+A_{q-4}]\nonumber\\
&=&\frac{1}{z_+}\bigg(\frac{1}{4z_+}\bigg)^2[A_{q+4}-2A_q+A_{q-4}]\bigg[1
-2\frac{A_{q+2}}{A_q}+2\frac{A_{q+4}}{A_q}-\cdots\bigg]\nonumber\\
&=&
\frac{1}{z_+}\bigg(\frac{1}{4z_+}\bigg)^2[A_{q-4}
-2A_{q-2}+ 0 +2A_{q+2}-A_{q+4} +\cdots ].
\label{C.10}
\end{eqnarray}
Finally with  Eqs.~(\ref{C.7}) and  (\ref{C.9b})
we obtain
\begin{eqnarray}
&& \frac{A_q}{(z^2-z^2_+)^2}
=\frac{z^2A_q}{(z^2-z^2_+)^2}\,\frac{1}{z^2}\nonumber\\
&=&\bigg(\frac{1}{4z_+}\bigg)^2[A_{q+4}-2A_q+A_{q-4}]\frac{1}{z^2_+}\bigg[1
-4\frac{A_{q+2}}{A_q}+8\frac{A_{q+4}}{A_q}-12\frac{A_{q+6}}{A_q}
+\cdots \bigg]\nonumber\\
&=&\bigg(\frac{1}{2z_+}\bigg)^4[A_{q-4}-4A_{q-2}+6A_q-4A_{q+2}
+A_{q+4}+\cdots ].
\label{C.11}
\end{eqnarray}
Inserting expressions 
(\ref{C.7}), (\ref{C.10}) and (\ref{C.11})
 into Eq.~(\ref{C.3}), 
we obtain\footnote{\scriptsize The reader
may observe the similarity with the corresponding 
coefficients in the simpler case of the cosine
potential, cf. Ref. \cite{5}. Infinite
series like here for solutions of type $A$
 arise in the more complicated
cases, e.g. in the corresponding
treatment of the elliptic potential
(Lam${\acute e}$ equation); cf. Ref. \cite{27}. } 
\begin{eqnarray}
A^{\prime\prime}_q&=& \bigg(\frac{1}{4z_+}\bigg)^2[ (q-1)(q-3)A_{q-4}
-4(q-1)^2A_{q-2} +2(3q^2+1)A_q\nonumber\\
&& -4(q+1)^2A_{q+2} 
+(q+1)(q+3)A_{q+4}\cdots ].
\label{C.12}
\end{eqnarray}
Hence the first approximation of $A$, 
$$
A^{(0)}=A_q,
$$
leaves uncompensated on the right hand side of Eq.~(\ref{91a})
the contribution
\begin{eqnarray}
R^{(0)}_q&=&-\frac{\sqrt 2}{2c}\bigg[A^{\prime\prime}_q+
\frac{\triangle}{h_{+}^4}A_q\bigg]\nonumber\\
&=&-\frac{\sqrt{2}c}{2^3h^4}[(q,q-4)A_{q-4}+(q,q-2)A_{q-2} 
+(q,q)A_q+(q,q+2)A_{q+2}
\nonumber\\
&& \qquad  +
(q,q+4)A_{q+4}+(q,q+6)A_{q+6}+\cdots],
\label{C.13}
\end{eqnarray}
where the lowest coefficients have been determined above as
\begin{eqnarray}
(q,q\mp 4)&=& (q\mp 1)(q\mp 3), \;\;\;
(q,q\mp 2)=-4(q\mp 1)^2,\nonumber\\
(q,q)&=&2(3q^2+1)+16z^2_+\frac{\triangle}{h_{+}^4}
=2(3q^2+1)+\frac{2}{c^2}\triangle.
\label{C.14}
\end{eqnarray}

It is now clear  how  the calculation of higher order 
contributions proceeds  in our standard way.  In particular the dominant
approximation of $\triangle$ is obtained by setting $(q,q)=0$, i.e.
\begin{eqnarray*}
{\triangle}&=& -c^2(3q^2+1),\nonumber\\
E(q,h^2)&=&-\frac{h^8}{2^5c^2} +\frac{1}{\sqrt{2}}q h^2
 -\frac{c^2(3q^2+1)}{2h^4}
+O\bigg(\frac{1}{h^6}\bigg).
\end{eqnarray*}
Since
$$
{\cal D}_qA_q=0, \;\; {\cal D}_{q+2i}A_{q+2i}=0 
\;\;\; {\rm and}\;\;\;
{\cal D}_{q+2i}={\cal D}_q+2iz_+,
$$
we have 
$$
{\cal D}_q\bigg(\frac{A_{q+2i}}{-2iz_+}\bigg)=A_{q+2i},
$$
except, of course, for $i=0$. The first approximation $A^{(0)}=A_q$
leaves uncompensated on the right hand side of Eq.~(\ref{91a})
the contribution $R^{(0)}_q$. Terms $\mu A_{q+2i}$ in this may
therefore be eliminated by adding to $A^{(0)}$ the contribution
$A^{(1)}$ given by
\begin{eqnarray}
A^{(1)}&=&\bigg(-\frac{\sqrt{2}c}{2^3h^4}\bigg)\bigg[
\frac{(q,q-4)}{8z_+}A_{q-4}+\frac{(q,q-2)}{4z_+}A_{q-2}
\nonumber\\
&& + \frac{(q,q+2)}{-4z_+}A_{q+2}
+\frac{(q,q+4)}{-8z_+}A_{q+4}+\cdots\bigg].
\label{C.15}
\end{eqnarray}
The sum $A=A^{(0)}+ A^{(1)}$ then  represents a solution to that order
provided the sum of terms in $A_q$ in 
$R^{(0)}_q$ and $R^{(1)}_q$ is set equal to zero,
where $R^{(1)}_q$ is the sum of terms left uncompensated
by $A^{(1)}$, i.e.
\begin{equation}
R^{(1)}_q=\bigg(-\frac{\sqrt{2}{c}}{2^3h^4}\bigg)\bigg[
\frac{(q,q-4)}{8z_+}R^{(0)}_{q-4}+\frac{(q,q-2)}{4z_+}R^{(0)}_{q-2}
+\cdots\bigg].
\label{C.16}
\end{equation}
This coefficient of $A_q$ set equal to zero yields to that order
the following equation
\begin{eqnarray*}
0&=&\bigg(-\frac{\sqrt{2}{c}}{2^3h^4}\bigg)(q,q)
+ \bigg(-\frac{\sqrt{2}{c}}{2^3h^4}\bigg)^2\bigg[
\frac{(q,q-4)(q-4,q)}{8z_+}
\nonumber\\
&& +\frac{(q,q-2)(q-2,q)}{4z_+}
 +\frac{(q,q+2)(q+2,q)}{-4z_+}
\nonumber\\
&& \qquad \quad  +\frac{(q,q+4)(q+4,q)}{-8z_+}\bigg]+\cdots,
\end{eqnarray*}
which reduces to
\begin{equation}
0=2(3q^2+1)+\frac{2}{c^2}\triangle
+\frac{\sqrt{2}c^2}{2h^6}q(17q^2+19),
\label{C.17}
\end{equation}
thus yielding the next approximation of $\triangle$.

\setcounter{equation}{0}
\renewcommand{\theequation}{D.\arabic{equation}}

\section{Calculation of 
eigenvalues with solutions of type 
${\boldsymbol B}$ (double well)}

Equation (\ref{C.1}) or (\ref{C.17})  can also be
obtained in conjunction with the solutions of
types $B$ or $C$.  The initial step is to use the
recurrence relation of parabolic cylinder functions
$D_{\nu}(w)$ to reexpress the right hand side of 
Eq.~(\ref{85a}) as a linear combination of
functions $B_q(w)$ defined by Eq.~(\ref{98a}).
Thus with the recurrence relation
\begin{equation}
wD_{\nu}(w)=D_{\nu+1}(w)+\nu D_{\nu-1}(w)
\label{D.1}
\end{equation}
and the expression (\ref{98a}), i.e.
\begin{equation}
B_q(w)=\frac{D_{\frac{1}{2}(q-1)}(w)}
{[\frac{1}{4}(q-1)]!2^{\frac{1}{4}(q-1)}},
\label{D.2}
\end{equation}
we obtain
\begin{equation}
wB_q(w)=\sqrt{2}
\frac{[\frac{1}{4}(q+1)]!}{[\frac{1}{4}(q-1)]!}B_{q+2}
+
\frac{[\frac{1}{4}(q-3)]!}{[\frac{1}{4}(q-5)]!}B_{q-2}.
\label{D.3}
\end{equation}

The procedure is then similar to that in the case
of the solution of type $A$.

\setcounter{equation}{0}
\renewcommand{\theequation}{E.\arabic{equation}}

\section{Recalculation of 
tunneling deviation using WKB solutions}

Here we
determine the tunneling deviation  $q-q_0$ of $q$
 from an odd 
integer $q_0$, above  obtained as 
Eq.~(\ref{116}),
by now using  the periodic  WKB solutions
below the turning points.

 The turning points at $z_0$ and $z_1$
on either side of the minimum at $z_+$
are given by Eqs.~(\ref{119}) and (\ref{120}).
We start from Eq.~(\ref{131}), i.e.
\begin{equation}
y_{\pm}(z)=\frac{1}{2}[y_A(z)\pm{\overline y}_A(z)]
=\frac{1}{2\gamma}y^{(l,z_0)}_{\rm WKB}(z)\pm
\frac{1}{2{\overline \gamma}}
{\overline y}^{(l,z_0)}_{\rm WKB}(z).
\label{E.1}
\end{equation}
Different from above we now continue the solutions
(in the sense of linearly matched WKB solutions)
across the turning point at $z_0$ in the direction
of the minimum of the potential at $z_+$. Then
($(r,z_0)$ meaning to the right of $z_0$
and note the asymmetric factor 
of $2$)\footnote{\scriptsize See Ref.~\cite{22}, p. 291, Eqs.~(21), (22)
or Ref.~\cite{26}, Vol. I, Sec. 6.2.4.}
\begin{eqnarray}
y_{\pm}(z)&=&
\frac{1}{2\gamma}y^{(r,z_0)}_{\rm WKB}(z)\pm
\frac{1}{2{\overline \gamma}}
{\overline y}^{(r,z_0)}_{\rm WKB}(z) 
=\bigg[\frac{1}{2}qh^2_+-\frac{1}{4}h^4_+U(z)\bigg]^{-1/4}
\nonumber\\
&& \times \bigg\{\frac{1}{\gamma}2\sin
\bigg[\int^z_{z_0}dz\bigg(\frac{1}{2}qh^2_+-\frac{1}{4}h^4_+U(z)
\bigg)^{1/2}+\frac{\pi}{4}\bigg]
\nonumber\\
&& \pm\frac{1}{\overline \gamma}\cos\bigg[
\int^z_{z_0}dz\bigg(\frac{1}{2}qh^2_+-\frac{1}{4}h^4_+U(z)
\bigg)^{1/2}+\frac{\pi}{4}\bigg]\bigg\}.
\label{E.2}
\end{eqnarray}
We also note that
\begin{eqnarray}
\frac{d}{dz}y_{\pm}(z)&\simeq & 
\bigg[\frac{1}{2}qh^2_+-\frac{1}{4}h^4_+U(z)\bigg]^{1/4}
\nonumber\\
&& \times \bigg\{\frac{1}{\gamma}2\cos
\bigg[\int^z_{z_0}dz\bigg(\frac{1}{2}qh^2_+-\frac{1}{4}h^4_+U(z)
\bigg)^{1/2}+\frac{\pi}{4}\bigg]
\nonumber\\
&& \mp \frac{1}{\overline \gamma}\sin\bigg[
\int^z_{z_0}dz\bigg(\frac{1}{2}qh^2_+-\frac{1}{4}h^4_+U(z)
\bigg)^{1/2}+\frac{\pi}{4}\bigg]\bigg\}.
\label{E.3}
\end{eqnarray}
We now apply the boundary conditions (\ref{105}) and
(\ref{106}) at the minimum
$z_+$  and obtain the conditions:
\begin{eqnarray}
0 &=& \frac{2}{\gamma}\sin
\bigg[\int^{z_+}_{z_0}dz\bigg(\frac{1}{2}qh^2_+-\frac{1}{4}h^4_+U(z)
\bigg)^{1/2}+\frac{\pi}{4}\bigg]
\nonumber\\
&& \pm\frac{1}{\overline \gamma}\cos\bigg[
\int^{z_+}_{z_0}dz\bigg(\frac{1}{2}qh^2_+-\frac{1}{4}h^4_+U(z)
\bigg)^{1/2}+\frac{\pi}{4}\bigg]
\label{E.4}
\end{eqnarray}
and
\begin{eqnarray}
0 &=& \frac{2}{\gamma}\cos
\bigg[\int^{z_+}_{z_0}dz\bigg(\frac{1}{2}qh^2_+-\frac{1}{4}h^4_+U(z)
\bigg)^{1/2}+\frac{\pi}{4}\bigg]
\nonumber\\
&& \mp\frac{1}{\overline \gamma}\sin\bigg[
\int^{z_+}_{z_0}dz\bigg(\frac{1}{2}qh^2_+-\frac{1}{4}h^4_+U(z)
\bigg)^{1/2}+\frac{\pi}{4}\bigg].
\label{E.5}
\end{eqnarray}
Hence
\begin{equation}
{\rm tan}
\bigg[\int^{z_+}_{z_0}dz\bigg(\frac{1}{2}qh^2_+-\frac{1}{4}h^4_+U(z)
\bigg)^{1/2}+\frac{\pi}{4}\bigg]\simeq\mp\frac{\gamma}
{2{\overline \gamma}}
\label{E.6}
\end{equation}
and
\begin{equation}
{\rm cot}
\bigg[\int^{z_+}_{z_0}dz\bigg(\frac{1}{2}qh^2_+-\frac{1}{4}h^4_+U(z)
\bigg)^{1/2}+\frac{\pi}{4}\bigg]\simeq\pm\frac{\gamma}
{2{\overline \gamma}}
\label{E.7}
\end{equation}
In the present considerations we approach the
minimum of the potential at $z_+$ by coming
from the left, i.e. from $z_0$.  We could, of course, 
approach the minimum also from the right, i.e.
from $z_1$.  Then at any point $z\in(z_0,z_1)$ we
expect\footnote{\scriptsize See e.g. Ref.~ \cite{26},
Sec. 6.2.6.}
\begin{equation}
|y^{(r,z_0)}_{\rm WKB}(z)|=
|y^{(l,z_1)}_{\rm WKB}(z)|, \;\;\;
|{\overline y}^{(r,z_0)}_{\rm WKB}(z)|=
|{\overline y}^{(l,z_1)}_{\rm WKB}(z)|.
\label{E.8}
\end{equation}
Choosing the point $z$ to be $z_+$, this implies
\begin{eqnarray}
&& \bigg|\begin{array}{c}
\sin \\  \cos
\end{array}
\bigg(\int^{z_+}_{z_0}dz\bigg[\frac{1}{2}qh^2_+-\frac{1}{4}h^4_+U(z)
\bigg]^{1/2}+\frac{\pi}{4}\bigg)\bigg|
\nonumber\\
&=&
\bigg|\begin{array}{c}
\sin \\  \cos
\end{array}
\bigg(\int^{z_1}_{z_+}dz\bigg[\frac{1}{2}qh^2_+-\frac{1}{4}h^4_+U(z)
\bigg]^{1/2}+\frac{\pi}{4}\bigg)\bigg|.
\label{E.9}
\end{eqnarray}
Thus e.g.
\begin{eqnarray*}
&&  \cos
\bigg[\int^{z_1}_{z_+}dz\bigg(\frac{1}{2}qh^2_+-\frac{1}{4}h^4_+U(z)
\bigg)^{1/2}+\frac{\pi}{4}\bigg]
\nonumber\\
&=&\cos\bigg[\int^{z_1}_{z_0}\cdots -\int^{z_+}_{z_0}+\frac{\pi}{4}\bigg]
 =
\cos\bigg[\int^{z_1}_{z_0}\cdots -\frac{\pi}{4}
 -\int^{z_+}_{z_0}+\frac{\pi}{2}\bigg]\nonumber\\
&=&
-\sin\bigg[\int^{z_1}_{z_0}\cdots -\frac{\pi}{4}
 -\int^{z_+}_{z_0}\cdots \bigg] =-(-1)^N
\cos\bigg[\int^{z_+}_{z_0}\cdots +\frac{\pi}{4}\bigg]
\end{eqnarray*}
provided the Bohr--Sommerfeld--Wilson
quantisation condition holds, i.e.
\begin{equation}
\int^{z_1}_{z_0} dz\bigg(\frac{1}{2}qh^2_+-\frac{1}{4}h^4_+U(z)\bigg)^{1/2}
=(2n+1)\frac{\pi}{2}, \;\;\; n=0,1,2,3,\ldots,
\label{E.10}
\end{equation}
where it is understood that (cf. Eq.~(83))  
$$
\frac{1}{2}qh^2_+\simeq E - V(z_{\pm}).
$$
Similarly under the same condition
$$
\sin\bigg[\int^{z_1}_{z_+}\cdots +\frac{\pi}{4}\bigg]=(-1)^N\sin
\bigg[\int^{z_+}_{z_0}\cdots +\frac{\pi}{4}\bigg].
$$
We rewrite the quantisation condition in the present context
and in view of the symmetry of the potential
 in the immediate vicinity of $z_+$ as
half  that of Eq.~(\ref{E.10}), i.e. 
\begin{equation}
\int^{z_+}_{z_0} dz
\bigg(\frac{1}{2}qh^2_+-\frac{1}{4}h^4_+U(z)\bigg)^{1/2}\simeq
q\frac{\pi}{4}.
\label{E.11}
\end{equation}
The integral on the left can be approximated by
$$
\frac{h^2_+}{2}\int^{z_+}_{z_0}dz
\bigg(\frac{2q}{h^2_+}-(z-z_+)^2\bigg)^{1/2}
=\frac{q}{2}{\sin}^{-1}\frac{z_+-z_0}{(2q/h^2_+)^{1/2}}=\frac{q}{4}\pi
$$
in agreement with the right hand side (the last step following
from Eqs.~(\ref{119})  and (\ref{120})).
We now see that Eqs.~(\ref{E.6}) and (\ref{E.7}) assume 
a form as in our perturbation theory, i.e. they become
\begin{equation}
{\rm tan}\bigg\{(q+1)\frac{\pi}{4}\bigg\}\simeq \mp\frac{\gamma}
{2{\overline \gamma}}, \;\;\; 
{\rm cot}\bigg\{(q+1)\frac{\pi}{4}\bigg\}\simeq \mp\frac{\gamma}
{2{\overline \gamma}}.
\label{E.12}
\end{equation}
The equations corresponding to these in  Ref. \cite{18}
are there described as WKB quantisation 
conditions of the double minimum potential. 
Since
\begin{eqnarray}
&& {\rm tan}\bigg\{(q+1)\frac{\pi}{4}\bigg\}=0\;\;\;{\rm for}
\;\;\;
q=q_0=3,7,11,\ldots,
\nonumber\\
&& 
{\rm cot}\bigg\{(q+1)\frac{\pi}{4}\bigg\}
=0\;\;\; {\rm for} \;\;\; q=q_0=1,5,9,\ldots,
\label{E.13}
\end{eqnarray}
we can expand the left hand sides about these  points and thus obtain
\begin{equation}
q-q_0\simeq \mp\frac{2\gamma}{\pi{\overline \gamma}}
\stackrel{(\ref{130})}{=}
\mp 4\sqrt{\frac{1}{2\pi}}\frac{(h^2_+)^{q_0/2}
(2z_+)^{q_0}}{[\frac{1}{2}(q_0-1)]!}e^{-\frac{1}{2}h^2_+z^2_+}
 \;\;
{\rm for} \;\; q_0=1,3,5,\ldots
\label{E.14}
\end{equation}
in agreement with Eq.~(\ref{116}).
We note here incidentally that this agreement demonstrates the
significance of the factor of $2$ in the first equality of  
Eq.~(\ref{E.14}) which results from the factor
of $2$ in front of the sine
in the WKB formula (\ref{E.2}).

\setcounter{equation}{0}
\renewcommand{\theequation}{F.\arabic{equation}}

\section{Evaluation of WKB exponential with  elliptic integral  }

Here we evaluate the elliptic integral $I_2(0)$ of Eq.~(\ref{135a}).

\subsection{Parameters and their expansions}

The expression to be evaluated is
\begin{equation}
I_2=\frac{2}{3G^2}(1+u)^{1/2}[E(k)-uK(k)].
\label{F.1}
\end{equation}
We have
\begin{equation}
k^2=\frac{1-u}{1+u},  \;\;\;
u= G\sqrt{2q},  \;\;\; q\simeq 2n+1. 
\label{F.2}
\end{equation}
Hence we obtain for $G$ close to zero:
\begin{equation}
k^2 = \frac{1-G\sqrt{2q}}{1+G\sqrt{2q}}
= 1-2G\sqrt{2q}+4G^2q-\cdots, \;\;\; k^2\sim 1,
\label{F.3}
\end{equation}
and
\begin{equation}
(1+u)^{1/2} 
= (1+G\sqrt{2q})^{1/2}= 1+G\sqrt{\frac{q}{2}}
-\frac{1}{4}G^2q+\cdots ,
\label{F.4}
\end{equation}
and
\begin{equation}
{k^{\prime}}^2 = 1-k^2
= 2G\sqrt{2q}-4G^2q+\cdots, \;\;\; 
{k^{\prime}}^2\sim 0.
\label{F.5}
\end{equation}
We now reexpress various quantities in terms of $u$.
Thus
\begin{equation}
{k^{\prime}}^2=2u-2u^2+\cdots,
\label{F.6}
\end{equation}
and hence
\begin{equation}
k^{\prime} = \sqrt{2u}(1-u)^{1/2}
= \sqrt{2u}\bigg(1-\frac{u}{2}+\cdots \bigg).
\label{F.7}
\end{equation}
The following expression appears  frequently
 in the  expansions of elliptic
integrals.  Therefore it is convenient to deal
with this here. We have
\begin{equation}
\frac{4}{k^{\prime}}=\frac{4}{\sqrt{2u}(1-\frac{u}{2}+\cdots )}
=\frac{4}{\sqrt{2u}}\bigg(1+\frac{u}{2}+\cdots \bigg).
\label{F.8}
\end{equation}
Hence
\begin{equation}
\ln\bigg(\frac{4}{k^{\prime}}\bigg)
=\ln\bigg[\frac{4}{\sqrt{2u}}\bigg(1+\frac{u}{2}-\cdots\bigg)\bigg]
\simeq  \ln\bigg(\frac{4}{\sqrt{2u}}\bigg)+\frac{u}{2}.
\label{F.9}
\end{equation}

\subsection{Evaluation of elliptic integrals}

Our next objective is the evaluation of the elliptic
integrals $E(k)$ and
$K(k)$  by expanding these in ascending powers of
$ {k^{\prime}}^2$, which is assumed to be small.
We obtain the expansions from Ref.~\cite{29}  as
\begin{equation}
E(k)=1+\frac{1}{2}\bigg[\ln\bigg(\frac{4}{k^{\prime}}\bigg)
-\frac{1}{2}\bigg]{k^{\prime}}^2
+\frac{3}{16}\bigg[\ln\bigg(\frac{4}{k^{\prime}}\bigg)
-\frac{13}{12}\bigg]{k^{\prime}}^4+\cdots
\label{F.10}
\end{equation}
and
\begin{equation}
K(k)=\ln\bigg(\frac{4}{k^{\prime}}\bigg)
+\frac{1}{4}\bigg[\ln\bigg(\frac{4}{k^{\prime}}\bigg)
-1\bigg]{k^{\prime}}^2+\cdots.
\label{F.11}
\end{equation}

Consider first $E(k)$:
\begin{eqnarray*}
E(k)
&=& 
1+\frac{1}{2}\bigg[\ln\bigg(\frac{4}{\sqrt{2u}}\bigg)
+\frac{u}{2}-\frac{1}{2}\bigg]2u(1-u)\nonumber\\
&& +\frac{3}{16}\bigg[\ln\bigg(\frac{4}{\sqrt{2u}}\bigg)
+\frac{u}{2}
-\frac{13}{12}\bigg]4u^2(1-u)^2+\cdots
\nonumber\\
&=&1+
\ln\bigg(\frac{4}{\sqrt{2u}}\bigg)\bigg\{u(1-u)+\frac{3}{4}u^2(1-u)^2\bigg\}
\nonumber\\
&&
+\frac{1}{2}(u-1)u(1-u)+\frac{3}{4}\frac{1}{2}\bigg(u
-\frac{13}{6}\bigg)u^2(1-u)^2+\cdots\nonumber\\
\end{eqnarray*}
We can rewrite this as
\begin{eqnarray}
E(k)& =&
1+
\ln\bigg(\frac{4}{\sqrt{2u}}\bigg)
\frac{u(1-u)}{4}\{4+3u(1-u)\}\nonumber\\
&&
+\frac{1}{2^4 3}u(u-1)(1-u)\{24-3u(6u-13)\}.
\label{F.12}
\end{eqnarray}

Analogously we have
\begin{eqnarray*}
uK(k)
&=& u\bigg[\ln\bigg(\frac{4}{\sqrt{2u}}\bigg)
+\frac{u}{2}\bigg]
+\frac{u}{4}\bigg[\ln\bigg(\frac{4}{\sqrt{2u}}\bigg)
+\frac{u}{2}-1\bigg]2u(1-u)\nonumber\\
&=&
\ln\bigg(\frac{4}{\sqrt{2u}}\bigg)\bigg\{u+\frac{1}{2}u^2(1-u)\bigg\}
+\frac{1}{2}u^2\bigg[1+(1-u)\bigg(\frac{u}{2}-1\bigg)\bigg]
\nonumber\\
&=& \ln\bigg(\frac{4}{\sqrt{2u}}\bigg)
\frac{1}{2}\{u^2(1-u)+2u\}
 +\frac{1}{4}u^2[2+(1-u)(u-2)]
\end{eqnarray*}
or
\begin{eqnarray}
uK(k)
&=&
 \ln\bigg(\frac{4}{\sqrt{2u}}\bigg)
\frac{u}{2}\{u(1-u)+2\}
 +\frac{1}{4}u^2[-u^2+3u]
\nonumber\\
&=&
\ln\bigg(\frac{4}{\sqrt{2u}}\bigg)
\frac{u}{2}\{u(1-u)+2\}
-\frac{1}{4}u^3(u-3).
\label{F.13}
\end{eqnarray}
With (\ref{F.12}) and (\ref{F.13}) we obtain
\begin{eqnarray}
&&E(k)-uK(k)\nonumber\\
&=&
1+\ln\bigg(\frac{4}{\sqrt{2u}}\bigg)
\bigg[\frac{1}{4}u(1-u)\{4+3u(1-u)\}
-\frac{1}{2}u\{u(1-u)+2\}\bigg]
\nonumber\\
&& +\frac{1}{2}u(u-1)(1-u)
\frac{1}{2^33}\{24-3u(6u-13)\}
+\frac{1}{4}u^3(u-3)\nonumber\\
&=&
1+\ln\bigg(\frac{4}{\sqrt{2u}}\bigg)\frac{u}{4}
[(1-u)\{4+3u(1-u)\}-2\{u(1-u)+2\}]
\nonumber\\
&&
+\frac{1}{2.4.6}u(u-1)(1-u)\{24-3u(6u-13)\}
+\frac{1}{4}u^3(u-3).
\label{F.14}
\end{eqnarray}
Now consider the last line here without the factor $2.4.6$
in the denominator and in the last step
 pick out the lowest order
terms in u (i.e. up to and including $u^2$):
\begin{eqnarray}
&&
u(u-1)(1-u)\{24-3u(6u-13)\}
+12u^3(u-3)\nonumber\\
&=& -u(1-u)^2\{24-18u^2+39u\}+12u^3(u-3)\nonumber\\
&=& -3u(1-u)^2\{8-6u^2+13u\}+12u^3(u-3)\nonumber\\
&\simeq &
-3u.8+6u^2.8 -3.13u^2\nonumber\\
&=& -3u.8 +9u^2.
\label{F.15}
\end{eqnarray}
Now consider the bracket $[...]$ in (\ref{F.14}), i.e.
\begin{equation}
 [(1-u)\{4+3u(1-u)\}-2\{u(1-u)+2\}]
= u(3u^2-4u-3).
\label{F.16}
\end{equation}
From (\ref{F.13}) with  (\ref{F.14}) and (\ref{F.15})  we now obtain
\begin{equation}
E(k)-uK(k)
\simeq 
1+\ln\bigg(\frac{4}{\sqrt{2u}}\bigg)\frac{1}{4}u^2(3u^2-4u-3)-
\frac{1}{2}u+\frac{9u^2}{2.4.6}.
\label{F.17}
\end{equation}

\subsection{Evaluation of $I_2$}

We now return to Eq.~(\ref{F.1}), i.e.
$$
I_2=\frac{2}{3G^2}(1+u)^{1/2}[E(k)-uK(k)].
$$
Inserting here from the above expansions the
contributions up to and including those
of order $u^2$, we obtain:
\begin{eqnarray}
I_2&\simeq &
\frac{2}{3G^2}\bigg(1+\frac{u}{2}-\frac{1}{8}u^2+\cdots\bigg)
\bigg[1-\ln\bigg(\frac{4}{\sqrt{2u}}\bigg)\frac{u^2}{4}(3+4u-3u^2)
\nonumber\\
&& \qquad
-\frac{1}{2}u+\frac{3u^2}{16} +O(u^3)\bigg].
\label{F.18}
\end{eqnarray}
Remembering that $u=G\sqrt{2q}$, this becomes
\begin{eqnarray}
I_2  &\simeq &
\frac{2}{3G^2}-\frac{u^2}{2G^2}
\ln\bigg(\frac{4}{\sqrt{2u}}\bigg)
+\frac{2}{3G^2}\bigg(-\frac{3u^2}{8}\bigg)
+\frac{u^2}{8G^2}
\nonumber\\
&=& \frac{2}{3G^2}-q\ln\bigg(\frac{4}{2^{1/2}G^{1/2}2^{1/4}q^{1/4}}
\bigg)-\frac{q}{4}\nonumber\\
&=& \frac{2}{3G^2}-\frac{q}{2}\ln\bigg(\frac{2^{5/2}}{Gq^{1/2}}\bigg)
-\frac{q}{4}.
\label{F.19}
\end{eqnarray}
Thus
\begin{eqnarray}
I_2 &\simeq &
\frac{2}{3G^2}+\frac{q}{2}\ln\bigg(\frac{G}{4}\bigg)
-\frac{q}{4}\ln\bigg(\frac{2}{q}
\bigg)-\frac{q}{4}\nonumber\\
&=&
\frac{2}{3G^2}+\frac{1}{2}(2n+1)\ln\bigg(\frac{G}{4}\bigg)
+\frac{1}{4}(2n+1)\ln\bigg(\frac{2n+1}{2}\bigg)
\nonumber\\
&& \qquad \qquad -\bigg(\frac{2n+1}{4}\bigg).
\nonumber\\
\label{F.20}
\end{eqnarray}

\end{appendix}

\end{document}